\documentclass[aps,prl,reprint,superscriptaddress,nofootinbib,longbibliography]{revtex4-2}

\usepackage{iftex}
\ifPDFTeX
  \usepackage[T1]{fontenc}
  \usepackage[utf8]{inputenc}
  \usepackage{lmodern}
\fi
\usepackage{amsmath,amssymb,mathtools,bm}
\usepackage{microtype}
\usepackage{xcolor}
\definecolor{darkred}{rgb}{0.55,0.0,0.0}
\usepackage{graphicx}
\usepackage{tikz}
\usetikzlibrary{arrows.meta,calc,decorations.pathmorphing,decorations.pathreplacing,positioning,fit}
\usepackage[colorlinks=true,allcolors=blue]{hyperref}
\usepackage{orcidlink}

\makeatletter
\newcommand{\EMappendix}[2]{%
  \setcounter{equation}{0}%
  \renewcommand{\theequation}{#1\arabic{equation}}%
  \@ifundefined{theHequation}{}{\renewcommand{\theHequation}{em#1.\arabic{equation}}}%
  \emph{Appendix #1: #2.---}}
\makeatother

\newif\ifincludesupplement
\includesupplementtrue

\DeclareMathOperator{\Tr}{Tr}
\DeclareMathOperator{\Fix}{Fix}

\newcommand{\Id}{\mathbf 1}
\newcommand{\dd}{\mathrm d}
\newcommand{\e}{\mathrm e}
\newcommand{\ii}{\mathrm i}
\newcommand{\ket}[1]{\lvert #1\rangle}
\newcommand{\bra}[1]{\langle #1\rvert}

\newcommand{\avg}[1]{\left\langle #1\right\rangle}
\newcommand{\cavg}[1]{\left\langle #1\right\rangle_{c}}

\newcommand{\nTL}{\mathfrak n}
\newcommand{\cT}{\mathcal T}

\newcommand{\bfm}{\mathbf m}
\newcommand{\bigO}[1]{O\!\left(#1\right)}

\begin{document}

\title{Statistical Mechanics of Non-Abelian Learnability Transitions}

\newcommand{\kitp}{
Kavli Institute for Theoretical Physics, University of California, Santa Barbara, California 93106, USA
}

\newcommand{\unige}{
  Department of Theoretical Physics, University of Geneva,
  24 quai Ernest-Ansermet, 1211 Gen\`eve, Switzerland
}

\newcommand{\gqc}{
  Geneva Quantum Center, University of Geneva
}

\author{Ruochen Ma \:\orcidlink{0000-0003-0640-2513}}
\affiliation{\kitp}

\author{Romain Vasseur \:\orcidlink{0000-0002-4636-4139}}
\affiliation{\unige}
\affiliation{\gqc}

\date{\today}

\begin{abstract}

Monitored many-body quantum systems can undergo sharp learnability transitions characterized by how much information can be learned by the observer. When the dynamics conserves a non-Abelian charge, such as an $SU(2)$ spin, understanding how the observer learns the total charge remains an outstanding problem. Unlike the Abelian case, where charge measurements on distinct sites commute, the $SU(2)$-symmetric readouts are noncommuting fusion measurements, making learning a genuinely quantum inference problem. We propose a theory of $1+1d$ monitored quantum dynamics with $SU(2)$ symmetry, and show that it can be described by an effective replicated loop model comprised of  a replica-pairing field and a diffusive ($z=2$) background sector that carries the $SU(2)$ charge and remains gapless throughout the phase diagram.
Our theory predicts that the ``spin-sharpening'' and entanglement transitions coincide as a {\em single} transition, which is supported by numerics. Ordering of the pairing field produces volume-law entanglement and hides the background sector from measurements, leading to a learning time of $t\sim L^{3}$ for the total spin.  When the pairing field disorders, the background sector alone gives logarithmic entanglement and a diffusive
learning time $t\sim L^{2}$.  
Our analysis is controlled by a
large-loop-fugacity expansion.

\end{abstract}

\maketitle

\emph{Introduction.---}Advances in noisy intermediate-scale quantum (NISQ)
simulators~\cite{Preskill2018,HuangKuengPreskill2020} pose a fundamental inferential question:  from local measurements and limited partial information, what can an observer learn about a many-body quantum system? The most prominent example is the learnability, or decoding, viewpoint of measurement-induced
phase transitions (MIPTs) in monitored circuits, where unitary dynamics are interspersed with
local
measurements~\cite{Li2018,Chan2019,Skinner2019,Szyniszewski2019,Li2019,Choi2020,GullansHuse2020,GullansHuse2020b,Zabalo2020,Bao2020,Jian2020,NahumRoy2021,Ippoliti2021,Lavasani2021,SangHsieh2021,Alberton2021,Buchhold2021,Zabalo2022,Fava2023}: states in the volume- and area-law phases have distinct entanglement scalings, and they are also distinct in whether the unitary dynamics can, or cannot, hide
the quantum information from the local
measurements~\cite{IppolitiKhemani2024,LiCrossEntropy2023,Dehghani2023}. (See
Refs.~\cite{Fisher2023,PotterVasseur2022,2026Vasseurnotes} for reviews.) When the dynamics preserves a global symmetry, the conserved charge becomes a natural target:
how quickly the record reads out its value changes parametrically with system size across the
charge-sharpening transition~\cite{Agrawal2022,BarrattLearn2022,AgrawalCollapse2024}. 

For an Abelian $U(1)$ charge, this learnability transition is by now well understood: exactly like the entanglement transition itself, charge sharpening admits a controlled statistical-mechanics
(stat-mech) description~\cite{Bao2020,Jian2020,Agrawal2022,BarrattCharge2022}, and it occurs
strictly inside the volume-law phase: the global charge becomes learnable at a measurement
rate well below the entanglement transition~\cite{Agrawal2022,BarrattCharge2022}. Physically, charge measurements on distinct sites commute and are simultaneously diagonal in a local basis, in which the charge dynamics reduces to a classical stochastic process for the charge
configuration that the record monitors~\cite{NahumJacobsen,Gopalakrishnan2025}. The charge-sharpening transition is then the pinning of inter-replica charge fluctuations in the stat-mech description~\cite{BarrattCharge2022}.

In contrast, learning a non-Abelian conserved charge, e.g., in an $SU(2)$-symmetric monitored circuit,
is far less understood. Numerically, the $SU(2)$-symmetric case behaves
differently from the $U(1)$ case~\cite{Majidy2023,Feng2025}: monitored spin-$1/2$ chains display a volume-law phase and a critical phase with logarithmic entanglement, separated by an
entanglement transition. One moreover finds a spin-sharpening transition, across which the
time to learn the total spin drops from $t\sim L^3$ in the spin-fuzzy phase to $t\sim L^2$ in
the spin-sharp phase. Strikingly, this spin-sharpening threshold is numerically
indistinguishable from the entanglement-transition threshold, $p_\sharp\approx p_c$~\cite{Majidy2023}, though whether the two genuinely coincide was left
open. The problem is nontrivial because the measurements allowed by a non-Abelian symmetry are fusion-channel measurements on bonds, which do not commute on overlapping
bonds. Each measurement therefore partially erases the information gathered by its
neighbors~\cite{MajidyNRP2023}, and no local basis diagonalizes the record. Largely for this
reason, the $SU(2)$ spin-sharpening transition has so far lacked a controlled theory: it is a
genuine quantum learning problem.

\begin{figure}[t]
\centering
\begin{tikzpicture}[xscale=0.44,yscale=0.34]
\node at (-1.6,4.9) {(a)};
\draw[->,line width=0.7pt] (-1.4,-0.5) -- (-1.4,4.4);
\node[left] at (-1.4,4.2) {$t$};
\foreach \x in {0,...,7}{\draw[line width=0.55pt,gray!60] (\x,-0.6) -- (\x,4.9);}
\foreach \y in {0,2.2}{\foreach \x in {0,2,4,6}{\draw[fill=gray!25,line width=0.6pt,rounded corners=1.2pt] (\x-0.32,\y) rectangle (\x+1.32,\y+0.85);}}
\foreach \y in {1.1,3.3}{\foreach \x in {1,3,5}{\draw[fill=gray!25,line width=0.6pt,rounded corners=1.2pt] (\x-0.32,\y) rectangle (\x+1.32,\y+0.85);}}
\draw[fill=red!14,draw=darkred,line width=0.7pt,rounded corners=1.2pt] (1.68,0) rectangle (3.32,0.85);
\draw[fill=red!14,draw=darkred,line width=0.7pt,rounded corners=1.2pt] (4.68,1.1) rectangle (6.32,1.95);
\draw[fill=red!14,draw=darkred,line width=0.7pt,rounded corners=1.2pt] (-0.32,2.2) rectangle (1.32,3.05);
\draw[fill=red!14,draw=darkred,line width=0.7pt,rounded corners=1.2pt] (2.68,3.3) rectangle (4.32,4.15);
\end{tikzpicture}

\vspace{2mm}

\begin{tikzpicture}[xscale=7.0,yscale=1.0,>=Stealth]
\node at (-0.06,1.55) {(b)};
\draw[->,line width=0.8pt] (0,0) -- (1.06,0);
\node[below] at (1.03,-0.03) {$p$};
\draw[line width=0.8pt] (0,0.05) -- (0,-0.05);
\node[below] at (0,-0.05) {$0$};
\draw[line width=0.9pt,darkred] (0.5,0.07) -- (0.5,-0.07);
\node[below,darkred] at (0.5,-0.05) {$p_{\sharp}$};
\node[align=center] at (0.235,0.85) {\small \textbf{spin fuzzy}\\[0.5pt]
\small $S_Q$ ordered\\[-0.5pt]
\small volume-law $S_A$\\[-0.5pt]
\small $t_{\rm learn}\sim L^{3}$};
\node[align=center] at (0.765,0.85) {\small \textbf{spin sharp}\\[0.5pt]
\small $S_Q$ disordered\\[-0.5pt]
\small $S_A \propto\log|A|$\\[-0.5pt]
\small $t_{\rm learn}\sim L^{2}$};
\draw[decorate,decoration={brace,mirror,amplitude=3.5pt}] (0.02,-0.5) -- (0.98,-0.5);
\node at (0.5,-0.78) {\small $z=2$ diffusive background sector};
\end{tikzpicture}
\caption{(a)~Brickwork monitored circuit. On each brick, an $SU(2)$-symmetric unitary $U(\theta)$ acts with probability $1-p$ (gray), or the two-spin fusion channel $\{s,t\}$ is projectively measured with probability $p$ (red). (b)~Phase diagram of the replicated loop model. The replica-pairing field $\sigma\in S_Q$ orders at small $p$ (spin-fuzzy phase) and disorders at large $p$ (spin-sharp phase). A $z=2$ diffusive background sector is present throughout: it mediates the ferromagnetic interaction that orders $\sigma$, remains critical in the sharp phase, and sets the diffusive learning time.}
\label{fig:overview}
\end{figure}
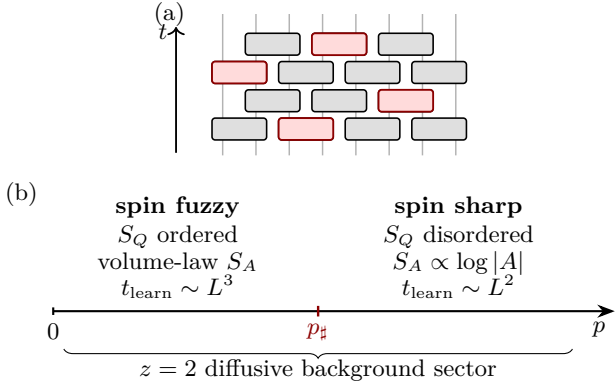

In this work, we develop a replica stat-mech theory for this non-Abelian learning problem. Via
a replica trick, the $SU(2)$ monitored circuit maps onto an interacting Temperley--Lieb (TL)
loop model with two coupled sectors: a discrete field of replica pairings, akin to the order
parameter of the MIPT in circuits without a global
symmetry~\cite{Bao2020,Jian2020,ZhouNahum2019}, and a diffusive ($z=2$) background sector that
carries the conserved spin quantum number and stays gapless throughout the phase diagram. This gaplessness explains the survival of nontrivial
dynamics even in the measurement-only limit. Within a controlled large-loop-fugacity
expansion, integrating out the background sector induces a ferromagnetic coupling between the
pairing variables that is short-ranged in the renormalization-group (RG) sense, and the
spin-sharpening transition is the order-disorder transition of this pairing sector. We argue
that the pairing order does double duty [Fig.~\ref{fig:overview}]: when ordered, it
produces volume-law entanglement and hides the diffusive background from the
record, so the total spin is learned only in a time $t_{\rm learn}\sim L^3/p$; when
disordered, the still-critical background sector produces logarithmic entanglement and the spin is learned in
$t_{\rm learn}\sim L^2$, the Thouless time of the diffusive sector. We thus propose that the
entanglement and spin-sharpening transitions of the $SU(2)$ chain are a single
transition, consistent with the numerics of Ref.~\cite{Majidy2023}. We further support this prediction by adding strongly symmetric decoherence, which removes the entanglement transition, and numerically observing that the signatures of the sharpening transition also disappear.

\emph{Model and diagnostics.---}Consider a periodic chain of an even number $L$ of spin-$1/2$ moments evolving under a brickwork circuit [Fig.~\ref{fig:overview}(a)]. On each brick we either apply an $SU(2)$-symmetric two-qubit unitary (probability $1-p$) or projectively measure the two-spin fusion channel (probability $p$). Up to an overall phase, the most general such unitary gate is $U(\theta)=P_s+\e^{\ii\theta}P_t$, where $P_s$ and $P_t$ are the projectors onto the singlet and triplet channels, and $\theta$ is drawn uniformly on $[0,2\pi)$. A measurement projects the two spins onto one of these two channels and records the outcome $m=s,t$. Every gate and measurement commutes with the total spin $\vec S=\frac12\sum_i\vec\sigma_i$, whose Casimir $\vec S^2$ has eigenvalues $S(S+1)$.

For a fixed circuit realization, namely the choice of unitary gates and measurement locations, an outcome history $\bfm$ defines an unnormalized trajectory $\rho_\bfm=K_\bfm\rho_0K_\bfm^\dagger$ with Born probability $p_\bfm=\Tr\rho_\bfm$, where $\rho_0$ is the initial state and $K_\bfm$ is the product of the applied gates and the projectors onto the observed outcomes. We study two diagnostics. The first is the trajectory-averaged $n$th R\'enyi entropy $S_A^{(n)}$ of a region $A$, computed via the replica trick~\cite{Bao2020,Jian2020} with $Q=nk+1$ replicas, $k\to0$; see the Supplemental Material (SM)~\cite{SuppMat} for the complete stat-mech treatment. The second is the learnability of the total spin from the record, quantified by the classical fidelity between the record distributions conditioned on two total-spin sectors:
\begin{equation}
F_t(S,S')
=
\mathbb{E}_{\rm circ}
\sum_{\bfm}
\left[p_{\bfm}(S,t)\,p_{\bfm}(S',t)\right]^{1/2}.
\label{eq:F-L}
\end{equation}
Here, $p_\bfm(S,t)$ is the probability of record $\bfm$ up to time $t$ given an initial state in sector $S$, and $\mathbb E_{\rm circ}$ averages over circuit realizations. The two sectors are asymptotically distinguishable when $F_t\to0$, and are identified with $O(1)$ confidence once $1-F_t=O(1)$~\cite{wilde2013quantum}. For analytic tractability we also use the R\'enyi generalizations of the fidelity, $F^{(q)}_t=Z^{(q)}_{S,S'}/\sqrt{Z^{(q)}_{S,S}\,Z^{(q)}_{S',S'}}$, built from the record overlaps $Z^{(q)}_{S,S'}=\mathbb E_{\rm circ}\sum_\bfm[p_\bfm(S,t)p_\bfm(S',t)]^q$.

\emph{Loop model and replica pairings.---}Following the stat-mech framework for monitored circuits~\cite{Nahum2017,ZhouNahum2019,Bao2020,Jian2020,2026Vasseurnotes}, we compute both diagnostics by contracting a replicated tensor network on the circuit spacetime with $Q$ replicas, each made of a forward and a backward branch. Each gate is a vertex and each qubit worldline a link, and the local tensor at a vertex is the circuit average of its unitary and measurement contributions, weighted by $1-p$ and $p$. The two contributions act differently: averaging the unitary phase $\theta$ enforces an equal number of singlet projectors in the forward and backward branches, whereas a projective measurement forces all $2Q$ branches to carry the same outcome. Expanding every fusion-outcome projector in the Temperley--Lieb generator $E_i=\nTL P_{s,i}$, with $E_i^2=\nTL E_i$, turns the contraction into the partition function of a fully packed, interacting loop model~\cite{TemperleyLieb1971}: the worldlines fragment into loops, each weighted by $\nTL=2$ in every branch~\cite{SuppMat}. Because the equal singlet counts in the forward and backward branches can be matched in many ways, the unitary contribution at each vertex decomposes into a sum over the pairings of backward to forward branches. A discrete ``spin'' $\sigma_v\in S_Q$, $\sigma_v:B\to F$, therefore emerges at every vertex of the stat-mech model. The partition function is invariant under the replica symmetry $S_Q^F\times S_Q^B$ that permutes the forward and backward branches independently~\cite{bao2021symmetry}. In the volume-law phase this symmetry breaks spontaneously to the diagonal, and the pairing $\sigma_v$ is the resulting order parameter.

To make the model tractable we analytically continue to large fugacity $\nTL$ at fixed integer $Q>1$, keeping the same local loop interaction. For integer $\nTL$ this is realized by $SU(\nTL)$ fundamental/anti-fundamental representations on odd and even sites, respectively, and the physical circuit is recovered at $\nTL=2$. Expanding the local tensor $\cT_v$ at each vertex in $1/\nTL$, we find the leading $\sigma$ dependence at $O(\nTL^{-2})$:
\begin{equation}
\begin{split}
\cT_v &= \sum_{\sigma_v\in S_Q}\cT_v(\sigma_v),\\
\cT_v(\sigma_v)
&=
\cT_{\rm bg}
+
g\sum_{a,b=1}^{Q}M_{ab}(\sigma_v)\,E^a_vE^{\bar b}_v
+\bigO{\nTL^{-3}},
\end{split}
\label{eq:sep-L}
\end{equation}
with $M_{ab}(\sigma)=\delta_{a,\sigma(b)}-1/Q$, $g=(1-p)/[(Q-1)!\,\nTL^2]$, and $E^a_v$ ($E^{\bar b}_v$) the unnormalized projector onto the $SU(\nTL)$ singlet of the two sites acted on by gate $v$ in forward branch $a$ (backward branch $\bar b$). Here $\cT_{\rm bg}$ is the $\sigma$-independent background loop dynamics, and $M_{ab}$ mediates the coupling between the pairing sector and the loops. Physically, the large-$\nTL$ continuation makes singlet fusion rare: two adjacent $SU(\nTL)$ spins fuse into the singlet ($\dim=1$) with probability $1/\nTL^{2}$, and into the adjoint ($\dim=\nTL^{2}-1$; the triplet at $\nTL=2$) with probability $1-1/\nTL^{2}$. This $O(1/\nTL^{2})$ rarity is what allows us to identify the leading couplings and makes the separation into replica-pairing and loop sectors controlled. We return to the limitations of this expansion in the discussion.

\emph{Background sector and induced interaction.---}At the measurement-only point $p=1$, up to $o(\nTL^{-2})$ corrections, the local tensor $\cT_{\rm bg}$ factorizes over the $2Q$ branches. In a continuous-time description, each branch then evolves in imaginary time under the ferromagnetic TL chain Hamiltonian $H_{\rm bg}=J\sum_x E_x$, $J>0$, which at $\nTL=2$ reduces to the Heisenberg ferromagnet. At long times, the dynamics of the background sector is governed by the low-lying spectrum of $H_{\rm bg}$, which is described by a spin-wave theory (with a small regulator lifting zero modes that do not continue to the physical value $\nTL=2$ \cite{SuppMat}): all modes disperse quadratically, $\omega\sim k^2$, so the loop background is a $z=2$ diffusive critical sector. Moving away from $p=1$ couples the branches at $O(\nTL^{-2})$, but only through operators that are irrelevant at this $z=2$ Gaussian fixed point, so the background remains $2Q$ weakly interacting spin-wave theories at low energies throughout our treatment.

Integrating out the background sector then generates an effective interaction in the pairing sector via a cumulant expansion. The first cumulant vanishes by the $S_Q^F\times S_Q^B$ symmetry of the background, and the second cumulant gives
\begin{equation}
S_{\rm eff}[\sigma]
=
-\frac{g^2}{2}\sum_{v,w}A(v,w)\left[\Fix(\sigma_w^{-1}\sigma_v)-1\right]+\cdots,
\label{eq:Seff-L}
\end{equation}
where $\Fix(\sigma)$ counts fixed points in the permutation $\sigma$, and $A(v,w)=C_E(v,w)^2\ge0$ is the squared connected correlator of the singlet density: the interaction is ferromagnetic. From the $z=2$ spin-wave theory, $A$ decays algebraically, but fast enough for the induced interaction to be summable \cite{SuppMat}. For a discrete ferromagnet in the two-dimensional spacetime, the pairing sector orders at small $p$ (large $g$) and disorders at large $p$~\cite{Peierls1936,Dobrushin1968}. We also show that all higher $\sigma$-dependent vertices contribute only at $O(\nTL^{-5})$, so Eq.~\eqref{eq:Seff-L} is the controlled leading induced interaction.

The spin-sharpening transition is identified with the order-disorder transition of the
pairing variable $\sigma$. Since the $\sigma$--spin-wave coupling is irrelevant at the $z=2$
fixed point, the critical theory factorizes in the infrared. Moreover, $A(x,\tau)$ decays fast enough that the induced interaction acts as a short-range one at the transition, by the Sak criterion~\cite{FisherMaNickel1972,Sak1973,SuppMat}. We thus conjecture that the transition is in the universality class of
the generic measurement-induced transition of monitored circuits without a global symmetry ~\cite{ZhouNahum2019,Bao2020,Jian2020}, tensored with the spectator critical
background sector.

\emph{Entanglement scaling.---}In the stat-mech language, $S_A^{(n)}$ is the replica-limit free-energy cost of twisted boundary conditions on $A$~\cite{Bao2020,Jian2020}. Two contributions arise, one from each sector. In the $S_Q$-ordered (fuzzy) phase, the boundary twist forces a domain wall in $\sigma$ with a finite line tension, giving a volume-law entropy. In the disordered (sharp) phase the twist costs no line tension in the pairing sector, but it still changes the boundary condition seen by the critical background sector, giving a logarithmic entanglement. The presence of this critical background throughout our framework accounts for the numerical observation of Ref.~\cite{Majidy2023} that the $SU(2)$-symmetric monitored circuit has no area-law phase.

Interestingly, the spin sector of $\rho_0$ selects the relevant ground state of $H_{\rm bg}$, which in turn sets the coefficient of the logarithmic entanglement contributed by the background. For simplicity, here we state the result at the physical value $\nTL = 2$. If $\rho_0$ carries a fixed total spin $S=O(1)$ (e.g., $S=0$, as in the numerics of Ref.~\cite{Majidy2023}), every replica branch is confined to the spin-$S$ sector throughout the dynamics. Within this sector, the ground state of the ferromagnetic chain is a Bloch-wall-like spiral texture~\cite{sutherland1995low,dhar2000bloch}, whose semiclassical representative is a unit vector winding once around the chain, at wavevector $q_0 = \frac{2\pi}{L}$ (End Matter). The late-time entanglement entropy of the background sector then comes from the Goldstone modes about this texture~\cite{MetlitskiGrover2011}, giving at $\nTL=2$~\cite{SuppMat}
\begin{equation}
S_A^{\rm bg}\big|_{S=0}=
\begin{cases}
\log|A|+O(1), & |A|\ll L^{2/3},\\[2pt]
\tfrac32\log L+O(1), & |A|=O(L),
\end{cases}
\label{eq:S0-L}
\end{equation}
independent of the R\'enyi index $n$. Physically, each broken generator contributes $1/2$ to the coefficient. A small region sees an almost uniform texture and detects only two broken generators, whereas a region of order the system size resolves the slow spiral winding and detects that all three $SU(2)$ generators are broken. When instead $\rho_0$ is the maximally mixed state, the late-time dynamics is governed by the uniform ferromagnet, which breaks only two generators, and the background sector gives $S_A^{\rm bg}=\log|A|+O(1)$ at $\nTL=2$.

\emph{Learning times.---}Two spin sectors become distinguishable at the time $t_{\rm learn}$ at which the record first carries an $O(1)$ amount of information about the sector, i.e., at which $-\log F^{(q)}_t$ becomes of order unity. This quantity separates, at any time $t$, into two pieces,
\begin{equation}
-\log F^{(q)}_t=\Delta\,t+\big[\mathcal T(0)-\mathcal T(t)\big],
\label{eq:F-split-L}
\end{equation}
which encode two ways the record can learn the spin. The first is a steady accumulation of information at rate $\Delta$, set by the leading eigenvalues of the time-direction transfer matrix of the replicated dynamics. The second is a transient part, which saturates to a constant $\mathcal{T}(0)$ after the Thouless time of the diffusive background (End Matter). Importantly, both terms vanish when the partition function $Z^{(q)}_{S,S'}$ factorizes over the replicas, since the fidelity [Eq.~\eqref{eq:F-L}] measures exactly the inter-replica correlations generated by the sum over the shared measurement record.

In the fuzzy ($S_Q$-ordered) phase, scrambling dominates. The final-time boundary condition acts as a symmetry-breaking field that selects the identity pairing between replicas~\cite{li2024statistical}, each forward branch glued to its own backward partner into the maximally mixed state of its spin sector. The leading eigenvector of the transfer matrix then factorizes over the replicas, and the transient vanishes, $\mathcal{T}(0) = 0$ up to $O(1/L)$ corrections: the $S_Q$ order hides the diffusive background from the record \cite{SuppMat}. The record therefore learns only through the first term in Eq.~\eqref{eq:F-split-L}. Perturbation theory about the unitary-only ($p=0$) limit gives, at first order in $p$,
\begin{equation}
\Delta\propto pL\,(r_S-r_{S'})^2,
\label{eq:bias-L}
\end{equation}
where $r_S$ is the expectation of a local swap operator in sector $S$. Physically, a symmetric local measurement distinguishes two sectors only through the Casimir density $S(S+1)/L^2$, so that $r_S-r_{S'}=O(L^{-2})$ for $S,S'=O(1)$. Resolving such a difference requires $O(L^4)$ outcomes by the central limit theorem, and the dynamics accumulates $O(pL)$ of them per unit time, so that $t_{\rm learn}\sim\Delta^{-1}\sim L^3/p$~\cite{Majidy2023}.

In the sharp phase, by contrast, the pairing field is disordered, and the remaining background sector undergoes an imaginary-time evolution with $z=2$. Learning then follows a cooling picture: a mode is resolved when it has frozen and contributes to the outcome statistics deterministically. The record thus resolves the background from short wavelengths to long, and the total spin, the uniform ($k=0$) component of the spin texture, is resolved once the diffusive dynamics has explored the whole system at the Thouless time $t\sim L^2$. Formally, the partition function no longer factorizes over the replicas, which are strongly correlated through the shared measurement record, and the transient is generically of order one~\cite{SuppMat}. The record then resolves the two sectors once the transient has saturated, giving $t_{\rm learn}\sim L^{2}$. Our large-$\nTL$ framework controls the phase structure and the diffusive time scale, but it does not yet determine $\mathcal{T}(0)$, which we leave to future work.

The order in the $S_Q$ pairing sector therefore simultaneously gives rise to the volume-law entanglement of the fuzzy phase and hides the diffusive sector from the observer. We thus propose that the entanglement transition and the spin-sharpening transition are a single transition, in marked contrast to the $U(1)$ circuit, where charge sharpening occurs strictly inside the volume-law phase~\cite{Agrawal2022,BarrattCharge2022}.

\emph{Adding decoherence.---}We support our conclusion that the two transitions coincide by introducing strongly symmetric decoherence into the dynamics. On each brick, with a small probability $p_d$, the fusion channel is measured and the outcome is discarded. Such decoherence explicitly breaks the replica symmetry $S_Q^F\times S_Q^B$ to its diagonal subgroup~\cite{Weinstein2022}, and hence removes the $S_Q$ ordering transition, which is the entanglement transition of the circuit with decoherence. Our theory predicts that the sharpening transition disappears with it, and that at every $p$ the spin is learned slowly, as in the fuzzy phase, where the replica symmetry is spontaneously broken. Note that when the sharpening transition is separate from the entanglement transition, as in a $U(1)$ circuit, a sharpening transition remains under strongly symmetric decoherence~\cite{Singh2026}.

\begin{figure}[t]
\centering
\includegraphics[width=\columnwidth]{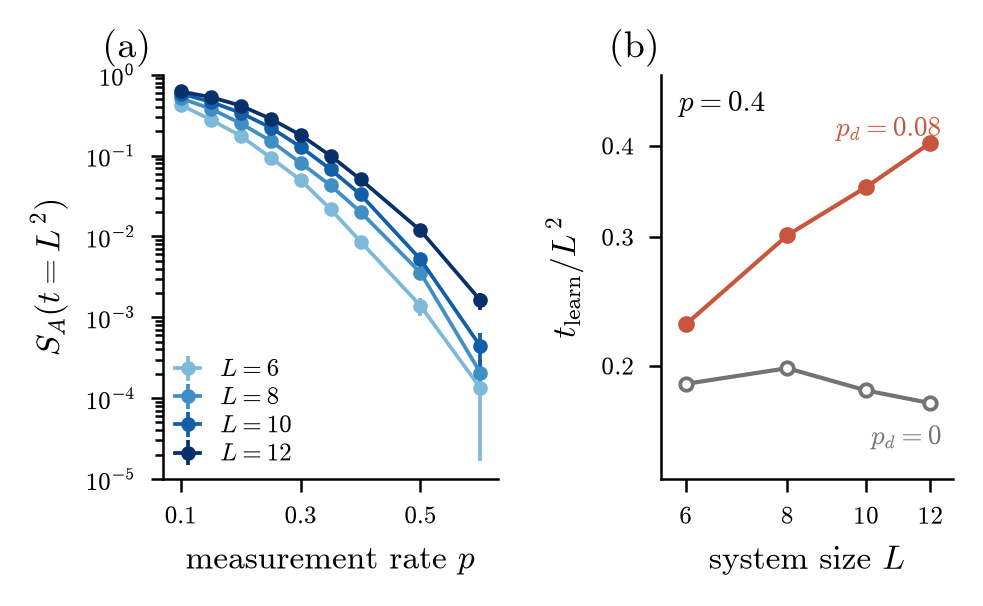}
\caption{Adding strongly symmetric decoherence ($p_d=0.08$). (a)~The curves of the ancilla entropy at $t=L^2$ do not cross. (b)~At $p=0.4$, the learning time with decoherence (filled) grows faster than $L^2$, the scaling without decoherence.}
\label{fig:decoherence}
\end{figure}

We test this prediction numerically with the protocol of Ref.~\cite{Majidy2023}. An ancilla qubit is entangled with two system states of total spin $S=0$ and $S=1$, and its trajectory-averaged entropy $S_A(t)$ measures the information about the spin not yet learned. Without decoherence, the curves $S_A(t=L^2)$ for different $L$ cross at $p_\sharp\approx0.28$~\cite{Majidy2023}, signaling the sharpening transition. With $p_d=0.08,$ there is no crossing at any $p$ where $S_A(t=L^2)$ can be resolved [Fig.~\ref{fig:decoherence}(a)]. At $p=0.4$, inside the sharp phase of the circuit without decoherence, we find the learning time $t_{\rm learn}\sim L^{2.81(5)}$, against $t_{\rm learn}\sim L^{1.91(8)}$ without decoherence [Fig.~\ref{fig:decoherence}(b); End Matter]. Both signatures of the sharp phase are therefore absent.

\emph{Discussion.---}We have developed a theory of how an observer learns a
non-Abelian charge in the $SU(2)$-symmetric monitored circuit, controlled by a large-$\nTL$ expansion. A single order parameter, the replica pairing, governs both the entanglement and the learnability of the total spin. The theory accounts for the key numerical observations of
Ref.~\cite{Majidy2023}: a volume-law phase and a critical phase with logarithmic entanglement,
no area-law phase, and learning times $t_{\rm learn}\sim L^3$
and $L^2$ in the two phases. It predicts that the entanglement and spin-sharpening
transitions coincide, $p_\sharp=p_c$, which our numerics with strongly symmetric decoherence support.

Our theory rests on two assumptions. The first is that the results of the large-loop-fugacity expansion are continuously connected to $\nTL=2$. For the background sector we can check this directly. At $\nTL=2$ and for any integer $Q>1$, the leading eigenvectors of the time-direction transfer matrix at $p=1$ are exactly the ferromagnetic states of the $2Q$ branches, and the symmetry allows only irrelevant couplings between the branches, which formally remain irrelevant as $Q\to1$~\cite{SuppMat}. The second assumption is that the replica limit $Q\to1$ commutes with the thermodynamic limit. This is the more subtle one. As stated above, when the thermodynamic limit is taken first, the background sector flows to $2Q$ weakly coupled branches at long distances for every $Q>1$, and formally also at $Q\to 1$. This is plausible in the fuzzy phase, where scrambling dominates and the measurements do not impose strong correlations between the replicas. In the measurement-dominated spin-sharp phase, by contrast, our theory might miss important physics. Indeed, the spin sharpening observed numerically in Ref.~\cite{Majidy2023} shows no sign of saturation after $t\sim L^2$ in the sharp phase. Capturing this behavior analytically is left for future study.

Another natural next step is to go beyond the leading order in $1/\nTL$, which would help pin down the true universality of the spin-sharpening transition. More broadly, our framework extends naturally to higher spin representations, to other non-Abelian groups, and in principle to anyonic fusion categories, where the loop fugacity need not be an integer. An extension of our theory to monitored anyon chains~\cite{MonitoredAnyonsInPrep} would naturally predict a sharp phase with dynamical exponent $z=1$, as the anyonic analogs of the Heisenberg ferromagnet describing the background sector are conformal field theories~\cite{PRLanyons}.

Going beyond monitored circuits, a system undergoing dephasing that respects a strong symmetry can show a strong-to-weak spontaneous symmetry breaking (SWSSB) transition~\cite{Lessa2025,Sala2024}. For a $U(1)$ charge, SWSSB in two dimensions appears closely related to charge sharpening in $(1{+}1)$d monitored dynamics: both are governed by the same replica effective field theory and Kosterlitz--Thouless-type criticality~\cite{BarrattCharge2022,Hauser2026}. For $SU(2)$ the two appear to be distinct, since space and time scale differently in the monitored setting ($z=2$). We leave a more in-depth study of their relation to future work.

\begin{acknowledgments}
We thank Ehud Altman, Yimu Bao, Sarang Gopalakrishnan, Xiaozhou Feng, Andreas Ludwig, Adam Nahum, Rushikesh Patil, Pablo Sala, Vlad Temkin, Simon Trebst, Cenke Xu, Yiqing Zhou, Guoyi Zhu for helpful discussions and/or collaborations on related topics. Claude Opus 5 and 5.5 were used for assistance in numerical simulations, technical derivations, and editing of the text. The authors take full responsibility for the content of this work. This research was supported in part by grant NSF PHY-2309135 (RM), the Mitchel Postdoctoral Scholar Career Development Fund (RM), the Simons Investigator program through an award to C. Xu (RM), the Swiss National Science Foundation (RV, grant 10008234) and the Foundation for the University of Geneva (RV). 
\end{acknowledgments}

\bibliography{references}

@article{Li2018,
  author = {Li, Yaodong and Chen, Xiao and Fisher, Matthew P. A.},
  title = {Quantum Zeno effect and the many-body entanglement transition},
  journal = {Phys. Rev. B},
  volume = {98},
  pages = {205136},
  year = {2018},
  doi = {10.1103/PhysRevB.98.205136}
}

@article{Weinstein2022,
  title = {Measurement-Induced Power-Law Negativity in an Open Monitored Quantum Circuit},
  author = {Weinstein, Zack and Bao, Yimu and Altman, Ehud},
  journal = {Phys. Rev. Lett.},
  volume = {129},
  pages = {080501},
  year = {2022},
  doi = {10.1103/PhysRevLett.129.080501},
  eprint = {2202.12905},
  archivePrefix = {arXiv}
}

@article{HuangKuengPreskill2020,
  author = {Huang, Hsin-Yuan and Kueng, Richard and Preskill, John},
  title = {Predicting many properties of a quantum system from very few measurements},
  journal = {Nat. Phys.},
  volume = {16},
  pages = {1050},
  year = {2020},
  doi = {10.1038/s41567-020-0932-7},
  eprint = {2002.08953},
  archivePrefix = {arXiv}
}

@article{NahumJacobsen,
  title = {Bayesian critical points in classical lattice models},
  author = {Nahum, Adam and Jacobsen, Jesper Lykke},
  journal = {Phys. Rev. B},
  volume = {112},
  issue = {23},
  pages = {235113},
  numpages = {57},
  year = {2025},
  month = {Dec},
  publisher = {American Physical Society},
  doi = {10.1103/7dpt-d4s5},
  url = {https://link.aps.org/doi/10.1103/7dpt-d4s5},
  eprint = {2504.01264},
  archivePrefix = {arXiv}  
}

@misc{MonitoredAnyonsInPrep,
  author = {Zhi, Zhaohui and Han, Bo and Vasseur, Romain and Ludwig, Andreas W. W. and Trebst, Simon and Zhu, Guo-Yi},
  title = {Measurement-induced criticality for non-{A}belian anyons: The monitored golden chain},
  year = {2026},
  eprint = {2610.18965},
  archivePrefix = {arXiv},
  primaryClass = {cond-mat.str-el}
}

@article{AgrawalCollapse2024,
  author = {Agrawal, Utkarsh and Lopez-Piqueres, Javier and Vasseur, Romain and Gopalakrishnan, Sarang and Potter, Andrew C.},
  title = {Observing Quantum Measurement Collapse as a Learnability Phase Transition},
  journal = {Phys. Rev. X},
  volume = {14},
  pages = {041012},
  year = {2024},
  doi = {10.1103/PhysRevX.14.041012},
  eprint = {2311.00058},
  archivePrefix = {arXiv}
}

@article{LiCrossEntropy2023,
  author = {Li, Yaodong and Zou, Yijian and Glorioso, Paolo and Altman, Ehud and Fisher, Matthew P. A.},
  title = {Cross Entropy Benchmark for Measurement-Induced Phase Transitions},
  journal = {Phys. Rev. Lett.},
  volume = {130},
  pages = {220404},
  year = {2023},
  doi = {10.1103/PhysRevLett.130.220404},
  eprint = {2209.00609},
  archivePrefix = {arXiv}
}

@article{Dobrushin1968,
  author = {Dobrushin, R. L.},
  title = {The description of a random field by means of conditional probabilities and conditions of its regularity},
  journal = {Theory Probab. Appl.},
  volume = {13},
  pages = {197},
  year = {1968},
  doi = {10.1137/1113026}
}

@article{Peierls1936,
  author = {Peierls, R.},
  title = {On Ising's model of ferromagnetism},
  journal = {Math. Proc. Cambridge Philos. Soc.},
  volume = {32},
  pages = {477},
  year = {1936},
  doi = {10.1017/S0305004100019174}
}

@article{ha2024measurement,
  title={Measurement-induced phase transitions in systems with diffusive dynamics},
  author={Ha, Hyunsoo and Pandey, Akshat and Gopalakrishnan, Sarang and Huse, David A},
  journal={Physical Review B},
  volume={110},
  number={14},
  pages={L140301},
  year={2024},
  publisher={APS}
}

@misc{SuppMat,
  note = {See Supplemental Material for the complete statistical-mechanics treatment and the detailed derivations quoted in the main text}
}

@article{Singh2026,
  author = {Singh, Hansveer and Vasseur, Romain and Potter, Andrew C. and Gopalakrishnan, Sarang},
  title = {Mixed-state learnability transitions in monitored noisy quantum dynamics},
  journal = {Phys. Rev. B},
  volume = {113},
  pages = {054305},
  year = {2026},
  eprint = {2503.10308},
  archivePrefix = {arXiv}
}

@article{Dehghani2023,
  author = {Dehghani, Hossein and Lavasani, Ali and Hafezi, Mohammad and Gullans, Michael J.},
  title = {Neural-network decoders for measurement induced phase transitions},
  journal = {Nat. Commun.},
  volume = {14},
  pages = {2918},
  year = {2023},
  doi = {10.1038/s41467-023-37902-1},
  eprint = {2204.10904},
  archivePrefix = {arXiv}
}

@article{Fisher2023,
  author = {Fisher, Matthew P. A. and Khemani, Vedika and Nahum, Adam and Vijay, Sagar},
  title = {Random Quantum Circuits},
  journal = {Annu. Rev. Condens. Matter Phys.},
  volume = {14},
  pages = {335--379},
  year = {2023},
  doi = {10.1146/annurev-conmatphys-031720-030658},
  eprint = {2207.14280},
  archivePrefix = {arXiv}
}

@article{Lessa2025,
  author = {Lessa, Leonardo A. and Ma, Ruochen and Zhang, Jian-Hao and Bi, Zhen and Cheng, Meng and Wang, Chong},
  title = {Strong-to-Weak Spontaneous Symmetry Breaking in Mixed Quantum States},
  journal = {PRX Quantum},
  volume = {6},
  pages = {010344},
  year = {2025},
  doi = {10.1103/PRXQuantum.6.010344},
  eprint = {2405.03639},
  archivePrefix = {arXiv}
}

@article{Preskill2018,
  author = {Preskill, John},
  title = {Quantum Computing in the {NISQ} era and beyond},
  journal = {Quantum},
  volume = {2},
  pages = {79},
  year = {2018},
  doi = {10.22331/q-2018-08-06-79},
  eprint = {1801.00862},
  archivePrefix = {arXiv}
}

@article{Sala2024,
  author = {Sala, Pablo and Gopalakrishnan, Sarang and Oshikawa, Masaki and You, Yizhi},
  title = {Spontaneous strong symmetry breaking in open systems: Purification perspective},
  journal = {Phys. Rev. B},
  volume = {110},
  pages = {155150},
  year = {2024},
  doi = {10.1103/PhysRevB.110.155150},
  eprint = {2405.02402},
  archivePrefix = {arXiv}
}

@misc{Hauser2026,
  author = {Hauser, Jacob and Su, Kaixiang and Ha, Hyunsoo and Lloyd, Jerome and Kiely, Thomas G. and Vasseur, Romain and Gopalakrishnan, Sarang and Xu, Cenke and Fisher, Matthew P. A.},
  title = {Strong-to-Weak Symmetry Breaking in Open Quantum Systems: From Discrete Particles to Continuum Hydrodynamics},
  year = {2026},
  eprint = {2602.16045},
  archivePrefix = {arXiv}
}

@article{Nahum2017,
  author = {Nahum, Adam and Ruhman, Jonathan and Vijay, Sagar and Haah, Jeongwan},
  title = {Quantum Entanglement Growth under Random Unitary Dynamics},
  journal = {Phys. Rev. X},
  volume = {7},
  pages = {031016},
  year = {2017},
  doi = {10.1103/PhysRevX.7.031016},
  eprint = {1608.06950},
  archivePrefix = {arXiv}
}

@article{Chan2019,
  author = {Chan, Amos and Nandkishore, Rahul M. and Pretko, Michael and Smith, Graeme},
  title = {Unitary-projective entanglement dynamics},
  journal = {Phys. Rev. B},
  volume = {99},
  pages = {224307},
  year = {2019},
  doi = {10.1103/PhysRevB.99.224307},
  eprint = {1808.05949},
  archivePrefix = {arXiv}
}

@article{Szyniszewski2019,
  author = {Szyniszewski, Marcin and Romito, Alessandro and Schomerus, Henning},
  title = {Entanglement transition from variable-strength weak measurements},
  journal = {Phys. Rev. B},
  volume = {100},
  pages = {064204},
  year = {2019},
  doi = {10.1103/PhysRevB.100.064204},
  eprint = {1903.05452},
  archivePrefix = {arXiv}
}

@article{Choi2020,
  author = {Choi, Soonwon and Bao, Yimu and Qi, Xiao-Liang and Altman, Ehud},
  title = {Quantum Error Correction in Scrambling Dynamics and Measurement-Induced Phase Transition},
  journal = {Phys. Rev. Lett.},
  volume = {125},
  pages = {030505},
  year = {2020},
  doi = {10.1103/PhysRevLett.125.030505},
  eprint = {1903.05124},
  archivePrefix = {arXiv}
}

@article{GullansHuse2020b,
  author = {Gullans, Michael J. and Huse, David A.},
  title = {Scalable Probes of Measurement-Induced Criticality},
  journal = {Phys. Rev. Lett.},
  volume = {125},
  pages = {070606},
  year = {2020},
  doi = {10.1103/PhysRevLett.125.070606},
  eprint = {1910.00020},
  archivePrefix = {arXiv}
}

@article{Zabalo2020,
  author = {Zabalo, Aidan and Gullans, Michael J. and Wilson, Justin H. and Gopalakrishnan, Sarang and Huse, David A. and Pixley, J. H.},
  title = {Critical properties of the measurement-induced transition in random quantum circuits},
  journal = {Phys. Rev. B},
  volume = {101},
  pages = {060301},
  year = {2020},
  doi = {10.1103/PhysRevB.101.060301},
  eprint = {1911.00008},
  archivePrefix = {arXiv}
}

@article{Ippoliti2021,
  author = {Ippoliti, Matteo and Gullans, Michael J. and Gopalakrishnan, Sarang and Huse, David A. and Khemani, Vedika},
  title = {Entanglement Phase Transitions in Measurement-Only Dynamics},
  journal = {Phys. Rev. X},
  volume = {11},
  pages = {011030},
  year = {2021},
  doi = {10.1103/PhysRevX.11.011030},
  eprint = {2004.09560},
  archivePrefix = {arXiv}
}

@article{Lavasani2021,
  author = {Lavasani, Ali and Alavirad, Yahya and Barkeshli, Maissam},
  title = {Measurement-induced topological entanglement transitions in symmetric random quantum circuits},
  journal = {Nature Physics},
  volume = {17},
  pages = {342--347},
  year = {2021},
  doi = {10.1038/s41567-020-01112-z},
  eprint = {2004.07243},
  archivePrefix = {arXiv}
}

@article{SangHsieh2021,
  author = {Sang, Shengqi and Hsieh, Timothy H.},
  title = {Measurement-protected quantum phases},
  journal = {Phys. Rev. Research},
  volume = {3},
  pages = {023200},
  year = {2021},
  doi = {10.1103/PhysRevResearch.3.023200},
  eprint = {2004.09509},
  archivePrefix = {arXiv}
}

@article{NahumRoy2021,
  author = {Nahum, Adam and Roy, Sthitadhi and Skinner, Brian and Ruhman, Jonathan},
  title = {Measurement and Entanglement Phase Transitions in All-To-All Quantum Circuits, on Quantum Trees, and in {L}andau-{G}insburg Theory},
  journal = {PRX Quantum},
  volume = {2},
  pages = {010352},
  year = {2021},
  doi = {10.1103/PRXQuantum.2.010352},
  eprint = {2009.11311},
  archivePrefix = {arXiv}
}

@article{Alberton2021,
  author = {Alberton, Ossian and Buchhold, Michael and Diehl, Sebastian},
  title = {Entanglement Transition in a Monitored Free-Fermion Chain: From Extended Criticality to Area Law},
  journal = {Phys. Rev. Lett.},
  volume = {126},
  pages = {170602},
  year = {2021},
  doi = {10.1103/PhysRevLett.126.170602},
  eprint = {2005.09722},
  archivePrefix = {arXiv}
}

@article{Buchhold2021,
  author = {Buchhold, Michael and Minoguchi, Yuki and Altland, Alexander and Diehl, Sebastian},
  title = {Effective Theory for the Measurement-Induced Phase Transition of {D}irac Fermions},
  journal = {Phys. Rev. X},
  volume = {11},
  pages = {041004},
  year = {2021},
  doi = {10.1103/PhysRevX.11.041004},
  eprint = {2102.08381},
  archivePrefix = {arXiv}
}

@article{Zabalo2022,
  author = {Zabalo, Aidan and Gullans, Michael J. and Wilson, Justin H. and Vasseur, Romain and Ludwig, Andreas W. W. and Gopalakrishnan, Sarang and Huse, David A. and Pixley, J. H.},
  title = {Operator Scaling Dimensions and Multifractality at Measurement-Induced Transitions},
  journal = {Phys. Rev. Lett.},
  volume = {128},
  pages = {050602},
  year = {2022},
  doi = {10.1103/PhysRevLett.128.050602},
  eprint = {2107.03393},
  archivePrefix = {arXiv}
}

@incollection{PotterVasseur2022,
  author = {Potter, Andrew C. and Vasseur, Romain},
  title = {Entanglement Dynamics in Hybrid Quantum Circuits},
  booktitle = {Entanglement in Spin Chains: From Theory to Quantum Technology Applications},
  editor = {Bayat, Abolfazl and Bose, Sougato and Johannesson, Henrik},
  publisher = {Springer International Publishing},
  address = {Cham},
  pages = {211--249},
  year = {2022},
  doi = {10.1007/978-3-031-03998-0_9},
  eprint = {2111.08018},
  archivePrefix = {arXiv}
}

@article{Fava2023,
  author = {Fava, Michele and Piroli, Lorenzo and Swann, Tobias and Bernard, Denis and Nahum, Adam},
  title = {Nonlinear Sigma Models for Monitored Dynamics of Free Fermions},
  journal = {Phys. Rev. X},
  volume = {13},
  pages = {041045},
  year = {2023},
  doi = {10.1103/PhysRevX.13.041045},
  eprint = {2302.12820},
  archivePrefix = {arXiv}
}

@article{Affleck1985,
  author = {Affleck, Ian},
  title = {Large-{$n$} Limit of {$SU(n)$} Quantum ``Spin'' Chains},
  journal = {Phys. Rev. Lett.},
  volume = {54},
  pages = {966--969},
  year = {1985},
  doi = {10.1103/PhysRevLett.54.966}
}

@article{MajidyNRP2023,
  author = {Majidy, Shayan and Braasch, William F. and Lasek, Aleksander and Upadhyaya, Twesh and Kalev, Amir and Yunger Halpern, Nicole},
  title = {Noncommuting conserved charges in quantum thermodynamics and beyond},
  journal = {Nature Reviews Physics},
  volume = {5},
  pages = {689--698},
  year = {2023},
  doi = {10.1038/s42254-023-00641-9},
  eprint = {2306.00054},
  archivePrefix = {arXiv}
}

@article{Skinner2019,
  author = {Skinner, Brian and Ruhman, Jonathan and Nahum, Adam},
  title = {Measurement-Induced Phase Transitions in the Dynamics of Entanglement},
  journal = {Phys. Rev. X},
  volume = {9},
  pages = {031009},
  year = {2019},
  doi = {10.1103/PhysRevX.9.031009}
}

@article{Li2019,
  author = {Li, Yaodong and Chen, Xiao and Fisher, Matthew P. A.},
  title = {Measurement-driven entanglement transition in hybrid quantum circuits},
  journal = {Phys. Rev. B},
  volume = {100},
  pages = {134306},
  year = {2019},
  doi = {10.1103/PhysRevB.100.134306}
}

@article{li2024statistical,
  title={Statistical mechanics model for {C}lifford random tensor networks and monitored quantum circuits},
  author={Li, Yaodong and Vasseur, Romain and Fisher, Matthew PA and Ludwig, Andreas WW},
  journal={Phys. Rev. B},
  volume={109},
  number={17},
  pages={174307},
  year={2024},
  doi={10.1103/PhysRevB.109.174307},
  eprint={2110.02988},
  archivePrefix={arXiv}
}

@article{GullansHuse2020,
  author = {Gullans, Michael J. and Huse, David A.},
  title = {Dynamical Purification Phase Transition Induced by Quantum Measurements},
  journal = {Phys. Rev. X},
  volume = {10},
  pages = {041020},
  year = {2020},
  doi = {10.1103/PhysRevX.10.041020}
}

@article{Bao2020,
  author = {Bao, Yimu and Choi, Soonwon and Altman, Ehud},
  title = {Theory of the phase transition in random unitary circuits with measurements},
  journal = {Phys. Rev. B},
  volume = {101},
  pages = {104301},
  year = {2020},
  doi = {10.1103/PhysRevB.101.104301}
}

@article{Jian2020,
  author = {Jian, Chao-Ming and You, Yi-Zhuang and Vasseur, Romain and Ludwig, Andreas W. W.},
  title = {Measurement-induced criticality in random quantum circuits},
  journal = {Phys. Rev. B},
  volume = {101},
  pages = {104302},
  year = {2020},
  doi = {10.1103/PhysRevB.101.104302}
}

@article{ZhouNahum2019,
  author = {Zhou, Tianci and Nahum, Adam},
  title = {Emergent statistical mechanics of entanglement in random unitary circuits},
  journal = {Phys. Rev. B},
  volume = {99},
  pages = {174205},
  year = {2019},
  doi = {10.1103/PhysRevB.99.174205},
  eprint = {1804.09737},
  archivePrefix = {arXiv}
}

@article{IppolitiKhemani2024,
  author = {Ippoliti, Matteo and Khemani, Vedika},
  title = {Learnability transitions in monitored quantum dynamics via eavesdropper's classical shadows},
  journal = {PRX Quantum},
  volume = {5},
  pages = {020304},
  year = {2024},
  doi = {10.1103/PRXQuantum.5.020304},
  eprint = {2307.15011},
  archivePrefix = {arXiv}
}

@misc{2026Vasseurnotes,
       author = {{Vasseur}, Romain},
        title = "{Les Houches lectures on random quantum circuits and monitored quantum dynamics}",
         year = 2026,
          doi = {10.48550/arXiv.2602.17258},
archivePrefix = {arXiv},
       eprint = {2602.17258},
 primaryClass = {quant-ph}
}

@article{Agrawal2022,
  author = {Agrawal, Utkarsh and Zabalo, Aidan and Chen, Kun and Wilson, Justin H. and Potter, Andrew C. and Pixley, J. H. and Gopalakrishnan, Sarang and Vasseur, Romain},
  title = {Entanglement and Charge-Sharpening Transitions in {$U(1)$} Symmetric Monitored Quantum Circuits},
  journal = {Phys. Rev. X},
  volume = {12},
  pages = {041002},
  year = {2022},
  doi = {10.1103/PhysRevX.12.041002},
  eprint = {2107.10279},
  archivePrefix = {arXiv}
}

@article{BarrattCharge2022,
  author = {Barratt, Fergus and Agrawal, Utkarsh and Gopalakrishnan, Sarang and Huse, David A. and Vasseur, Romain and Potter, Andrew C.},
  title = {Field Theory of Charge Sharpening in Symmetric Monitored Quantum Circuits},
  journal = {Phys. Rev. Lett.},
  volume = {129},
  pages = {120604},
  year = {2022},
  doi = {10.1103/PhysRevLett.129.120604},
  eprint = {2111.09336},
  archivePrefix = {arXiv}
}

@article{PRLanyons,
  title = {Interacting Anyons in Topological Quantum Liquids: The Golden Chain},
  author = {Feiguin, Adrian and Trebst, Simon and Ludwig, Andreas W. W. and Troyer, Matthias and Kitaev, Alexei and Wang, Zhenghan and Freedman, Michael H.},
  journal = {Phys. Rev. Lett.},
  volume = {98},
  issue = {16},
  pages = {160409},
  numpages = {4},
  year = {2007},
  month = {Apr},
  publisher = {American Physical Society},
  doi = {10.1103/PhysRevLett.98.160409},
  url = {https://link.aps.org/doi/10.1103/PhysRevLett.98.160409},
  eprint = {cond-mat/0612341},
  archivePrefix = {arXiv}
}

@article{BarrattLearn2022,
  author = {Barratt, Fergus and Agrawal, Utkarsh and Potter, Andrew C. and Gopalakrishnan, Sarang and Vasseur, Romain},
  title = {Transitions in the Learnability of Global Charges from Local Measurements},
  journal = {Phys. Rev. Lett.},
  volume = {129},
  pages = {200602},
  year = {2022},
  doi = {10.1103/PhysRevLett.129.200602},
  eprint = {2206.12429},
  archivePrefix = {arXiv}
}

@article{Majidy2023,
  author = {Majidy, Shayan and Agrawal, Utkarsh and Gopalakrishnan, Sarang and Potter, Andrew C. and Vasseur, Romain and Yunger Halpern, Nicole},
  title = {Critical phase and spin sharpening in {$SU(2)$}-symmetric monitored quantum circuits},
  journal = {Phys. Rev. B},
  volume = {108},
  pages = {054307},
  year = {2023},
  doi = {10.1103/PhysRevB.108.054307},
  eprint = {2305.13356},
  archivePrefix = {arXiv}
}

@article{Feng2025,
  author = {Feng, Xiaozhou and Fishchenko, Nadezhda and Gopalakrishnan, Sarang and Ippoliti, Matteo},
  title = {Charge and Spin Sharpening Transitions on Dynamical Quantum Trees},
  journal = {Quantum},
  volume = {9},
  pages = {1692},
  year = {2025},
  doi = {10.22331/q-2025-04-07-1692},
  eprint = {2405.13894},
  archivePrefix = {arXiv}
}

@article{Gopalakrishnan2025,
  title = {Monitored Fluctuating Hydrodynamics},
  author = {Gopalakrishnan, Sarang and McCulloch, Ewan and Vasseur, Romain},
  journal = {Phys. Rev. X},
  volume = {16},
  issue = {1},
  pages = {011024},
  numpages = {17},
  year = {2026},
  month = {Feb},
  publisher = {American Physical Society},
  doi = {10.1103/295c-lj1w},
  url = {https://link.aps.org/doi/10.1103/295c-lj1w}
}

@article{FisherMaNickel1972,
  author = {Fisher, Michael E. and Ma, Shang-keng and Nickel, B. G.},
  title = {Critical Exponents for Long-Range Interactions},
  journal = {Phys. Rev. Lett.},
  volume = {29},
  pages = {917--920},
  year = {1972},
  doi = {10.1103/PhysRevLett.29.917}
}

@article{Sak1973,
  author = {Sak, J.},
  title = {Recursion Relations and Fixed Points for Ferromagnets with Long-Range Interactions},
  journal = {Phys. Rev. B},
  volume = {8},
  pages = {281--285},
  year = {1973},
  doi = {10.1103/PhysRevB.8.281}
}

@misc{MetlitskiGrover2011,
  author = {Metlitski, Max A. and Grover, Tarun},
  title = {Entanglement Entropy of Systems with Spontaneously Broken Continuous Symmetry},
  year = {2011},
  eprint = {1112.5166},
  archivePrefix = {arXiv},
  primaryClass = {cond-mat.str-el}
}

@article{Bykov2012,
  author = {Bykov, Dmitri},
  title = {Haldane limits via Lagrangian embeddings},
  journal = {Nuclear Physics B},
  volume = {855},
  number = {1},
  pages = {100--127},
  year = {2012},
  doi = {10.1016/j.nuclphysb.2011.10.005},
  eprint = {1104.1419},
  archivePrefix = {arXiv},
  primaryClass = {hep-th}
}

@article{WamerAffleck2020,
  author = {Wamer, Kyle and Affleck, Ian},
  title = {Flag manifold sigma models from {SU}($n$) chains},
  journal = {Nuclear Physics B},
  volume = {959},
  pages = {115156},
  year = {2020},
  doi = {10.1016/j.nuclphysb.2020.115156},
  eprint = {2007.01912},
  archivePrefix = {arXiv},
  primaryClass = {cond-mat.str-el}
}

@book{wilde2013quantum,
  author = {Wilde, Mark M.},
  title = {Quantum Information Theory},
  publisher = {Cambridge University Press},
  address = {Cambridge},
  year = {2013},
  doi = {10.1017/CBO9781139525343}
}

@article{TemperleyLieb1971,
  author = {Temperley, H. N. V. and Lieb, E. H.},
  title = {Relations between the `percolation' and `colouring' problem and other graph-theoretical problems associated with regular planar lattices: some exact results for the `percolation' problem},
  journal = {Proceedings of the Royal Society of London A},
  volume = {322},
  pages = {251--280},
  year = {1971},
  doi = {10.1098/rspa.1971.0067}
}

@article{bao2021symmetry,
  title={Symmetry enriched phases of quantum circuits},
  author={Bao, Yimu and Choi, Soonwon and Altman, Ehud},
  journal={Annals of Physics},
  volume={435},
  pages={168618},
  year={2021},
  doi={10.1016/j.aop.2021.168618},
  eprint={2102.09164},
  archivePrefix={arXiv}
}

@article{sutherland1995low,
  title={Low-lying eigenstates of the one-dimensional {H}eisenberg ferromagnet for any magnetization and momentum},
  author={Sutherland, Bill},
  journal={Phys. Rev. Lett.},
  volume={74},
  number={5},
  pages={816},
  year={1995},
  doi={10.1103/PhysRevLett.74.816}
}

@article{dhar2000bloch,
  title={{B}loch walls and macroscopic string states in {B}ethe's solution of the {H}eisenberg ferromagnetic linear chain},
  author={Dhar, Abhishek and Shastry, B Sriram},
  journal={Phys. Rev. Lett.},
  volume={85},
  number={13},
  pages={2813},
  year={2000},
  doi={10.1103/PhysRevLett.85.2813}
}

@article{popkov2005logarithmic,
  title={Logarithmic divergence of the block entanglement entropy for the ferromagnetic {H}eisenberg model},
  author={Popkov, Vladislav and Salerno, Mario},
  journal={Phys. Rev. A},
  volume={71},
  number={1},
  pages={012301},
  year={2005},
  doi={10.1103/PhysRevA.71.012301}
}

\vspace{6pt}
\begin{center}\textbf{End Matter}\end{center}

\EMappendix{A}{Background sector and its entanglement}The initial state selects which ground state of the background sector governs the late-time dynamics. If $\rho_0$ carries a fixed total spin $S$, every branch stays in the spin-$S$ sector, and the lowest state of the ferromagnetic chain in this sector is a Bloch-wall-like spiral. For $S=O(1)$, its semiclassical representative is a unit vector winding once around the chain~\cite{sutherland1995low,dhar2000bloch},
\begin{equation}
\vec n_0(x)=(\sin\theta\cos q_0x,\ \sin\theta\sin q_0x,\ \cos\theta),
\quad q_0=\frac{2\pi}{L},
\label{eq:spiral-L}
\end{equation}
where $\cos\theta=2S/L$. The slow winding is the origin of the two regimes in Eq.~\eqref{eq:S0-L}, as only a region of order $L$ resolves that all three generators of $SU(2)$ are broken.

When instead $\rho_0$ is the maximally mixed state, all spin sectors are present, and the late-time dynamics is governed by the overall ground state of $H_{\rm bg}$, a uniform ferromagnet. For general $\nTL$ its ground-state manifold is the flag manifold $\mathcal F_{1,1;\nTL}=SU(\nTL)/[U(1)\times U(1)\times SU(\nTL-2)]$~\cite{Affleck1985,Bykov2012,WamerAffleck2020,SuppMat}, which reduces to the familiar $\mathbb{CP}^1$ ferromagnet at $\nTL=2$. The twisted free-energy cost then gives
\begin{equation}
S_A^{\rm bg}=d_{\mathcal F}\log|A|+O(1),
\qquad d_{\mathcal F}=2\nTL-3,
\label{eq:log-L}
\end{equation}
independent of the R\'enyi index. Each real coordinate of the ground-state manifold contributes $\tfrac12$ to the coefficient, which is the standard Goldstone-mode contribution to entanglement~\cite{MetlitskiGrover2011,popkov2005logarithmic}, although here the entropy is computed as a boundary-twist correlator. At $\nTL=2$ this gives $d_{\mathcal F}=1$, as quoted in the main text.

\EMappendix{B}{The two parts of the record fidelity}We define here the two parts of Eq.~\eqref{eq:F-split-L}. The record overlap $Z^{(q)}_{\bm S}$, with $\bm S=(S,S)$, $(S',S')$ or $(S,S')$ as in the main text, is a matrix element of the replicated dynamics between the initial state of the $2q$ replicas and the final-time boundary state $\langle e|$, which implements the trace that glues each forward branch to its backward partner. The label $\bm S$ gives the total spins of the two groups of $q$ replicas~\cite{SuppMat}. At long times $Z^{(q)}_{\bm S}$ is dominated by the leading eigenvalue $\lambda_{\bm S}$ of the time-direction transfer matrix. The steady rate is then
\begin{equation}
\Delta=\tfrac12\big(\log\lambda_{S,S}+\log\lambda_{S',S'}\big)-\log\lambda_{S,S'},
\label{eq:Delta-def}
\end{equation}
and the transient $\mathcal T(0)-\mathcal T(t)$ collects the contributions above the leading eigenvalue. Since $\mathcal T(t)$ decays to zero on the inverse gap of the $z=2$ spin waves, the transient saturates to $\mathcal T(0)$ once $t\gtrsim L^2$.

The saturation value is fixed by the leading eigenvector $|b_{\bm S}\rangle$ in the sector with spin assignment $\bm S$~\cite{SuppMat},
\begin{equation}
\begin{aligned}
\mathcal T(0)&=-\log\frac{a_0^{(S,S')}}{\sqrt{a_0^{(S,S)}\,a_0^{(S',S')}}},\\
a_0^{(S,S')}&=\langle e|b_{S,S'}\rangle\langle b_{S,S'}|\rho_{0,S}^{\otimes q}\otimes\rho_{0,S'}^{\otimes q}\rangle,
\end{aligned}
\label{eq:T0-L}
\end{equation}
and similarly for $a_0^{(S,S)}$ and $a_0^{(S',S')}$, where $\rho_{0,S}$ denotes the initial state in the spin-$S$ sector. Both the initial and the final states are products over the $2q$ replicas. If the leading eigenvector is also such a product, as in the fuzzy phase, each $a_0$ is a product of one factor per replica. These factors cancel in the ratio, and $\mathcal T(0)=0$. The pairing order thereby hides the diffusive background from the record. The corrections to this cancellation, and the argument that $\mathcal T(0)$ is generically of order one in the sharp phase, are given in the SM~\cite{SuppMat}.

\EMappendix{C}{Numerical protocol and learning times}We describe here how the data of Fig.~\ref{fig:decoherence} are obtained. Following Ref.~\cite{Majidy2023}, an ancilla qubit is initially entangled with two states of the chain: a product of nearest-neighbor singlets, with $S=0$, and the same state with one singlet replaced by the $m=0$ triplet, with $S=1$. The chain is first scrambled for $L^2$ steps with unitary gates only (one time step consists of two brickwork layers), and then evolved for $L^2$ steps with measurements and decoherence. At each time step and on each bond, the fusion channel is measured and recorded with probability $p$, measured and discarded with probability $p_d$, and otherwise a gate $U(\theta)$ is applied. The ancilla entropy $S_A$, averaged over circuit realizations and over measurement records with their Born weights, is used as the order parameter of the spin-sharpening transition~\cite{GullansHuse2020b,Agrawal2022,Majidy2023}. It measures how uncertain the observer remains about the total spin of the chain. We average $S_A$ over 4000 samples for each $p$ and each $L=6,8,10,12$.

The learning time is defined as follows. As the observer learns the spin during the dynamics, we expect the ancilla entropy to decay exponentially at late times, $S_A(t)\propto e^{-t/t_{\rm learn}}$. We extract $t_{\rm learn}$ from a linear fit of $\ln S_A$ in the window $0.01<S_A<0.3$. This excludes early times, where $S_A$ is still close to its initial value $\ln2\approx0.69$, and late times, where the relative statistical error of our numerical data becomes large. The exponents quoted in the main text are power-law fits of $t_{\rm learn}$ over the four sizes.

\ifincludesupplement
\clearpage
\onecolumngrid
\newpage
\setcounter{equation}{0}
\setcounter{figure}{0}
\setcounter{table}{0}
\setcounter{section}{0}
\setcounter{secnumdepth}{3}
\renewcommand{\theequation}{S\arabic{equation}}
\renewcommand{\thefigure}{S\arabic{figure}}
\renewcommand{\thesection}{S\Roman{section}}
\makeatletter
\@ifundefined{theHequation}{}{\renewcommand{\theHequation}{sm.\arabic{equation}}}
\@ifundefined{theHfigure}{}{\renewcommand{\theHfigure}{sm.\arabic{figure}}}
\@ifundefined{theHsection}{}{\renewcommand{\theHsection}{sm.\arabic{section}}}
\makeatother
\begin{center}
{\large\bfseries Supplemental Material for ``Statistical Mechanics of Non-Abelian Learnability Transitions''}\\[6pt]
Ruochen Ma$^{1}$ and Romain Vasseur$^{2,3}$\\[4pt]
{\itshape $^{1}$\kitp}\\[2pt]
{\itshape $^{2}$\unige}\\[2pt]
{\itshape $^{3}$\gqc}
\end{center}
\vspace{4pt}
\DeclareRobustCommand{\chgcolor}{\color{red}}
\providecommand{\chg}[1]{{\chgcolor #1}}
\providecommand{\chgon}{\begingroup\chgcolor}
\providecommand{\chgoff}{\endgroup}

This Supplemental Material derives the results quoted in the main text and in the End Matter. Section~\ref{sec:model} defines the circuit and the two diagnostics. Section~\ref{sec:loop-model} maps the replicated dynamics exactly onto a loop model. Section~\ref{sec:separation} develops the expansion at large loop fugacity. It separates the model into a pairing sector and a background sector, shows that the background sector consists of $2Q$ ferromagnetic branches with $z=2$ spin waves, derives the interaction $A(v,w)$ that the background induces between the pairings, and checks the ferromagnetic background exactly at $\nTL=2$ for integer $Q>1$. Section~\ref{sec:twist} obtains the phases of the pairing sector, the entanglement scaling in each phase, and the universality class of the transition. Section~\ref{sec:learning} derives the learning times in the two phases. 

\section{Circuit model and diagnostics}
\label{sec:model}

Consider a periodic chain of an even number $L$ of spin-$1/2$ moments evolving under a brickwork circuit.  On each brick, we apply either an $SU(2)$-symmetric two-qubit unitary, with probability $1-p$, or a projective measurement of the two-spin fusion channel, with probability $p$.  Up to an overall phase, the most general such unitary gate is
\begin{equation}
U(\theta)=P_s+\e^{\ii\theta}P_t,
\qquad \theta\in[0,2\pi),
\label{eq:gate}
\end{equation}
where $P_s$ and $P_t$ are the projectors onto the singlet and triplet channels.  The measurement projects the two spins onto one of these two fusion channels. An illustration of the brickwork circuit is shown in Fig.~\ref{fig:tensor-network}. One time step consists of two layers of the brickwork, which we also call one period. The first layer acts on all bonds \((2i,2i+1)\), and the second acts on all bonds \((2i-1,2i)\). This dynamics preserves the global $SU(2)$ spin-rotation symmetry generated by
\begin{equation}
S^\alpha=\frac12\sum_{i=1}^{L}\sigma_i^\alpha,
\qquad \alpha=x,y,z.
\end{equation}
The spin-squared operator $\vec{S}^2$ has eigenvalues $S(S+1)$ labeled by the total spin quantum number $S$.

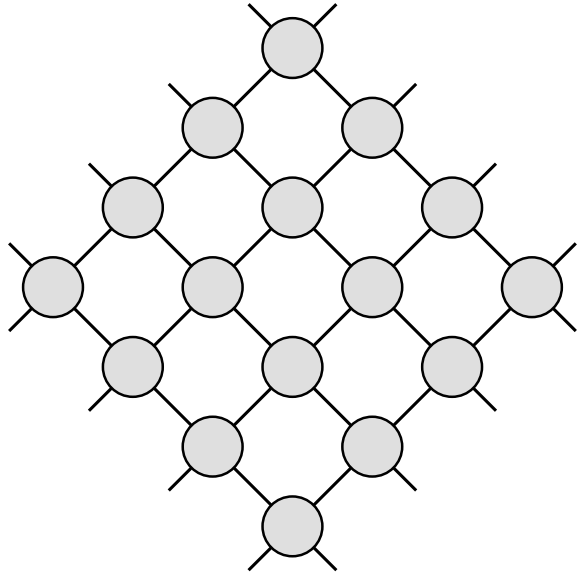
\begin{figure}[b]
\centering
\resizebox{0.42\textwidth}{!}{%
\begin{tikzpicture}[
    tensor/.style={
        circle,
        draw=black,
        fill=gray!25,
        line width=0.9pt,
        minimum size=7.5mm,
        inner sep=0pt
    },
    bond/.style={
        black,
        line width=1.1pt
    }
]

\foreach \name/\x/\y in {
    A/0/3,
    B/-1/2, C/1/2,
    D/-2/1, E/0/1, F/2/1,
    G/-3/0, H/-1/0, I/1/0, J/3/0,
    K/-2/-1, L/0/-1, M/2/-1,
    N/-1/-2, O/1/-2,
    P/0/-3
}{
    \coordinate (\name) at (\x,\y);
}

\draw[bond] (A)--(B);
\draw[bond] (A)--(C);

\draw[bond] (B)--(D);
\draw[bond] (B)--(E);
\draw[bond] (C)--(E);
\draw[bond] (C)--(F);

\draw[bond] (D)--(G);
\draw[bond] (D)--(H);
\draw[bond] (E)--(H);
\draw[bond] (E)--(I);
\draw[bond] (F)--(I);
\draw[bond] (F)--(J);

\draw[bond] (G)--(K);
\draw[bond] (H)--(K);
\draw[bond] (H)--(L);
\draw[bond] (I)--(L);
\draw[bond] (I)--(M);
\draw[bond] (J)--(M);

\draw[bond] (K)--(N);
\draw[bond] (L)--(N);
\draw[bond] (L)--(O);
\draw[bond] (M)--(O);

\draw[bond] (N)--(P);
\draw[bond] (O)--(P);

\draw[bond] (A)--++(-0.55,0.55);
\draw[bond] (A)--++(0.55,0.55);

\draw[bond] (B)--++(-0.55,0.55);
\draw[bond] (C)--++(0.55,0.55);

\draw[bond] (D)--++(-0.55,0.55);
\draw[bond] (F)--++(0.55,0.55);

\draw[bond] (G)--++(-0.55,0.55);
\draw[bond] (G)--++(-0.55,-0.55);
\draw[bond] (J)--++(0.55,0.55);
\draw[bond] (J)--++(0.55,-0.55);

\draw[bond] (K)--++(-0.55,-0.55);
\draw[bond] (M)--++(0.55,-0.55);

\draw[bond] (N)--++(-0.55,-0.55);
\draw[bond] (O)--++(0.55,-0.55);

\draw[bond] (P)--++(-0.55,-0.55);
\draw[bond] (P)--++(0.55,-0.55);

\foreach \name in {A,B,C,D,E,F,G,H,I,J,K,L,M,N,O,P}{
    \node[tensor] at (\name) {};
}

\end{tikzpicture}%
}
\caption{Spacetime tensor network for the replicated monitored dynamics.}
\label{fig:tensor-network}
\end{figure}

For a given realization of the unitary gates and measurement data in spacetime, namely the measurement locations and outcomes, the monitored dynamics can be described in terms of unnormalized quantum trajectories
\(\rho_{\bfm}=K_{\bfm}\rho_0 K_{\bfm}^{\dagger}\), where \(\rho_0\) is the initial state. Here \(\bfm=\{m(x,t)\}\) denotes the measurement-outcome history for a fixed spacetime configuration of unitary choices and measurement locations. The Kraus operator \(K_{\bfm}\) is built from the corresponding sequence of random unitary gates and, at the measured spacetime bricks, the projector \(P_s\) or \(P_t\) onto the observed singlet or triplet outcome. We will study the entanglement properties of each trajectory state, and then average over outcome histories with their Born probabilities \(p_{\bfm}=\Tr(\rho_{\bfm})\). Finally, we average the resulting quantities over the ensemble of quantum circuits.

We quantify the entanglement of the trajectory ensemble by the averaged R\'enyi entropy of a subregion \(A\), defined as
\begin{equation}
    S_A^{(n)}
    =
    \mathbb{E}_{\rm circ}
    \sum_{\bfm} p_{\bfm}\frac{1}{1-n}
    \log\!\left[
    \frac{\Tr(\rho_{A,\bfm}^n)}{\Tr(\rho_{\bfm})^n}
    \right],
    \label{eq:Renyientropy}
\end{equation}
where \(\mathbb{E}_{\rm circ}\) denotes the average over the externally sampled randomness in the circuit realizations, including the random unitary gates and the stochastic choice, on each spacetime brick, between applying a unitary gate with probability \(1-p\) and performing a measurement with probability \(p\). \(\rho_{A,\bfm}=\Tr_{\bar A}(\rho_{\bfm})\) is the reduced state on subregion \(A\) for a trajectory of a particular circuit realization with measurement-outcome history \(\bfm\). Following Refs.~\cite{Bao2020,Jian2020}, we perform the average over logarithms using the replica trick. We introduce the replica cyclic permutation operator on subregion \(A\),
\begin{equation}
\begin{split}
    T_{n,A}
    &=
    \prod_{i\in A} T_{h,i}
    \\
    T_{h,i}
    &=
    \sum_{s_i^1,\ldots,s_i^n}
    |s_i^{h(1)}s_i^{h(2)}\cdots s_i^{h(n)}\rangle
    \langle s_i^1s_i^2\cdots s_i^n| ,
\end{split}
    \label{eq:cyclic-permutation}
\end{equation}
where \(\{|s_i\rangle\}\) is an orthonormal basis of the physical Hilbert space at site \(i\), and
\[ |s_i^1s_i^2\cdots s_i^n\rangle \equiv |s_i^1\rangle \otimes |s_i^2\rangle \otimes\cdots\otimes |s_i^n\rangle\]
is the corresponding basis in the \(n\)-replica Hilbert space. Here \(h=(12\cdots n)\) is the cyclic permutation in the permutation group \(S_n\). The averaged R\'enyi entropy can then be written as
\begin{equation}
    S_A^{(n)}
    =
    \lim_{k\to 0}
    \frac{1}{k(1-n)}
    \sum_{\bfm}
    \left[Z_A(\bfm)-Z_0(\bfm)\right],
    \label{eq:renyiTN}
\end{equation}
where
\begin{equation}
    \begin{split}
        Z_A(\bfm)
        &=
        \mathbb{E}_{\rm circ}
        \Tr\!\left(
        \rho_{\bfm}^{\otimes nk+1}
        T_{n,A}^{\otimes k}
        \right),\\
        Z_0(\bfm)
        &=
        \mathbb{E}_{\rm circ}
        \Tr\!\left(
        \rho_{\bfm}^{\otimes nk+1}
        \right).
    \end{split}
    \label{eq:replicatrick}
\end{equation}
As the notation suggests, \(Z_A\) and \(Z_0\) will later be identified with partition functions of the statistical-mechanics model, with boundary conditions that differ in region \(A\). We write \(Z_A(Q;n)=\sum_{\bfm}Z_A(\bfm)\) and \(Z_0(Q;n)=\sum_{\bfm}Z_0(\bfm)\) for their sums over records. We also denote the total number of replicas in Eq.~\eqref{eq:replicatrick} by \(Q=nk+1\), where the additional replica comes from the Born probability factor in Eq.~\eqref{eq:Renyientropy}.

The learnability of the total-spin quantum number from the measurement record can be quantified by the classical fidelity between the measurement-record distributions associated with two different total-spin sectors. Let \(p_{\bfm}(S,t)\) denote the Born probability of observing a spacetime measurement record \(\bfm\) up to time \(t\), given that the initial state lies in the total-spin sector \(S\). We define
\begin{equation}
F_t(S,S')
=
\mathbb{E}_{\rm circ}
\sum_{\bfm}
\left[p_{\bfm}(S,t)p_{\bfm}(S',t)\right]^{1/2}.
\label{eq:F-main}
\end{equation}
The two spin sectors \(S\) and \(S'\) are asymptotically distinguishable from the measurement record when \(F_t(S,S')\to 0\) \cite{wilde2013quantum}. For analytic tractability, we also consider the R\'enyi generalization
\begin{equation}
F_t^{(q)}(S,S')
=
\frac{
Z_{S,S'}^{(q)}(t)
}{
\sqrt{
Z_{S,S}^{(q)}(t)Z_{S',S'}^{(q)}(t)
}
},
\end{equation}
where
\begin{equation}
Z_{S,S'}^{(q)}(t)
\equiv
\mathbb{E}_{\rm circ}
\sum_{\bfm}
\left[p_{\bfm}(S,t)p_{\bfm}(S',t)\right]^q .
\label{eq:replicatedoverlap}
\end{equation}
For \(q=1/2\), this reduces to the classical fidelity in Eq.~\eqref{eq:F-main}, since \(Z_{S,S}^{(1/2)}=Z_{S',S'}^{(1/2)}=1\) for normalized record distributions. For integer \(q\), \(Z^{(q)}_{S,S'}\) is computed with \(2q\) replicas, of which \(q\) start in sector \(S\) and \(q\) in sector \(S'\).

\section{Statistical mechanics mapping}
\label{sec:loop-model}

We now derive the mapping from the \(SU(2)\)-symmetric random monitored circuit to a statistical-mechanics (stat-mech) loop model. To compute Eqs.~\eqref{eq:replicatrick} and~\eqref{eq:replicatedoverlap}, we introduce \(Q\) replicas, each made of a forward and a backward branch. Each brick of the circuit becomes a vertex in the resulting stat-mech model. At each vertex we independently average over the random unitary gate and sum over the measurement outcomes. We denote the forward branches by \(a\in F=\{1,\ldots,Q\}\) and the backward branches by \(\bar b\), with \(b\in B=\{1,\ldots,Q\}\). For \(\theta\) in Eq.~\eqref{eq:gate} distributed uniformly over \([0,2\pi)\), averaging over the phase enforces equality between the number of singlet projectors in the forward and backward branches, or equivalently between the number of triplet projectors. Thus
\begin{equation}
\begin{split}
    \cT_u
    &\equiv
    \mathbb{E}_{U}
    \left[(U\otimes U^*)^{\otimes Q}\right] \\
    &=
    \sum_{\{f^a,f^{\bar b}=s,t\}}
    \delta_{N_s^F,N_s^B}
    \prod_{a\in F}P_{f^a}^{a}
    \prod_{b\in B}P_{f^{\bar b}}^{\bar b},
\end{split}
\label{eq:Tu}
\end{equation}
where
\begin{equation}
    N_s^F=\sum_{a\in F}\delta_{f^a,s},
    \qquad
    N_s^B=\sum_{b\in B}\delta_{f^{\bar b},s},
\end{equation}
and \(\mathbb{E}_{U}\) denotes the average over the \(SU(2)\)-symmetric unitaries parametrized by \(\theta\). On the other hand, a projective measurement forces all \(2Q\) branches to carry the same outcome,
\begin{equation}
\cT_m=\prod_{\alpha=1}^{2Q}P_s^\alpha+
\prod_{\alpha=1}^{2Q}P_t^\alpha.
\label{eq:Tm}
\end{equation}
The averaged R\'enyi entropy in Eq.~\eqref{eq:renyiTN} can then be expressed as the contraction of a spacetime tensor network, as depicted in Fig.~\ref{fig:tensor-network},
\begin{equation}
\begin{aligned}
    S_A^{(n)}
    =
    \lim_{k\to 0}
    \frac{1}{k(1-n)}&[
    \langle T_{n,A}^{\otimes k}|
    \prod_v \cT_v
    |\rho_0^{\otimes Q}\rangle
    -
    \langle e|
    \prod_v \cT_v
    |\rho_0^{\otimes Q}\rangle ].
\end{aligned}
\label{eq:statmech}
\end{equation}
Here we have used the standard state--operator mapping to represent operators as states in a doubled Hilbert space. The bra \(\langle e|\) denotes the identity permutation, corresponding to the trace with untwisted boundary conditions. The local tensor at each spacetime vertex $v$ is the average of the unitary and measurement tensors, weighted by their respective probabilities,
\begin{equation}
    \cT_v=(1-p)\cT_u+p\cT_m .
    \label{eq:localtensorp}
\end{equation}

The tensor-network contraction can equivalently be understood as the partition function of a loop model. Define
\begin{equation}
    E_i=\nTL P_{s,i}=\ket{\epsilon_i}\bra{\epsilon_i},
    \qquad
    \ket{\epsilon_i}
    =
    \ket{\uparrow_i\downarrow_{i+1}}
    -
    \ket{\downarrow_i\uparrow_{i+1}},
    \label{eq:e-def}
\end{equation}
where \(\ket{\epsilon_i}\) is an unnormalized singlet state on bond \((i,i+1)\), of squared norm \(\nTL=2\). The triplet projector on the same bond is then
\begin{equation}
    P_{t,i}=1-\frac{E_i}{\nTL} .
    \label{eq:nTL}
\end{equation}
The operators \(\{E_i\}\) obey the Temperley--Lieb (TL) algebra,
\begin{equation}
\begin{split}
    E_i^2&=\nTL E_i,
    \qquad
    E_i E_{i\pm1}E_i=E_i, \\
    E_i E_j&=E_j E_i,
    \qquad |i-j|>1,
\end{split}
\label{eq:TL-rel}
\end{equation}
with loop weight \(\nTL=2\). Therefore, expanding the tensor-network contraction in Eq.~\eqref{eq:statmech} in the diagrammatic representation of the TL algebra gives a fully packed loop model~\cite{TemperleyLieb1971}. To see this, we regard the qubit worldlines between gates as links, and assign a two-state spin index to each link.\footnote{
The spin index on a link can be identified with the original qubit Hilbert
space after a particle--hole conjugation on one sublattice.  Equivalently, on
the even sublattice we use the conjugated basis
\(\ket{\tilde\uparrow}=\ket{\downarrow}\) and
\(\ket{\tilde\downarrow}=-\ket{\uparrow}\), so that the bond singlet becomes
\(\ket{\epsilon}=\ket{\uparrow\tilde\uparrow}
+\ket{\downarrow\tilde\downarrow}\). In this basis, the unnormalized bond-singlet state is represented diagrammatically by the cup connectivity in Fig.~\ref{fig:TL-local-patterns}.} The tensor at each vertex is then a linear combination of tensor products, over the \(2Q\) branches, of the two local connectivities \(1\) and \(E_i\) of the links, as shown in Fig.~\ref{fig:TL-local-patterns}. The spin index is conserved along each continuous line. After tracing over the spin indices on the links, each closed loop contributes a factor \(\nTL=2\).

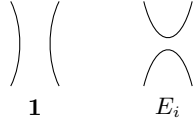
\begin{figure}[t]
    \centering
    \begin{tikzpicture}[scale=0.8,baseline=(current bounding box.center)]
        \draw (0,0).. controls (0.2,0.4) and (0.2,1.0) ..(0,1.4);
        \draw (0.8,0).. controls (0.6,0.4) and (0.6,1.0) ..(0.8,1.4);
        \node at (0.4,-0.35) {\(\mathbf 1\)};

        \begin{scope}[xshift=2.2cm]
            \draw (0,0) .. controls (0.2,0.8) and (0.6,0.8) .. (0.8,0);
            \draw (0,1.4) .. controls (0.2,0.6) and (0.6,0.6) .. (0.8,1.4);
            \node at (0.4,-0.35) {\(E_i\)};
        \end{scope}
    \end{tikzpicture}
    \caption{The two elementary TL connectivities appearing in the local
    tensor: the identity connectivity and the cup-cap connectivity \(E_i\).
    Expanding the tensor network in these local connectivities produces a
    fully packed loop configuration, with each closed loop weighted by
    \(\nTL=2\).}
    \label{fig:TL-local-patterns}
\end{figure}

For later convenience, we express the local tensor, which determines the vertex weight in the loop model, in the connectivity basis. We expand the projectors \(P_f\), with \(f=s,t\), in terms of the two local connectivities \(\eta=0,1\):
\begin{equation}
    P_f=\sum_{\eta=0,1}C_f(\eta)E^\eta,
    \qquad
    E^0=\mathbf 1,\quad E^1=E ,
\end{equation}
with
\begin{equation}
    C_s(0)=0,
    \quad C_s(1)=\frac{1}{\nTL},
    \qquad
    C_t(0)=1,
    \quad C_t(1)=-\frac{1}{\nTL} .
    \label{eq:C-physical}
\end{equation}
The unitary tensor in Eq.~\eqref{eq:Tu} contains an equal number of singlet fusions in the forward and backward branches. Therefore, for each term, the backward fusion pattern can be matched to the forward fusion pattern by a permutation. We define the pairing permutation as a map
\begin{equation}
    \sigma\in S_Q:\; B\longrightarrow F .
    \label{eq:sigma-direction}
\end{equation}
Choose a forward branch pattern \(\bm f=(f_a)_{a\in F}\), and denote by $N_s=\sum_{a\in F}\delta_{f_a,s}$ the number of singlet fusions in this pattern. The backward branch is then obtained by assigning to the backward branch \(\bar b\) the channel \(f_{\sigma(b)}\). Thus the unitary tensor can be written as a sum over pairing permutations,
\begin{equation}
\begin{split}
    \cT_u
    &= \sum_{\sigma\in S_Q}
    \sum_{\bm \eta}
    U_\sigma(\bm\eta) \mathcal E(\bm\eta),\\
    \mathcal E(\bm\eta) &= \prod_{a=1}^{Q}(E^a)^{\eta^a} \prod_{b=1}^{Q}(E^{\bar b})^{\eta^{\bar b}},\\
    U_\sigma(\bm\eta)
    &=
    \sum_{\bm f}
    \frac{1}{N_s!\,(Q-N_s)!}
    \prod_{a\in F}C_{f_a}(\eta^a)
    \prod_{b\in B}C_{f_{\sigma(b)}}(\eta^{\bar b}) .
\end{split}
\label{eq:U-sigma}
\end{equation}
Here $\bm\eta=\{\eta^a,\eta^{\bar b}\}_{a,b=1}^{Q}$ specifies the loop connectivity in all \(2Q\) branches, and the denominator in the third line arises from the stabilizer subgroup
\(S_{N_s}\times S_{Q-N_s}\), which leaves \(\bm f\) invariant. On the other hand, the measurement tensor is \(\sigma\)-independent:
\begin{equation}
    \cT_m
    =
    \sum_{\bm\eta}
    T_m(\bm\eta)\,
    \mathcal E(\bm\eta),
\end{equation}
with
\begin{equation}
    T_m(\bm\eta)
    =
    \frac{1}{\nTL^{2Q}}\delta_{N_\eta,2Q}
    +
    \left(-\frac{1}{\nTL}\right)^{N_\eta},
    \qquad
    N_\eta=\sum_{\alpha=1}^{2Q}\eta^\alpha .
    \label{eq:mtensorloop}
\end{equation}
Here the first term is the contribution from the singlet measurement outcome, while the second term is the contribution from the triplet outcome.

For the moment, ignoring boundary conditions, we define the partition function of the loop model by summing over all possible connectivity configurations:\begin{equation}
\begin{split}
Z_Q(p) = \Tr\prod_v \cT_v &= \sum_{\{\bm \eta_v\}} \Tr  \prod_v \sum_{\sigma_v\in S_Q} [ \frac{p}{Q!}T_m(\bm\eta_v) + (1-p)U_{\sigma_v}(\bm\eta_v) ] \mathcal E_v(\bm\eta_v) ,
\end{split}
\label{eq:Z-loop-physical}
\end{equation}
where \(\Tr\) denotes the trace over the spin indices on the links. This model has a nontrivial interaction at each vertex.

\section{A large loop-fugacity expansion}
\label{sec:separation}

The loop model at \(\nTL=2\) has no small parameter, and is therefore strongly interacting at the lattice scale. To obtain one, we note that the relations of Sec.~\ref{sec:loop-model} involve the loop fugacity only through \(E_i^2=\nTL E_i\), \(P_s=E_i/\nTL\) and \(P_t=1-E_i/\nTL\), together with the coefficients in Eqs.~\eqref{eq:C-physical} and~\eqref{eq:mtensorloop}. We keep these relations and treat \(\nTL\) as a free parameter, which we take to be large. The physical circuit is recovered at \(\nTL=2\).
For integer \(\nTL\), one concrete realization of this \(\nTL\)-generalization is to place an \(SU(\nTL)\) fundamental representation on one sublattice, say the even sublattice, and an \(SU(\nTL)\) anti-fundamental representation on the other sublattice. In this realization, \(E_i\) is the unnormalized projector onto the singlet channel on a nearest-neighbor bond.

The idea is as follows. Motivated by the theory of monitored circuits without symmetry~\cite{Bao2020,Jian2020,2026Vasseurnotes}, and by the form of Eq.~\eqref{eq:Z-loop-physical} as a sum over permutations, we separate the partition function into two coupled sectors. The pairing sector consists of the variables \(\sigma_v\in S_Q\) on each vertex \(v\). The background sector consists of the spin indices on the links. The effective actions of both sectors and their coupling are organized as an expansion in \(1/\nTL\). At leading order the background sector is a \(z=2\) spin-wave theory. Integrating it out generates an effective coupling between the pairing variables, which then undergo an order-disorder transition as \(p\) is tuned. This transition will later be identified with the spin-sharpening transition.

More precisely, the partition function in Eq.~\eqref{eq:Z-loop-physical} is invariant under an
\(S_Q^F\times S_Q^B\) symmetry, which independently permutes the forward and the backward branches~\cite{bao2021symmetry}. In the spin-fuzzy phase, where the observer learns the total spin only slowly, this symmetry is spontaneously broken to the diagonal subgroup \(S_Q^{\rm diag}\). The order-parameter space is then the discrete coset
\begin{equation}
    \frac{S_Q^F\times S_Q^B}{S_Q^{\rm diag}}
    \simeq S_Q,
\end{equation}
which is parameterized by our pairing variable \(\sigma_v\).

We proceed in the following order. Section~\ref{subsec:sector-separation} states the vertex expansion and the resulting effective action for the pairings. Section~\ref{app:tensor-expansion} derives the expansion. Section~\ref{sec:background} analyzes the background sector, and Sec.~\ref{app:kernel-A} computes the induced interaction. Section~\ref{sec:exact-ferro} checks the ferromagnetic background exactly at \(\nTL=2\).

\subsection{\texorpdfstring{{ The two sectors and the induced ferromagnetic interaction}}{The two sectors and the induced ferromagnetic interaction}}
\label{subsec:sector-separation}

As shown in Sec.~\ref{app:tensor-expansion}, after expanding the local tensors in Eqs.~\eqref{eq:U-sigma} and~\eqref{eq:mtensorloop} in the local connectivity basis, the leading term that depends on the choice of \(\sigma\) at a vertex appears at order \(O(1/\nTL^2)\):
\begin{equation}
\begin{split}
    &\cT_v = \sum_{\sigma_v\in S_Q} \cT_v(\sigma_v) \\
    &\cT_v(\sigma_v)
    =
    \cT_{\mathrm{bg}}
    +
    g\sum_{a,b=1}^Q
    M_{ab}(\sigma_v)\chi_{ab}(v)
    +
    \bigO{\nTL^{-3}},
    \\
    &\chi_{ab}(v)
    =
    E_v^aE_v^{\bar b},
    \qquad
    g=\frac{1-p}{(Q-1)!\,\nTL^2}.
\end{split}
\label{eq:T-separation}
\end{equation}
Here
\begin{equation}
    \cT_{\mathrm{bg}}
    =
    \frac{1}{Q!}\left[
    \prod_{\alpha=1}^{2Q}
    \left(1-\frac{E^\alpha}{\nTL}\right)
    +
    \frac{1-p}{\nTL^2}
    \sum_{a,b=1}^Q
    \chi_{ab}
    \right]
    \label{eq:T-bg}
\end{equation}
describes the pairing-independent background sector. The operator \(\chi_{ab}(v)\) places a singlet at vertex \(v\) on the forward branch \(a\) and on the backward branch \(\bar b\). The pairing enters only through the tensor
\begin{equation}
    M_{ab}(\sigma)
    =
    \delta_{a,\sigma(b)}-\frac1Q .
    \label{eq:M-def}
\end{equation}
Under permutations of the forward and backward branches,
\begin{equation}
    M_{ab}(\sigma)
    \mapsto
    M_{ab}(g_F^{-1}\sigma g_B)
    =
    M_{g_F a,g_B b}(\sigma).
    \label{eq:standardrep}
\end{equation}
Thus \(M_{ab}\) has no replica-symmetric part, \(\sum_a M_{ab}=\sum_b M_{ab}=0\), which is the property used below. In the language of group theory, it transforms in the standard representation of both \(S_Q^F\) and \(S_Q^B\). It mediates the coupling between the pairing sector and the background sector.

The effective model of the pairings emerges after integrating out the background sector. For a fixed spacetime configuration \(\{\sigma_v\}\), let
\begin{equation}
    Z[\sigma]
    =
    \Tr \prod_v \cT_v(\sigma_v).
\end{equation}
The full partition function in Eq.~\eqref{eq:Z-loop-physical} is then recovered by summing over all configurations \(\{\sigma_v\}\). The induced interaction between the pairing variables is obtained from the standard cumulant expansion,
\begin{align}
    -S_{\rm eff}[\sigma]
    ={}&
    g\sum_{v,a,b}
    M_{ab}(\sigma_v)
    \avg{\chi_{ab}(v)}_{\rm bg} +
    \frac{g^2}{2}
    \sum_{v,w}
    \sum_{a,b,c,d}
    M_{ab}(\sigma_v)M_{cd}(\sigma_w)
    \avg{\chi_{ab}(v)\chi_{cd}(w)}_{{\rm bg},c}
    +\cdots .
    \label{eq:cumulant}
\end{align}
Here \(\avg{\cdots}_{{\rm bg},c}\) denotes the connected correlation function with respect to the background sector, and we keep terms up to \(O(\nTL^{-4})\). Orders in \(1/\nTL\) count the explicit powers in the vertex weights of Eq.~\eqref{eq:T-separation}, with the correlators of \(E\) in the background sector counted as order one. The precise definition of these correlators is given in Sec.~\ref{app:tensor-expansion}.

Symmetry simplifies the effective action. As shown in Sec.~\ref{sec:background}, the background sector consists of \(2Q\) effectively decoupled branches of spin waves. In the bulk, away from the initial and final times, it is therefore invariant under \(S_Q^F\times S_Q^B\), and
\begin{equation}
    \avg{\chi_{ab}(v)}_{\mathrm{bg}}=\mu,
\end{equation}
independent of \(a,b\). The final-time boundary \(\langle e|\) glues each forward branch to its own backward partner and is not invariant. It acts as a boundary field on the pairings, which we use in Sec.~\ref{sec:learning}. Therefore, the \(O(g)\) term in Eq.~\eqref{eq:cumulant} vanishes after contraction with \(M_{ab}\). Similarly, the correlator in the second term of Eq.~\eqref{eq:cumulant} can be expanded in terms of the four-index invariant tensors of \(S_Q^F\times S_Q^B\) as
\begin{align}
    \avg{\chi_{ab}(v)\chi_{cd}(w)}_{{\rm bg},c}
    ={}&
    A(v,w)\delta_{ac}\delta_{bd}
    +B(v,w)\delta_{ac} +
    C(v,w)\delta_{bd}
    +D(v,w).
    \label{eq:rank-four}
\end{align}
After contraction with \(M_{ab}(\sigma_v)M_{cd}(\sigma_w)\), only the first term survives, giving
\begin{equation}
    S_{\rm eff}[\sigma]
    =
    -\frac{g^2}{2}
    \sum_{v,w}
    A(v,w)
    \left[
    \Fix(\sigma_w^{-1}\sigma_v)-1
    \right]
    +\cdots .
    \label{eq:Seff-main}
\end{equation}
Here \(\Fix(\sigma)\) denotes the number of fixed points of \(\sigma\in S_Q\). Thus \(\Fix(\sigma_w^{-1}\sigma_v)\) counts the backward branches that are paired with the same forward branch at \(v\) and at \(w\). We show in Sec.~\ref{app:kernel-A} that \(A(v,w)\ge0\). The induced interaction therefore favors equal pairings and is ferromagnetic. This is the leading interaction in the pairing sector.  It is generated by the \(O(\nTL^{-2})\) term in the local tensor and appears at \(O(\nTL^{-4})\).  In Sec.~\ref{app:tensor-expansion}, we also justify that keeping only the \(O(\nTL^{-2})\) term in the local tensor is sufficient for identifying this leading effective interaction for \(\{\sigma_v\}\).

\subsection{\texorpdfstring{{ Vertex expansion to order $\nTL^{-2}$ and why it suffices}}{Vertex expansion to order 1/n2 and why it suffices}}
\label{app:tensor-expansion}
\label{app:large-n}

We now derive Eq.~\eqref{eq:T-separation} and show that higher orders in \(1/\nTL\) do not change the leading effective action. Each \(E\) at a vertex costs a factor \(1/\nTL\) [Eq.~\eqref{eq:C-physical}]. The pairing permutation relates forward to backward branches, so a term can depend on \(\sigma\) only if it contains an \(E\) on a forward branch and an \(E\) on a backward branch. The leading \(\sigma\)-dependent term therefore comes from the unitary tensor \(U_\sigma(\bm\eta)\) at order \(O(\nTL^{-2})\). Here we display the tensor through this order, which corresponds to two \(E\)-type connectivities at a vertex \(v\).  These can be
\begin{enumerate}
\item one \(E\) on a forward branch and one \(E\) on a backward branch, i.e.,
\begin{equation}
 \eta^a=\eta^{\bar b}=1,
\end{equation}
with all other \(\eta\)'s equal to \(0\);
\item both \(E\)'s on forward branches or both on backward branches.
\end{enumerate}
For the moment, let us focus on the unitary tensor and analyze the
contribution from case (1).  To obtain a nonzero contribution, the
structure of the projector \(P_f\) [Eq.~\eqref{eq:nTL}] forces
\begin{equation}
 f_c=t,\quad \forall c\neq a,
\qquad
 f_{\sigma(d)}=t,\quad \forall d\neq b.
\end{equation}
There are two possibilities.  The first is \(a\neq \sigma(b)\).  Then all
\(f_i\) must be in the \(t\) channel, giving the vertex weight
\begin{equation}
 U_{\sigma}\big[\eta^a=\eta^{\bar b}=1,\ a\neq \sigma(b)\big]
 =
 \frac{1}{Q!\,\nTL^2}.
\end{equation}
The second possibility is \(a=\sigma(b)\).  Then \(f_a\) can be either
\(t\) or \(s\), giving the vertex weight
\begin{equation}
 U_{\sigma}\big[\eta^a=\eta^{\bar b}=1,\ a=\sigma(b)\big]
 =
 \frac{1}{Q!\,\nTL^2}
 +
 \frac{1}{(Q-1)!\,\nTL^2}.
\end{equation}
Combining the two cases, we have
\begin{equation}
 U_{\sigma}\big[\eta^a=\eta^{\bar b}=1\big]
 =
 \frac{1}{Q!\,\nTL^2}
 +
 \frac{\delta_{a,\sigma(b)}}{(Q-1)!\,\nTL^2}.
\end{equation}
One can carry out the analysis in the same way for case (2), as well as
for the measurement tensor.  Summing over all possible connectivities, we
find that, for fixed \(Q>1\), up to order \(O(\nTL^{-2})\), the local
tensor at each spacetime vertex can be written as
\begin{equation}
\begin{split}
    \cT & = \sum_{\sigma\in S_Q} \cT(\sigma),\\
    \cT(\sigma)
    &=
    \frac{1}{Q!}\Bigg\{
    1-\frac1{\nTL}
    \Big(\sum_{a=1}^{Q} E^a+\sum_{b=1}^{Q} E^{\bar b}\Big)
    \\
    &\quad
    +\frac{1}{\nTL^2}
    \Big[
    \sum_{1\le a<b\le Q} E^aE^b
    +\sum_{1\le a<b\le Q} E^{\bar a} E^{\bar b}
    +\sum_{a,b=1}^Q E^a E^{\bar b}
    +Q(1-p)\sum_{a=1}^{Q} E^{\sigma(a)} E^{\bar a}
    \Big]\Bigg\}
    +O\Big(\frac1{\nTL^3}\Big),
\end{split}
\label{eq:fullvertexn}
\end{equation}
which is Eq.~\eqref{eq:T-separation} of Sec.~\ref{subsec:sector-separation}.

We now define the correlators with respect to the background sector that appear in Eq.~\eqref{eq:cumulant}.  Let
\begin{equation}
Z_{\rm bg}
=
\Tr\prod_v\cT_{\rm bg}
\end{equation}
be the partition function of the background sector alone.  Then
\begin{equation}
\avg{\chi_{ab}(v)}_{\rm bg}
=
\frac{1}{Z_{\rm bg}}
\Tr\left[
\chi_{ab}(v)\prod_{u\ne v}\cT_{{\rm bg},u}
\right],
\end{equation}
and similarly
\begin{equation}
\avg{\chi_{ab}(v)\chi_{cd}(w)}_{{\rm bg},c}
=
\avg{\chi_{ab}(v)\chi_{cd}(w)}_{\rm bg}
-
\avg{\chi_{ab}(v)}_{\rm bg}
\avg{\chi_{cd}(w)}_{\rm bg}.
\end{equation}
Expanding \(Z[\sigma]=\Tr\prod_v\cT_v(\sigma_v)\) around \(Z_{\rm bg}\) to \(O(g^2)\) and taking the logarithm gives the cumulant expansion in Eq.~\eqref{eq:cumulant}.

Finally, we justify that retaining terms up to \(O(\nTL^{-2})\) at each vertex
is sufficient to identify the leading contribution in \(1/\nTL\) to the
effective action for the pairing variables.  As shown in the
analysis leading to Eq.~\eqref{eq:fullvertexn}, the first
\(\sigma\)-dependent term in the tensor appears at order \(O(\nTL^{-2})\)
through \(M_{ab}(\sigma)\chi_{ab}\), and $M_{ab}$ transforms in the standard
representation of both \(S_Q^F\) and \(S_Q^B\).  Its second cumulant gives an
\(O(\nTL^{-4})\) contribution to \(S_{\mathrm{eff}}[\sigma]\).  We now show that every higher-order \(\sigma\)-dependent local term contributes only at order \(O(\nTL^{-5})\) or smaller. The reason is the same as for the first cumulant above. The background treats all branches alike, so a single \(\sigma\)-dependent vertex averages to a constant that does not depend on \(\sigma\).

Consider a \(\sigma\)-dependent term at order \(O(\nTL^{-k})\) in the
unitary tensor in Eq.~\eqref{eq:U-sigma}, with \(k\ge 3\).  Such a term
must contain at least one forward \(E\) insertion and at least one backward
\(E\) insertion, since the \(\sigma\) dependence arises precisely from the
pairing between forward and backward branches.  Thus we may label such a
term by two subsets \(I,J\subset\{1,\dots,Q\}\), with
\begin{equation}
    |I|=r,\qquad |J|=s,\qquad r,s\ge 1,\qquad r+s=k,
\end{equation}
and write the corresponding \(E\)-connectivity operator as
\begin{equation}
    \chi_{I,J}(v)
    =
    \left(\prod_{a\in I} E_v^a \right)
    \left(\prod_{b\in J}E^{\bar b}_v\right).
\end{equation}
Let \(K_{I,J}(\sigma)\) denote its coefficient in the local tensor in
Eq.~\eqref{eq:T-separation}.  For fixed \(r,s\), define
\begin{equation}
    X_{r,s}:=
    \left\{
    (I,J):
    I\subset \{1,\dots,Q\},\ |I|=r;\
    J\subset \{1,\dots,Q\},\ |J|=s
    \right\}.
\end{equation}
This is the set of all subsets with the correct numbers of \(E\) insertions
in the forward and backward branches.  The only property we need for the
later argument is the covariance of \(K_{I,J}(\sigma)\) under independent
permutations of the forward and backward branches, as an analogue of
Eq.~\eqref{eq:standardrep}:
\begin{equation}
K_{g_F I,\; g_B J}\!\left(g_F \sigma g_B^{-1}\right)
=
K_{I,J}(\sigma),
\qquad
g_F\in S_Q^F,
\quad
g_B\in S_Q^B,
\end{equation}
which follows from the definition of the pairing in
Eq.~\eqref{eq:sigma-direction}.  Now define the average over all pairs
\((I,J)\in X_{r,s}\) at fixed \(\sigma\),
\begin{equation}
    \bar K_{r,s}(\sigma):=
\frac{1}{|X_{r,s}|}
\sum_{(I,J)\in X_{r,s}}
K_{I,J}(\sigma).
\end{equation}
Then
\begin{equation}
    \begin{split}
        \bar K_{r,s}\left(g_F \sigma g_B^{-1}\right)
        &=
\frac{1}{|X_{r,s}|}
\sum_{(I,J)\in X_{r,s}}
K_{I,J}\left(g_F \sigma g_B^{-1}\right) \\
&=
\frac{1}{|X_{r,s}|}
\sum_{(I,J)\in X_{r,s}}
K_{g_F^{-1}I,\;g_B^{-1}J}(\sigma) \\
&=
\frac{1}{|X_{r,s}|}
\sum_{(I',J')\in X_{r,s}}
K_{I',J'}(\sigma)
=
\bar K_{r,s}(\sigma),
    \end{split}
\end{equation}
where we have relabeled \(I'=g_F^{-1}I\) and \(J'=g_B^{-1}J\).  Since
\((I,J)\mapsto (g_F^{-1}I,g_B^{-1}J)\) is just a permutation of
\(X_{r,s}\), the sum is unchanged.  Thus \(\bar K_{r,s}\) is invariant
under the full left-right action of \(S_Q^F\times S_Q^B\) on \(S_Q\).
Since this action is transitive, \(\bar K_{r,s}(\sigma)\) must be
independent of \(\sigma\):
\begin{equation}
    \bar K_{r,s}(\sigma)=\bar K_{r,s}.
\end{equation}
We can therefore decompose
\begin{equation}
    K_{I,J}(\sigma)
    =
    \bar K_{r,s}
    +
    M_{I,J}(\sigma),
\qquad
\sum_{(I,J)\in X_{r,s}} M_{I,J}(\sigma)=0,
\end{equation}
into the average component $\bar K_{r,s}$ and the components that transform in nontrivial representations of \(S_Q^F\times S_Q^B\). On the other hand, by the \(S_Q^F\times S_Q^B\) symmetry of the background
sector, the one-point function of \(\chi_{I,J}\) depends only on \(r\) and
\(s\):
\begin{equation}
    \langle \chi_{I,J}(v)\rangle_{\mathrm{bg}}=\mu_{r,s}(v),
\end{equation}
independent of the particular pair \((I,J)\in X_{r,s}\).  Therefore
\begin{equation}
    \sum_{(I,J)\in X_{r,s}}
M_{I,J}(\sigma_v)
\langle \chi_{I,J}(v)\rangle_{\mathrm{bg}}
=
\mu_{r,s}(v)
\sum_{(I,J)\in X_{r,s}}
M_{I,J}(\sigma_v)
=
0.
\end{equation}
So the first cumulant of every higher-order local term contributes only a
\(\sigma\)-independent constant.

It then follows that the leading nontrivial \(\sigma\) dependence comes
from the second cumulant of the \(O(\nTL^{-2})\) term in the vertex.  Any mixed
cumulant between the \(O(\nTL^{-2})\) vertex and an \(O(\nTL^{-k})\),
\(k\ge 3\), vertex is of order \(O(\nTL^{-k-2})\le O(\nTL^{-5})\), while
the second cumulant of two higher-order vertices is of order
\(O(\nTL^{-2k})\le O(\nTL^{-6})\).  Hence the leading \(\sigma\)-dependent
effective action is completely determined by the local tensor through
order \(O(\nTL^{-2})\).

\subsection{\texorpdfstring{{ The background sector: $2Q$ ferromagnetic branches with $z=2$ spin waves}}{The background sector: 2Q ferromagnetic branches with z=2 spin waves}}
\label{sec:background}

We now ask what the background sector is at long distances. Its dynamics is determined by the tensor in Eq.~\eqref{eq:T-bg}. An important observation is that at \(p=1\), the measurement-only limit, the tensor factorizes branch by branch up to \(o(\nTL^{-2})\) corrections, so that the \(2Q\) branches decouple.\footnote{The neglected piece is the term of order \(\nTL^{-2Q}\) in which all \(2Q\) branches fuse into the singlet at once, which is \(o(\nTL^{-2})\) at any fixed \(Q>1\). Importantly, it is irrelevant at the $z=2$ Gaussian fixed point of the $2Q$ spin waves for any $Q\ge 1$. At any integer $Q>1$ and physical $\nTL=2$ the statement is exact: with this term included, the leading eigenvectors are still the ferromagnetic states of the $2Q$ branches (Sec.~\ref{sec:exact-ferro}).} At \(p<1\), \(\cT_{\mathrm{bg}}\) contains local forward-backward branch-coupling terms at order \(\nTL^{-2}\). We therefore first examine the limit \(p=1\). We then show that the coupling between forward and backward branches is irrelevant (Sec.~\ref{subsec:irrelevance}).

The factor \(1-E^\alpha/\nTL\) is the projector onto the non-singlet channel of a bond, which is the limit of \(\exp[-aE^\alpha]\) at large \(a\). A finite \(a\) leaves the same states invariant, namely those annihilated by \(E^\alpha\). For long-wavelength purposes, we can therefore replace the discrete local gate at each vertex by a softened continuous-time evolution,
\begin{equation}
    1-\frac{E^\alpha}{\nTL}
    \quad \Longrightarrow \quad
    \exp\!\left[-\delta\tau J E^\alpha\right].
\end{equation}
We define the transfer matrix of the background sector as the product of all tensors within one period, with all internal legs contracted, i.e.,
\begin{equation}
    T_{\rm bg}
    =
    \prod_{v\in {\rm period}} \cT_{\mathrm{bg},v}.
\end{equation}
At $p=1$, the transfer matrix of the background sector can be approximated by an imaginary-time evolution,
\begin{equation}
    T_{\mathrm{bg}}^{(p=1)}
    \quad \Longrightarrow \quad
    e^{-\delta\tau H_{\mathrm{bg}}},
\end{equation}
with
\begin{equation}
    H_{\mathrm{bg}}
    =
    J\sum_{\alpha=1}^{2Q}\sum_x
    E_x^\alpha,\quad J>0,
    \label{eq:Hbg}
\end{equation}
where \(E_x^\alpha\), \(\alpha=1,\ldots,2Q\), is the unnormalized singlet projector, or TL generator, on bond \((x,x+1)\).

The \(2Q\) branches in Eq.~\eqref{eq:Hbg} decouple, so we can analyze the dynamics in a single branch.  We convert the imaginary-time evolution into a coherent-state path integral. We group two sites into a unit cell, and place an \(SU(\nTL)\) fundamental coherent state \(z_j\in\mathbb C^{\nTL}\) on the even sublattice and an anti-fundamental coherent state \(w_j\in\mathbb C^{\nTL}\) on the odd sublattice:
\begin{equation}
    z_j^\dagger z_j=w_j^\dagger w_j=1,
\end{equation}
with \(z_j\) and \(w_j\) each defined up to a phase. As shown in Sec.~\ref{app:coherent-state}, the coherent-state path integral takes the form
\begin{equation}
\begin{split}
    e^{-H_{\rm bg}t}
    &\Longrightarrow
    \int\mathcal{D}[z,w]\, e^{-S_E[z,w]},\\
    S_E[z,w]
    &=
    S_B+\int d\tau\, H_{\mathrm{cl}}[z,w].
\end{split}
\label{eq:effectiveactionbg}
\end{equation}
The Euclidean Berry-phase term is
\begin{equation}
    S_B
    =
    \int d\tau
    \sum_j
    \left(
    z_j^\dagger\partial_\tau z_j
    -
    w_j^\dagger\partial_\tau w_j
    \right),
    \label{eq:Berry-alt}
\end{equation}
where the relative minus sign comes from the conjugate representation on the odd sublattice. The classical energy entering the path integral is
\begin{equation}
    H_{\rm cl}
    =
    J
    \sum_j
    \left[
    |w_j^\dagger z_j|^2
    +
    |w_j^\dagger z_{j+1}|^2
    \right].
    \label{eq:Hcl-alt}
\end{equation}
The classical ground-state configurations are obtained by minimizing this energy:
\begin{equation}
    w_j^\dagger z_j=0,
    \qquad
    w_j^\dagger z_{j+1}=0
    \qquad
    \text{for all }j .
    \label{eq:zero-mode-constraint}
\end{equation}

For the physical value \(\nTL=2\), the constraint in Eq.~\eqref{eq:zero-mode-constraint} is sufficient to force neighboring fundamental spins to align:
\begin{equation}
    z_{j+1}\parallel z_j .
\end{equation}
Similarly, \(w_j\) is fixed, up to a phase, by the condition that it is orthogonal to \(z_j\). Using the pseudoreality of the \(SU(2)\) fundamental representation, the anti-fundamental coherent state \(w_j\) can be identified with a fundamental coherent state aligned with \(z_j\). The low-energy manifold is therefore the usual ferromagnetic \(\mathbb{CP}^1\) manifold.

However, for \(\nTL>2\), the same constraint is underdetermined. Given two generic neighboring vectors \(z_j\) and \(z_{j+1}\), the allowed anti-fundamental vector can be chosen in
\begin{equation}
    w_j
    \in
    \left[
    \mathrm{span}(z_j,z_{j+1})
    \right]^\perp,
\end{equation}
whose complex dimension is \(\nTL-2\). Therefore, spatially varying textures of the \(z_j\)'s can remain at exactly zero energy after choosing appropriate \(w_j\)'s locally. Thus, for \(\nTL>2\), the classical Hamiltonian contains exact local zero modes, which are absent at the physical value \(\nTL=2\). We therefore regard these zero modes as artifacts of the specific \( SU(\nTL) \) generalization, rather than physical degrees of freedom continuously connected to the original qubit system.

We therefore introduce a small \(SU(\nTL)\)-invariant regulator that lifts these zero modes. A natural choice is a same-sublattice ferromagnetic stiffness,
\begin{equation}
\begin{aligned}
    H_{\rm lift}
    =
    \lambda
    \sum_j
    \left[
    1-\mathsf P^F_{j,j+1}
    +
    1-\mathsf P^{\bar F}_{j,j+1}
    \right],
    \qquad
    \lambda>0 ,
\end{aligned}
\label{eq:Hlift}
\end{equation}
where \(\mathsf P^F\) swaps two neighboring fundamental sites on the even sublattice, and \(\mathsf P^{\bar F}\) swaps two neighboring anti-fundamental sites on the other sublattice. One circuit realization of this regulator is to add weak, random $SU(\nTL)$-symmetric swap gates $e^{-i\phi\,\mathsf P_{j,j+1}}$ between neighboring \emph{same-sublattice} sites (even-even or odd-odd), with $\phi$ small and random: averaging over $\phi$ generates Eq.~\eqref{eq:Hlift} with $\lambda\propto\overline{\phi^{2}}$. In the path integral, the term in Eq.~\eqref{eq:Hlift} gives an additional contribution to the classical energy,
\begin{equation}
    H_{{\rm lift},{\rm cl}}
    =
    \lambda
    \sum_j
    \left[
    1-|z_j^\dagger z_{j+1}|^2
    +
    1-|w_j^\dagger w_{j+1}|^2
    \right].
    \label{eq:Hlift-cl}
\end{equation}
Physically, this term selects the physical low-energy configurations as those with smooth textures, which are continuously connected to the physical \(\nTL=2\) problem. The correct order of limits in our theory is therefore as follows: we first compute physical quantities at finite regulator \(\lambda\) and generic \(\nTL\), then analytically continue to the physical value \(\nTL=2\), and only at the end take \(\lambda\to0\).

\subsubsection{\texorpdfstring{{ Fixed total spin: a slowly winding spiral with $z=2$ spin waves}}{Fixed total spin: a slowly winding spiral with z=2 spin waves}}
\label{sec:fixedS}

Now suppose the initial state carries a fixed total spin $S=O(1)$, the situation relevant for the numerics of Ref.~\cite{Majidy2023}, where $S=0$. Since every gate and every measurement commute with the total-spin Casimir, each branch is confined for all times to the spin-$S$ sector. The long-time dynamics in the background sector is then governed by the low-lying states of $H_{\rm bg}$ in Eq.~\eqref{eq:Hbg}, restricted to this sector.

It is illuminating to first examine the physical case \(\nTL=2\), to understand the ground state and its low-lying excitations. At \(\nTL=2\), it is convenient to write the spin-coherent-state path integral in terms of a unit vector $\vec n = z^\dagger \vec \sigma z$. At \(p=1\), where the branches decouple, the effective action of a single branch in Eq.~\eqref{eq:effectiveactionbg} becomes
\begin{equation}
S=\int dt\,dx\,\Big[\,s\,(\cos\theta-1)\,\partial_t\varphi
-\frac{\rho_s}{2}\,(\partial_x\vec n)^2\,\Big],
\label{eq:LLaction}
\end{equation}
where $(\theta,\varphi)$ are the polar coordinates of $\vec n$, $s=\frac{1}{2}$ is the spin per site, and $\rho_s$ is a non-universal stiffness. Here the first term is the Berry phase term, whereas the second is the energy. Fixing the total spin of a branch constrains its textures:
\begin{equation}
    s \int dx\, \vec{n}(x) = \vec S_{\rm tot}.
    \label{eq:n2-spin-constraint}
\end{equation}
Minimizing the energy at fixed total spin then selects the minimal-winding cone,
\begin{equation}
\vec n_0(x)=(\sin\theta\cos q_0x,\ \sin\theta\sin q_0x,\ \cos\theta),
\label{eq:spiral}
\end{equation}
with $q_0=\frac{2\pi}{L}$, the smallest winding compatible with the periodicity of $\vec n$, and the tilt fixed by the constraint, $\cos\theta=S/(sL)=2S/L$. The singlet sector is the equatorial case $\theta=\frac{\pi}{2}$. In fact, the lowest eigenstates of the ferromagnetic Hamiltonian [Eq.~\eqref{eq:Hbg}] at fixed total spin are known \cite{sutherland1995low,dhar2000bloch}: they are Bethe strings containing a macroscopic number of magnons, identified in Ref.~\cite{dhar2000bloch} as generalized quantum Bloch wall states. Ref.~\cite{dhar2000bloch} also constructed a semiclassical representative of these states, of which the cone texture in Eq.~\eqref{eq:spiral} is the fixed-sector case. Its classical energy reproduces the sector ground-state energy at leading order in $1/L$, $E_0(S)=\frac{2\pi^2 \rho_s}{L}\sin^2\theta$, reducing to $E_0(S{=}0)=\frac{2\pi^2 \rho_s}{L}$ in the singlet sector.

For $S=O(1)$ the tilt is $O(1/L)$ and affects the fluctuation analysis only at relative order $1/L^{2}$. We therefore set $\theta=\frac{\pi}{2}$ below. We now expand about the saddle in Eq.~\eqref{eq:spiral} to deduce the excitation spectrum within the $S=0$ sector. We parametrize the fluctuations using two real fields \(\xi\) and \(\zeta\):
$\vec n=
\xi \hat z+
\zeta \left(\hat z \times \vec n_0\right)+
\left(1-\frac{\xi^2+\zeta^2}{2}\right)\vec n_0$. The effective action in Eq.~\eqref{eq:effectiveactionbg} to quadratic order becomes
\begin{equation}
    \mathcal L = s \xi \partial_t \zeta - \frac{\rho_s}{2} [(\partial_x \xi)^2 +(\partial_x \zeta)^2 - q_0^2 \xi^2],
    \label{eq:spiral-quadratic}
\end{equation}
in which the Berry phase term pairs the two fields canonically. The spin-wave excitations above the $S=0$ saddle therefore have energies
\begin{equation}
    \omega(k) = \frac{\rho_s}{s}\sqrt{k^2(k^2-q_0^2)}, \qquad k=\frac{2\pi m}{L}, \quad m \in \mathbb{Z}.
    \label{eq:spiral-dispersion}
\end{equation}
The three zero modes, \(m=0,\pm 1\), correspond to global rotations of the saddle-point texture. This is because the spiral saddle breaks all three generators of the \(SU(2)\) spin-rotation symmetry, and thus produces three Goldstone zero modes. On the other hand, crucially, for \(k\gg q_0\), the finite-size level structure is governed by a spin-wave theory with dynamical critical exponent \(z=2\). This spin-wave
theory is what we use in Sec.~\ref{app:kernel-A} to evaluate the induced interaction
\(A(v,w)\) of Eq.~\eqref{eq:Seff-main}.

The conclusion generalizes to arbitrary \(\nTL\). In the coherent-state variables, the global \(SU(\nTL)\) charge reads
\begin{equation}
    \avg{Q^A}
    =
    \sum_j
    \left(
    z_j^\dagger T^A z_j
    -
    w_j^\dagger T^A w_j
    \right),
\end{equation}
the relative sign arising because the odd sublattice carries the anti-fundamental representation. Since the \(T^A\) span all traceless Hermitian matrices, and the trace of \(zz^\dagger-ww^\dagger\) vanishes identically, the singlet sector is characterized by the simple matrix constraint
\begin{equation}
    \int_0^L \dd x\; z(x)z(x)^\dagger
    =
    \int_0^L \dd x\; w(x)w(x)^\dagger :
    \label{eq:singlet-constraint}
\end{equation}
the two coherent-state lines must have equal spatial averages, while remaining orthogonal at every point by the ground-state condition Eq.~\eqref{eq:zero-mode-constraint}. Staying in the singlet sector therefore requires a texture. This is the general-\(\nTL\) version of the statement Eq.~\eqref{eq:n2-spin-constraint} that a ferromagnet must wind in order to have \(\vec S_{\rm tot}=0\).

The minimal texture rotates \(z\) into \(w\) within their common two-plane. Picking two orthonormal vectors \(e_1,e_2\), we take
\begin{equation}
    z(x)=\cos(qx)\,e_1+\sin(qx)\,e_2,
    \qquad
    w(x)=-\sin(qx)\,e_1+\cos(qx)\,e_2,
    \qquad
    q=\frac{\pi}{L},
    \label{eq:spiral-general-n}
\end{equation}
where the minimal winding \(q=\pi/L\) is fixed by periodicity up to gauge: the coherent-state variables are defined only up to a \(U(1)\) phase. The constraint~\eqref{eq:singlet-constraint} is satisfied exactly, since \(zz^\dagger-ww^\dagger=\cos(2qx)\,(e_1e_1^\dagger-e_2e_2^\dagger)+\sin(2qx)\,(e_1e_2^\dagger+e_2e_1^\dagger)\) integrates to zero around the periodic chain. A direct computation from Eqs.~\eqref{eq:Hcl-alt} and~\eqref{eq:Hlift-cl} gives the classical energy
\begin{equation}
    E_{\rm spiral}
    =
    \rho_{12}\,q^2L
    =
    \frac{\pi^2\rho_{12}}{L},
    \qquad
    \rho_{12}\equiv J+2\lambda .
    \label{eq:spiral-energy-n}
\end{equation}
At \(\nTL=2\), Eq.~\eqref{eq:spiral-general-n} precisely reduces to the spiral pattern in Eq.~\eqref{eq:spiral}, with winding \(q_0=2q\) and stiffness \(\rho_s=\rho_{12}/2\). For \(\nTL\ge3\), the auxiliary directions also admit global singlet textures, in which \(z\) and \(w\) rotate into each other by detouring through the orthogonal complement, i.e., the remaining \(\nTL-2\) directions, at cost \(O(\lambda/L)\). These are likewise smooth, slowly winding configurations with wavevector \(O(1/L)\), which is all we need for the analysis of the excitation spectrum. They do not exist at \(\nTL=2\), where there are no auxiliary directions. Under the prescribed order of limits they therefore do not contribute to the entanglement analysis.

The excitations above the spiral at general \(\nTL\) are again spin waves with \(z=2\). The fluctuations organize into canonically conjugate pairs, each governed by a Lagrangian of the same form as Eq.~\eqref{eq:spiral-quadratic}. One pair rotates the spiral within its \((e_1,e_2)\) plane. This is the \((\xi,\zeta)\) pair of the \(\nTL=2\) analysis, with stiffness of order \(\rho_{12}\) and the dispersion of Eq.~\eqref{eq:spiral-dispersion}. For \(\nTL\ge3\), rotations of \(z\) and of \(w\) into the \((\nTL-2)\)-dimensional orthogonal complement contribute \(4(\nTL-2)\) additional real fields, with the regulator stiffness \(\rho_\perp=\lambda\). They again disperse quadratically. There are \(4\nTL-5\) zero modes, exactly the number of \(SU(\nTL)\) generators broken by the spiral saddle.

\subsubsection{Irrelevance of the inter-branch coupling at \texorpdfstring{$p<1$}{p<1}}
\label{subsec:irrelevance}

We now ask whether the coupling between branches at \(p<1\) changes this picture. The branch-coupling term in Eq.~\eqref{eq:T-bg} is a product of local singlet projectors on a forward branch and a backward branch:
\begin{equation}
    \frac{1-p}{\nTL^2}
    \sum_{a,b}E^aE^{\bar b}
    =
    (1-p)
    \sum_{a,b}P_s^aP_s^{\bar b}.
    \label{eq:replicacoupling}
\end{equation}
The structure of this coupling in spin-wave variables is the same independent of the spin sector: in an ordered background whose saddle winds at wavevector \(q_0\), the singlet density expands as a constant, plus a term linear in the spin waves with coefficient \(O(q_0)\), plus terms quadratic and higher. The branch coupling therefore starts at bilinear order with a coefficient \(\propto q_0^2\), and continues at quartic order, which is the leading term in a uniform background where \(q_0=0\).

For simplicity we showcase this analysis in the \(S=0\) sector at \(\nTL=2\). The general-\(\nTL\) analysis follows without change. The singlet projector on a bond becomes $P_s \propto (\partial_x \vec n)^2$ in the unit-vector representation. Expanding about the spiral saddle point, Eq.~\eqref{eq:spiral}, and using \(\partial_x\vec n_0=q_0\,(\hat z\times\vec n_0)\),
\begin{equation}
    P_s^a\;\propto\;q_0^2+2q_0\,\partial_x\zeta_a
    +\big[(\partial_x\xi_a)^2+(\partial_x\zeta_a)^2\big]+\cdots ,
    \label{eq:Ps-spiral-expansion}
\end{equation}
so that the leading nonconstant inter-branch coupling is
\begin{equation}
\delta S^{\rm lead}_{p<1}\;\propto\;(1-p)\,q_0^2
\sum_{a,b}\int dt\,dx\;\partial_x\zeta_a\,\partial_x\zeta_{\bar b},
\label{eq:interbranch-bilinear}
\end{equation}
(the cross term linear in a single branch integrates to zero around
the periodic chain). Since its coefficient carries two powers of $q_0=2\pi/L$, this bilinear coupling vanishes in the thermodynamic limit. Another way of understanding it is that \(q_0\) carries the dimension of a momentum, so the term requires a coefficient of negative scaling dimension, making it irrelevant. Multiplying the quadratic terms of Eq.~\eqref{eq:Ps-spiral-expansion} between two branches gives, in addition, the quartic coupling
\begin{equation}
\delta S^{(4)}_{p<1}\;\propto\;(1-p)
\sum_{a,b}\int dt\,dx\;
\big[(\partial_x\xi_a)^2+(\partial_x\zeta_a)^2\big]
\big[(\partial_x\xi_{\bar b})^2+(\partial_x\zeta_{\bar b})^2\big],
\label{eq:interbranch-quartic}
\end{equation}
which survives at \(q_0=0\) and is the coupling displayed in Eq.~\eqref{eq:replica-coupling-leading} below for the uniform background. Carrying four fields and four gradients, it is strongly irrelevant at the \(z=2\) Gaussian fixed point by power counting. This is our central conclusion for the background sector: at order \(O(1/\nTL^2)\), the background sector is described by \(2Q\) decoupled copies of a \(z=2\) spin-wave theory, and the perturbation away from the measurement-only limit \(p=1\) does not modify its infrared behavior. Section~\ref{sec:exact-ferro} confirms that the ferromagnetic background indeed continues to the physical value \(\nTL=2\) for integer \(Q>1\). The induced interaction between the pairing variables therefore retains its diffusive form [Eq.~\eqref{eq:z=2interaction} below]. We will see shortly that this conclusion also holds when the initial state does not carry a fixed total spin, for instance a maximally mixed state.

\subsubsection{Maximally mixed initial states: spin waves on the flag manifold}
\label{subsec:uniformbg}

When the initial state is maximally mixed, \(\rho_0\propto\Id\), all total-spin sectors contribute, and the late-time dynamics at $t\gtrsim L^2$ is governed by the overall ground-state manifold of \(H_{\rm bg}+H_{\rm lift}\) and the excitations above it. With the regulator, the zero-energy configurations obey
\begin{equation} z_{j+1}\parallel z_j, \qquad w_{j+1}\parallel w_j, \qquad w^\dagger z=0 .
\end{equation}
A ground state is therefore specified by an ordered pair of orthogonal directions \((z,w)\) in \(\mathbb C^{\nTL}\), each defined up to a phase. The group \(SU(\nTL)\) rotates any such pair into any other. The transformations that leave a given pair unchanged are the two phases and the rotations of the remaining \(\nTL-2\) directions. Thus the ground-state manifold is the partial flag manifold
\begin{equation}
\mathcal F_{1,1;\nTL} = \frac{SU(\nTL)}{U(1)\times U(1)\times SU(\nTL-2)} . \label{eq:partial-flag}
\end{equation}
At \(\nTL=2\), this reduces to the usual ferromagnetic \(\mathbb{CP}^1\) manifold.

The excitations above the ground state are magnon-like. Setting the unit-cell spacing to one, we define the covariant derivatives
\begin{equation}
    D_x z=\partial_x z-z(z^\dagger\partial_x z),
    \qquad
    D_x w=\partial_x w-w(w^\dagger\partial_x w),
\end{equation}
and the projector onto the orthogonal complement of the two-plane spanned by \(z\) and \(w\),
\begin{equation}
    P_\perp=1-zz^\dagger-ww^\dagger .
\end{equation}
Expanding Eqs.~\eqref{eq:Hcl-alt} and~\eqref{eq:Hlift-cl} to second order in gradients gives
\begin{equation}
\begin{split}
    \mathcal L
    ={}&
    i\left(z^\dagger\dot z-w^\dagger\dot w\right)
    -
    \rho_{12}|w^\dagger D_x z|^2
    \\
    &-
    \rho_\perp
    \left(
    |P_\perp D_x z|^2
    +
    |P_\perp D_x w|^2
    \right)
    +\cdots ,
\end{split}
\label{eq:regulated-flag-LL}
\end{equation}
where we have analytically continued to real time for the dispersion analysis, and
\begin{equation}
    \rho_{12}=J+2\lambda,
    \qquad
    \rho_\perp=\lambda .
\end{equation}
The terms proportional to \(\rho_\perp\) describe rotations of \(z\) and \(w\) into the \((\nTL-2)\)-dimensional orthogonal complement. These are precisely the zero modes lifted by the regulator. We parametrize
\begin{align}
    z&=
    \begin{pmatrix}
    \sqrt{1-|\pi|^2-\psi_z^\dagger\psi_z}\\
    \pi\\
    \psi_z
    \end{pmatrix},\nonumber
    \\
    w&=
    \begin{pmatrix}
    -\pi^*-\psi_z^\dagger\psi_w\\
    \sqrt{1-|\pi|^2-\psi_w^\dagger\psi_w}\\
    \psi_w
    \end{pmatrix},
    \label{eq:zw-magnon-param}
\end{align}
which describes small fluctuations around a reference ground state. Here the complex field \(\pi\) describes rotations of \(z\) into \(w\), while the two \((\nTL-2)\)-component fields \(\psi_z\) and \(\psi_w\) describe rotations of \(z\) and \(w\), respectively, into the orthogonal complement. This parametrization satisfies \(z^\dagger z=w^\dagger w=1\) and \(w^\dagger z=0\) to second order in the fields. Substituting Eq.~\eqref{eq:zw-magnon-param} into Eq.~\eqref{eq:regulated-flag-LL} gives
\begin{equation}
\begin{split}
    \mathcal L^{(2)}
    ={}&
    i\left[
    2\pi^*\dot\pi
    +
    \psi_z^\dagger\dot\psi_z
    -
    \psi_w^\dagger\dot\psi_w
    \right]
    \\
    &-
    \rho_{12}|\partial_x\pi|^2
    -
    \rho_\perp
    \left(
    |\partial_x\psi_z|^2
    +
    |\partial_x\psi_w|^2
    \right).
\end{split}
\label{eq:flag-quadratic-action}
\end{equation}
Therefore, the physical \(z\)-\(w\) rotation mode has
\begin{equation}
    \omega_\pi(k)=\frac{\rho_{12}}{2}k^2+\cdots,
\end{equation}
while the auxiliary modes \(\psi_z\) and \(\psi_w\) also have quadratic dispersion. At the physical value \(\nTL=2\), the fields \(\psi_z\) and \(\psi_w\) are absent, and the theory reduces to the usual \(\mathbb{CP}^1\) Landau-Lifshitz magnon theory.

Moving away from \(p=1\), the inter-branch coupling in Eq.~\eqref{eq:replicacoupling} can be expanded in the same variables. For a given branch \(\alpha\), in terms of the spin-wave fields, we have
\begin{align}
    P_s^\alpha
    \sim
    |w_{\alpha,j}^\dagger z_{\alpha,j+1}|^2
    \sim{}&
    |\partial_x\pi_\alpha|^2
    +
    2\,{\rm Re}
    \left[
    (\partial_x\pi_\alpha)^*
    \psi_{w,\alpha}^\dagger\partial_x\psi_{z,\alpha}
    \right]
    \nonumber\\
    &+
    |\psi_{w,\alpha}^\dagger\partial_x\psi_{z,\alpha}|^2
    +\cdots .
    \label{eq:Ps-expansion}
\end{align}
The leading inter-branch coupling is consequently
\begin{equation}
    \delta S_{p<1}^{\rm lead}
    \sim
    \sum_{a,b}
    \int dt\,dx\;
    |\partial_x\pi_a|^2
    |\partial_x\pi_{\bar b}|^2 .
    \label{eq:replica-coupling-leading}
\end{equation}
By simple power counting, this perturbation is irrelevant at the \(z=2\) Gaussian fixed point, as anticipated in Sec.~\ref{subsec:irrelevance}.

One can then derive the induced interaction between the pairing variables from the spin-wave effective action in Eq.~\eqref{eq:flag-quadratic-action}. As shown in Sec.~\ref{app:kernel-A}, we find
\begin{align}
A(x,\tau)
\sim
|\tau|^{-6}
\left(1-\frac{x^2}{2D|\tau|}\right)^4
\exp\!\left[-\frac{x^2}{D|\tau|}\right],
\label{eq:A-tail}
\end{align}
where \(x\) and \(\tau\) are the spatial and temporal separations between the two vertices \(v\) and \(w\), and \(D\sim J\) is a positive diffusion constant.  The induced interaction is non-negative and takes the \(z=2\) diffusive scaling form
\begin{equation}
    A(x,\tau)
    =
    |\tau|^{-6}\,
    \Phi\!\left(\frac{x^2}{D|\tau|}\right),
    \label{eq:z=2interaction}
\end{equation}
with exponential suppression outside the diffusive regime \(x^2\lesssim D|\tau|\). Summing over \(x\) at fixed \(\tau\) gives a factor of order \((D|\tau|)^{1/2}\), so the total weight at time separation \(\tau\) decays as \(|\tau|^{-11/2}\). The interaction is therefore summable, and its second moments in \(x\) and \(\tau\) are finite. The same form holds for initial states with fixed total spin, since the derivation uses only the \(z=2\) spin-wave structure [Sec.~\ref{app:kernel-A}].

\subsection{Derivation of the spin-coherent-state path integral}
\label{app:coherent-state}
\label{app:spinwave}

We derive the coherent-state action used in Sec.~\ref{sec:background}.  We place an $SU(\nTL)$ fundamental coherent state on one sublattice,
\begin{equation}
|z\rangle=\sum_{\alpha=1}^{\nTL}z_\alpha|\alpha\rangle,
\qquad z^\dagger z=1,
\end{equation}
and an anti-fundamental coherent state on the other,
\begin{equation}
|\bar w\rangle=\sum_{\alpha=1}^{\nTL}w_\alpha^*|\bar\alpha\rangle,
\qquad w^\dagger w=1,
\end{equation}
each defined up to a phase. Trotter decomposing \(\Tr e^{-H_{\rm bg} t}\) and inserting coherent-state resolutions of identity gives, with \(|\mathcal Z_\ell\rangle\) the product of the coherent states of all sites at time slice \(\ell\),
\begin{equation}
\langle \mathcal Z_{\ell+1}|e^{-\delta\tau H_{\rm bg}}|\mathcal Z_\ell\rangle
=\langle \mathcal Z_{\ell+1}|\mathcal Z_\ell\rangle
\exp\left[-\delta\tau
\frac{\langle \mathcal Z_{\ell+1}|H_{\rm bg}|\mathcal Z_\ell\rangle}
{\langle \mathcal Z_{\ell+1}|\mathcal Z_\ell\rangle}
+O(\delta\tau^2)\right].
\end{equation}
For smooth paths the overlap gives the Euclidean Berry term
\begin{equation}
S_B=\int d\tau\sum_j\left(z_j^\dagger\partial_\tau z_j-w_j^\dagger\partial_\tau w_j\right),
\end{equation}
where the relative minus sign arises because the two sites in a unit cell transform in conjugate representations.

The unnormalized singlet on a fundamental--anti-fundamental bond is
\begin{equation}
|\epsilon\rangle=\sum_{\alpha=1}^{\nTL}|\alpha\rangle|\bar\alpha\rangle,
\qquad E=|\epsilon\rangle\langle \epsilon|.
\end{equation}
Its expectation value in the coherent state is
\begin{equation}
\langle z,\bar w|E|z,\bar w\rangle=|w^\dagger z|^2 .
\end{equation}
Summing the contributions from all bonds gives the energy term in the coherent-state path integral, Eq.~\eqref{eq:Hcl-alt}. 

\subsection{\texorpdfstring{{ The induced interaction with a diffusive scaling form}}{The induced interaction: A=CE2 with a diffusive scaling form}}
\label{app:kernel-A}

Here we derive the induced interaction in Eq.~\eqref{eq:A-tail}.  The derivation below is carried out at \(p=1\), since perturbations away from this limit are irrelevant.  Let
\begin{equation}
    \bar E(v):=\avg{E_v},
    \qquad
    C_E(v,w):=\cavg{E_vE_w}
\end{equation}
denote the one-point function and connected correlator of the
unnormalized singlet projector in a single branch of the
background theory.  Since \(\chi_{ab}(v)=E_v^aE_v^{\bar b}\), the factorization over branches at $p=1$ gives
\begin{equation}
    \avg{\chi_{ab}(v)\chi_{cd}(w)}_{\rm bg}
    =
    \avg{E_v^aE_w^c}_{\rm bg}
    \avg{E_v^{\bar b}E_w^{\bar d}}_{\rm bg}
    =
    \left[\bar E(v)\bar E(w)+C_E(v,w)\delta_{ac}\right]
    \left[\bar E(v)\bar E(w)+C_E(v,w)\delta_{bd}\right].
\end{equation}
Subtracting the disconnected piece,
\begin{equation}
    \avg{\chi_{ab}(v)}_{\rm bg}
    \avg{\chi_{cd}(w)}_{\rm bg}
    =
    \bar E(v)^2\bar E(w)^2,
\end{equation}
we obtain
\begin{equation}
\begin{split}
    \avg{\chi_{ab}(v)\chi_{cd}(w)}_{{\rm bg},c}
    ={}&
    \left[\bar E(v)\bar E(w)+C_E(v,w)\delta_{ac}\right]
    \left[\bar E(v)\bar E(w)+C_E(v,w)\delta_{bd}\right]
    \\
    &-
    \bar E(v)^2\bar E(w)^2 .
\end{split}
\label{eq:chi-factorized-app}
\end{equation}
By definition, $A(v,w)$ is the coefficient of the $\delta_{ac}\delta_{bd}$ tensor, leading to
\begin{equation}
    A(v,w)=C_E(v,w)^2.
    \label{eq:A-CE2-app}
\end{equation}

It remains to estimate \(C_E\) from the \(z=2\) spin-wave theory. Let us first state a general property of correlators of this type. A connected correlator \(C_O(v,w)=\cavg{O_vO_w}\) is defined by inserting the two operators into the tensor-network contraction. The two insertions enter the contraction on an equal footing, so by definition \(C_O(v,w)=C_O(w,v)\), i.e., \(C_O(-x,-\tau)=C_O(x,\tau)\) as a function of the separation. Combined with the statistical invariance of the circuit ensemble under spatial reflection, this yields
\begin{equation}
    C_O(x,\tau)=C_O(x,-\tau)=C_O(-x,\tau),
    \label{eq:CE-even}
\end{equation}
the correlation is even in both of its arguments. In particular, the induced interaction obeys \(A(x,\tau)=A(x,-\tau)=A(-x,\tau)\).

From Eq.~\eqref{eq:Ps-expansion}, to leading order in the RG sense one has
\begin{equation}
    E(x,\tau)\propto |\partial_x\pi(x,\tau)|^2,
\end{equation}
where \(\pi\) is a spin-wave field governed by the quadratic Euclidean action
\begin{equation}
    S_{\rm sw}
    =
    \int d\tau\,dx\;
    \pi^*
    \left(
    \partial_\tau-D\partial_x^2
    \right)\pi ,
\end{equation}
with \(D=\rho_{12}/2\). At any occupancy \(\{n_k\}\) of the magnon modes, the Euclidean magnon propagator takes the form
\begin{equation}
    G(x,\tau)
    =
    \int\!\frac{\dd k}{2\pi}\,\e^{\ii kx}
    \left[
    (1+n_k)\,\Theta(\tau)\,\e^{-\omega_k\tau}
    +
    n_k\,\Theta(-\tau)\,\e^{\omega_k\tau}
    \right],
    \qquad
    \omega_k=Dk^2 ,
    \label{eq:G-occupied}
\end{equation}
which, at the level of the scaling form and up to a nonuniversal amplitude, reduces to the diffusion kernel\footnote{The computation in Eq.~\eqref{eq:diffusivekernel} takes $n_k$ smooth and $O(1)$ at small $k$. In fact, all conclusions on the phase diagram and on the universality of the phase transition (Secs.~\ref{sec:twist} and~\ref{sec:universality}) are insensitive to the detailed occupancy. Since the spin-wave density is finite (because the local Hilbert space dimension is finite), we have $\int\dd k\,n_k<\infty$. Thus the infrared part of the occupancy must be integrable, $n_k\lesssim k^{-a}$ with $a<1$. Repeating the power counting of Sec.~\ref{sec:universality}
then gives $A\sim|\tau|^{2a-6}\,\Phi\big(x^2/D|\tau|\big)$, shifting the exponents in Eq.~\eqref{eq:SRSak} to
$\lambda_\tau=\tfrac92-2a$ and $\lambda_x=9-4a$: the induced interaction remains on the
short-range side for all $a<1$.}
\begin{equation}
    K_D(x,\tau)
    :=
    \frac{1}{\sqrt{4\pi D|\tau|}}
    \exp\!\left[-\frac{x^2}{4D|\tau|}\right].
    \label{eq:diffusivekernel}
\end{equation}
A Wick contraction, where we keep only the connected piece, then gives
\begin{equation}
    C_E(x,\tau)
    \propto
    \left[\partial_x^2 K_D(x,\tau)\right]^2 .
    \label{eq:CE-derivative-app}
\end{equation}
A short calculation yields
\begin{equation}
    \partial_x^2 K_D(x,\tau)
    =
    \frac{1}{2D|\tau|}
    \left(
    \frac{x^2}{2D|\tau|}-1
    \right)
    K_D(x,\tau).
\end{equation}
Combining this with Eq.~\eqref{eq:A-CE2-app}, we obtain
\begin{equation}
    A(x,\tau)
    \sim
    |\tau|^{-6}
    \left(
    1-\frac{x^2}{2D|\tau|}
    \right)^4
    \exp\!\left[-\frac{x^2}{D|\tau|}\right],
\end{equation}
which is Eq.~\eqref{eq:A-tail}. The same derivation applies to initial states with fixed total spin: expanding about the spiral saddle of Sec.~\ref{sec:fixedS}, the singlet density acquires in addition a term linear in the spin waves with coefficient \(q_0\), whose contribution to \(C_E\) carries \(q_0^2=O(1/L^2)\) and is negligible. The leading contribution is again the quartic contraction above, so \(A(v,w)\) retains the diffusive form of Eq.~\eqref{eq:z=2interaction}, with modified nonuniversal constants. Only the diffusive scaling form of Eq.~\eqref{eq:z=2interaction} is universal and matters for the analysis of the phase structure in the main text.

\subsection{\texorpdfstring{{ Exact check at $\nTL=2$: ferromagnetic leading eigenvectors for integer $Q>1$}}{Exact check at n=2: ferromagnetic leading eigenvectors for integer Q>1}}
\label{sec:exact-ferro}

The analysis above relies on the expansion at large \(\nTL\). We now ask whether the ferromagnetic background survives at the physical value \(\nTL=2\). To do so, we look for the leading eigenvectors of the time-direction transfer matrix of the replicated dynamics, with no restriction on the total spin. At the measurement-only point \(p=1\) and for every integer \(Q>1\), they are exactly the ferromagnetic states of the \(2Q\) branches. The renormalization group then extends the result to \(p<1\).

We start at the measurement-only point \(p=1\). The tensor on bond \(x\) is then Eq.~\eqref{eq:Tm},
\begin{equation}
    h_x=\prod_{\alpha=1}^{2Q}P^\alpha_{s,x}+\prod_{\alpha=1}^{2Q}P^\alpha_{t,x},
    \label{eq:hx}
\end{equation}
which is a projector. The transfer matrix of one time step is the product of the \(h_x\) over the two layers. Since every factor is a projector, no eigenvalue exceeds one in modulus, and a state has eigenvalue one only if every bond leaves it unchanged, \(h_x|\psi\rangle=|\psi\rangle\) for all \(x\). Physically, on every bond the \(2Q\) branches must be all in the triplet channel or all in the singlet channel.

This condition forces every branch to be ferromagnetic. Let \(R^\alpha_x\) swap the two sites of bond \(x\) in branch \(\alpha\), so that \(P_t=(1+R)/2\) and \(P_s=(1-R)/2\). The condition says that all \(R^\alpha_x\) take the same value \(\pm1\) on bond \(x\), or equivalently
\begin{equation}
    R^\alpha_xR^\beta_x|\psi\rangle=|\psi\rangle
    \qquad\text{for all }x\text{ and all }\alpha\neq\beta .
    \label{eq:double-swap}
\end{equation}
The state is thus invariant under a swap performed in two branches at once. Now take three different branches \(\alpha,\beta,\gamma\), which requires \(Q\ge2\), and two neighboring bonds. Since \(R^2=1\) and swaps in different branches commute,
\begin{equation}
    \big(R^\alpha_xR^\beta_x\big)\big(R^\alpha_{x+1}R^\gamma_{x+1}\big)\big(R^\alpha_xR^\beta_x\big)\big(R^\alpha_{x+1}R^\gamma_{x+1}\big)
    =\big(R^\alpha_xR^\alpha_{x+1}\big)^2 .
    \label{eq:three-cycle}
\end{equation}
The swaps in branches \(\beta\) and \(\gamma\) cancel. What remains is a cyclic permutation of the three sites \(x,x+1,x+2\) in branch \(\alpha\) alone, and it leaves \(|\psi\rangle\) unchanged. Cyclic permutations of neighboring triples generate all even permutations of the \(L\) sites. A state that is invariant under all even permutations is the sum of a fully symmetric and a fully antisymmetric part. For \(L\ge3\) spins \(1/2\) the antisymmetric part vanishes, because more than two spins \(1/2\) have no fully antisymmetric state. Each branch is therefore in the multiplet of maximal spin \(L/2\). Every bond is then a triplet in every branch, and Eq.~\eqref{eq:double-swap} holds. The eigenvalue-one space is exactly the space of ferromagnetic states of the \(2Q\) branches, of dimension \((L+1)^{2Q}\). Exact diagonalization on periodic chains confirms this counting. We find \(256\), \(625\) and \(4096\) eigenvalue-one states for \((Q,L)=(2,3)\), \((2,4)\) and \((3,3)\), equal to \((L+1)^{2Q}\).

Away from \(p=1\) the unitary tensor couples the forward and the backward branches [Eq.~\eqref{eq:replicacoupling}]. The stability of the ferromagnetic background can be argued with the renormalization group. Each fusion-channel projector \(P^\alpha_f\) is invariant under a spin rotation of branch \(\alpha\) alone. A coupling between two branches is therefore a product of two rotation-invariant operators, one in each branch. In a uniform ferromagnet the lowest such operator besides the identity is the energy density, which is quadratic in the spin waves and carries two gradients [Eq.~\eqref{eq:Ps-spiral-expansion}]. The product of two of them has scaling dimension \(6\) at the \(z=2\) fixed point, against \(d+z=3\), and is irrelevant, in agreement with Sec.~\ref{subsec:irrelevance}. The ferromagnetic background therefore persists at \(p<1\). The counting does not depend on \(Q\), so the couplings formally remain irrelevant as \(Q\to1\).

We emphasize that our argument suggests that, for fixed integer \(Q>1\), the results of the large-\(\nTL\) analysis for the background sector remain valid at the physical value \(\nTL=2\). However, whether the leading eigenvector in the limit \(Q\to1\) (after all, we do not even have an operational way to define it when \(Q\) is not an integer) is a continuation of those at \(Q>1\) is a separate question, which remains open. This is why the main text states as a separate assumption that the replica limit \(Q\to1\) commutes with the thermodynamic limit.

\section{\texorpdfstring{{ Phases of the pairing sector and entanglement scaling}}{Phases of the pairing sector and entanglement scaling}}
\label{sec:twist}

We now ask which phases the pairing sector has and how each phase appears in the entanglement entropy. We first recall how the entropy in Eq.~\eqref{eq:statmech} is read in the stat-mech language~\cite{Bao2020,Jian2020}. We define the free energies with different boundary conditions as
\begin{equation}
    F_A(Q;n)=-\log Z_A(Q;n),
    \qquad
    F_0(Q;n)=-\log Z_0(Q;n),
\end{equation}
where \(Z_A\) is the partition function with the entanglement twist inserted on region \(A\), and \(Z_0\) is the untwisted partition function. Since
\begin{equation}
    Z_A(1;n)=Z_0(1;n)=1,
\end{equation}
the averaged R\'enyi entropy can be written as
\begin{equation}
\begin{split}
    S_A^{(n)}
    &=
    \lim_{k\to0}
    \frac{Z_A(1+nk;n)-Z_0(1+nk;n)}
    {k(1-n)}
    \\
    &=
    \frac{1}{n-1}
    \left.
    \partial_k
    \left[
    F_A(1+nk;n)-F_0(1+nk;n)
    \right]
    \right|_{k=0}.
\end{split}
\label{eq:renyi-free-energy-cost}
\end{equation}
Thus the R\'enyi entropy is mapped to the replica-limit free-energy cost of imposing twisted boundary conditions. Here the derivative is taken at fixed R\'enyi index \(n\), continuing in the number of replica groups \(k\). The von Neumann entropy is obtained after taking the further limit \(n\to1\). Specific to our model, the entanglement entropy receives two contributions in the sector-separated description. (1) First, there is the contribution from the \(S_Q\) pairing variables. The boundary gluing in region \(A\) favors the pairing \(h_n\), associated with the tensor \(T_{n,A}^{\otimes k}\), between forward and backward branches, while in \(\bar A\) it favors the identity pairing \(e\). Thus the entanglement entropy measures the free-energy cost of forcing a domain wall in the \(S_Q\) pairing variables between the \(h_n\)-pinned and \(e\)-pinned boundary segments~\cite{Bao2020,Jian2020}. (2) Second, our theory also contains a gapless background spin-wave sector. Even when the \(S_Q\) domain-wall tension vanishes, the boundary twist still changes the boundary condition seen by this critical background sector. The remaining entropy is therefore controlled by the free-energy response of the gapless diffusive sector to the entanglement boundary twist.

We now analyze the possible phases of the pairing variables, whose interaction is induced by the background sector as shown in Eq.~\eqref{eq:A-tail}. Although
algebraic, the induced interaction $A(x,\tau)$ is ferromagnetic and summable, and the
interaction energy is controlled by \(\Fix(\sigma_w^{-1}\sigma_v)\) with strength
\(g^{2}\propto(1-p)^{2}\). We therefore anticipate the phase diagram of the pairing sector, a
discrete ferromagnet in the two-dimensional spacetime, as follows:
\begin{itemize}
\item For sufficiently small \(p\) (large \(g\)), the pairing sector orders. A discrete ferromagnet in two dimensions orders at strong coupling because a domain wall costs an energy proportional to its length (the Peierls argument~\cite{Peierls1936}). A boundary twist on \(A\) then forces a domain wall with a finite line tension, and
\begin{equation}
    F_A-F_0\propto |A| .
\end{equation}
Translated back to the original circuit language, this gives a volume-law contribution to the entanglement entropy.

\item For sufficiently large \(p\) (small \(g\)), the pairing sector disorders. Because \(A\) is summable, the total coupling of one pairing variable to all the others is finite, and when it is weak enough it cannot order a discrete spin (Dobrushin's uniqueness condition~\cite{Dobrushin1968}). The boundary twist then has no line tension in the pairing sector. However, the still-critical background sector contributes a logarithmic entanglement, as discussed shortly.

\item The spin-sharpening transition is then identified with the order--disorder transition of the \(S_Q\) pairing variables. Its universality will be discussed in Sec.~\ref{sec:universality}.
\end{itemize}

\subsection{Entanglement scaling of the background sector}

We now ask what entanglement remains at large measurement rate, where the pairing sector is disordered. In random monitored circuits without symmetry~\cite{Bao2020,Jian2020}, or with only an Abelian \(U(1)\) symmetry~\cite{Agrawal2022}, the answer is an area law. Here the ferromagnetic background sector gives a logarithmic entanglement scaling. Here we first discuss the maximally mixed initial state, \(\rho_0\propto\Id\), as it is technically simpler. Initial states with fixed total spin are treated at the end of this section. We do the computation at \(p=1\), since perturbations away from this point are irrelevant. At \(p=1\), the background sector consists of \(2Q\) decoupled ferromagnetic branches. In the ground state each branch has an ordered orientation $(z_\alpha,w_\alpha)\in \mathcal F_{1,1;\nTL}$. At the final time, each unit cell in \(A\) glues the backward branch \(\bar b\) to the forward branch \(h(b)\), and each unit cell outside \(A\) glues \(\bar b\) to \(b\). Each gluing contributes the overlap of the two orientations. With \(|A|/2\) unit cells in \(A\), the twist on a long interval therefore produces the factor
\begin{equation}
\frac{Z_A[h]}{Z_0}
\sim
\frac{1}{Z_0}\int
\prod_{\alpha=1}^{2Q}
\dd\mu_{\mathcal F}(z_\alpha,w_\alpha)\;
\prod_{b\in B}
[ \mathcal{O}_{h(b)\bar b}
]^{|A|/2} [ \mathcal{O}_{b \bar b} ]^{|\bar A|/2},
\label{eq:flag-overlap}
\end{equation}
where $\mathcal{O}_{\alpha \beta} =
\left(z_{\alpha}^\dagger z_\beta\right)
\left(w_{\alpha}^\dagger w_\beta\right)$ and \(\dd\mu_{\mathcal F}\) is the measure on the flag manifold, and
\(h\) is the permutation imposed by the boundary gluing in region \(A\).

We now evaluate Eq.~\eqref{eq:flag-overlap} by expanding around the saddle in which all orientations connected by gluing coincide. Taking one cycle \(c=(c_1c_2\cdots c_\ell)\) of \(h\), the fluctuations around the saddle can be parametrized by \((\varphi_1,\ldots,\varphi_\ell)\) on the forward branches and \((\bar\varphi_1,\ldots,\bar\varphi_\ell)\) on the backward branches, with each \(\varphi\) (\(\bar\varphi\)) denoting local complex coordinates \(\varphi_{\alpha}^{I}\), \(I=1,\ldots,d_{\mathcal F}\), on \(\mathcal F_{1,1;\nTL}\) near the common orientation. Expanding the overlap \(\mathcal O_{\alpha\beta}\) to quadratic order in fluctuations gives the contribution from one cycle to $Z_A[h]$,
\begin{equation}
\begin{split}
I_\ell
=
\int \mathcal{D}[\varphi_m,\bar\varphi_m]\,
\exp [
-\frac{|A|}{4}
\sum_{m=1}^{\ell}
\left(\bar\varphi_m-\varphi_{m+1}\right)^2 \\
-\frac{|\bar A|}{4}
\sum_{m=1}^{\ell}
\left(\bar\varphi_m-\varphi_m\right)^2
],
\qquad
\varphi_{\ell+1}:=\varphi_1 .
\end{split}
\label{eq:cyclecontribution}
\end{equation}
Here we have absorbed the metric on \(\mathcal F_{1,1;\nTL}\) into the definition of \(\varphi\) and \(\bar\varphi\). We can first integrate out the \(\bar\varphi\) variables, leading to
\begin{equation}
\begin{split}
I_\ell
&\propto
\int \prod_{m=1}^{\ell} d\varphi_m\,
\exp\left[
-\frac{|A||\bar A|}{4L}
\sum_{m=1}^{\ell}
\left(\varphi_m-\varphi_{m+1}\right)^2
\right]
\\
&\propto
\left(
\frac{|A||\bar A|}{L}
\right)^{-d_{\mathcal F}(\ell-1)} ,
\end{split}
\label{eq:GaussianEE}
\end{equation}
where the last line follows because such a cycle contributes \(\ell-1\) independent relative modes, each of complex dimension \(d_{\mathcal F}\). Since the complex dimension of the flag manifold is
\begin{equation}
    d_{\mathcal F}
    =
    \dim_{\mathbb C}\mathcal F_{1,1;\nTL}
    =
    2\nTL-3,
\end{equation}
multiplying over all cycles gives
\begin{equation}
\frac{Z_A[h]}{Z_0}
\sim
\prod_{c\in h}
\left(
\frac{|A||\bar A|}{L}
\right)^{-d_{\mathcal F}(\ell_c-1)}
=
\left(
\frac{|A||\bar A|}{L}
\right)^{-d_{\mathcal F}[Q-C(h)]}.
\label{eq:zero-mode}
\end{equation}
Here \(C(h)\) is the number of cycles of \(h\).

For the R\'enyi twist \(h_n\), with \(Q=1+nk\), one has
\begin{equation}
    C(h_n)=1+k,
    \qquad
    Q-C(h_n)=k(n-1).
\end{equation}
Using Eq.~\eqref{eq:renyi-free-energy-cost}, this gives
\begin{equation}
    S_A^{(n),{\rm bg}}
    =
    d_{\mathcal F}\log |A|+O(1),
    \label{eq:renyi-log-main}
\end{equation}
for a complement with $|\bar A| = O(L)$. The von Neumann entropy is obtained by the further limit \(n\to1\), and gives the same logarithmic coefficient:
\begin{equation}
    S_A^{{\rm vN},{\rm bg}}
    =
    d_{\mathcal F}\log |A|+O(1).
    \label{eq:vn-log-main}
\end{equation}
Finally, taking the physical continuation \(\nTL\to2\), the flag manifold reduces to \(\mathbb{CP}^1\), so \(d_{\mathcal F}=1\). Thus
\begin{equation}
    S_A^{\rm bg}
    =
    \log |A|+O(1)
    \qquad
    (\nTL=2).
    \label{eq:log-main}
\end{equation}
The counting rule, namely that a real coordinate on the ground-state manifold gives half a logarithmic entanglement entropy, is the same as the standard rule for the Goldstone-mode contribution to entanglement entropy, as in Ref.~\cite{MetlitskiGrover2011,popkov2005logarithmic}.

We now revisit the entanglement entropy for initial states with a fixed total spin $S=0$, expanding about the spiral saddle of Sec.~\ref{sec:fixedS} (for $S=O(1)$ the tilt modifies the saddle only at $O(1/L)$, leaving the coefficients unchanged). As in Eq.~\eqref{eq:cyclecontribution}, one has
to compute the replica gluing factor when the branches in a cycle
fluctuate around the common-orientation saddle point. At the physical
value $\nTL=2$, the fluctuation is simply parametrized by a small
rotation $\vec\theta$ of the spiral saddle point $\vec n_0(x)$ about
an axis.  For two branches with a small relative rotation
$\vec\theta$, within a segment $A$ of $|A|$ consecutive sites, the
overlap factor is
\begin{equation}
W(\vec\theta)\sim
\exp\Big[-c\sum_{x\in A}\big|\vec\theta\times\vec n_0(x)\big|^2\Big],
\end{equation}
where $c$ is a nonuniversal constant.  A crucial difference from the
uniform ferromagnetic ground state is that $\vec n_0(x)$ now has a
spiral texture in space.  For an interval with $q_0|A|\ll1$ [so that
$\vec n_0(x)\simeq\vec n_0$ is almost uniform within it], we have
\begin{equation}
W(\vec\theta)\sim
\exp\Big[-c\,|A|\,\theta_z^2-c\,|A|\,\theta_m^2
-\tfrac{1}{12}\,c\,q_0^2|A|^3\,\theta_n^2\Big],
\end{equation}
where $\theta_z$, $\theta_m$, and $\theta_n$ are the relative rotation
angles about $\hat z$, $\hat z\times\vec n_0$, and $\vec n_0$,
respectively. A rotation about the local magnetization $\vec n_0$ contributes only at order $q_0^2|A|^3$, due to the slow winding.  The
integral over fluctuations [similar to Eq.~\eqref{eq:cyclecontribution}] is now over three
relative rotation angles: this is precisely due to the fact that the
spiral saddle point for $S=0$ breaks all three generators of the
$SU(2)$ spin symmetry.  Carrying out the Gaussian integral, with the
gluings in $A$ and $\bar A$ combining in series, one deduces the
entanglement entropy
\begin{equation}
S_A^{\rm bg}\big|_{S=0}=\log\frac{|A||\bar A|}{L}
+\frac12\log\Big[1+\tfrac{1}{12}\,q_0^2
\big(|A|^{-3}+|\bar A|^{-3}\big)^{-1}\Big]+O(1).
\end{equation}
The expansion used here is accurate for \(q_0|A|\ll1\). For \(|A|\) of order \(L\) it still gives the correct power of \(L\) for each of the three stiffnesses, which is all that enters the coefficient of the logarithm. This result has a clear physical interpretation.  (1)~For
$|A|\ll L^{2/3}$, $S_A^{\rm bg}|_{S=0}=\log|A|+O(1)$.  This is essentially
because, for small $|A|$, the local texture of $\vec n_0$ is almost
uniform: the region cannot see that the entire $SU(2)$ symmetry is
broken, and the entanglement entropy receives contributions from only
two of the three broken generators.  (2)~For $|A|=|\bar A|=O(L)$, the
region $A$ can finally detect that all three generators are broken.
The three real Goldstone directions, each contributing half a
logarithm, give $S_A^{\rm bg}=\tfrac32\log L+O(1)$ in this regime.

As in Eq.~\eqref{eq:zero-mode}, the gluing exponent is linear in \(Q-C(h_n)=k(n-1)\), so these coefficients are independent of the R\'enyi index, as quoted in the End Matter.

One can carry out the same computation at general \(\nTL\). The only distinction is the counting of broken generators. The uniform flag-manifold saddle breaks \(4\nTL-6\) generators, reproducing \(d_{\mathcal F}=2\nTL-3\) in Eq.~\eqref{eq:renyi-log-main}. The singlet-sector spiral of Eq.~\eqref{eq:spiral-general-n} breaks \(4\nTL-5=3+4(\nTL-2)\) of them, of which \(4\nTL-6\) are locally visible, while the single winding direction contributes only at order \(q_0^2|A|^3\), exactly as at \(\nTL=2\). The pair of coefficients \((1,\tfrac32)\) therefore generalizes to \((2\nTL-3,\;2\nTL-\tfrac52)\), with the same crossover at \(|A|\sim L^{2/3}\): at every \(\nTL\), fixing the total spin adds exactly half a logarithm for regions of order the system size.

\subsection{Universality class of the spin-sharpening transition}
\label{sec:universality}

We now deduce the universality of the spin-sharpening transition at leading order in the large-\(\nTL\) expansion. For this purpose, it is convenient to write the partition function as a coupled continuum theory, \begin{equation} Z = \sum_{\{\sigma\}} \int \mathcal{D}[\pi]\, e^{-S_{\rm sw}-S_{\sigma\pi}}, \end{equation} where \(S_{\rm sw}\) is the effective action of the background sector in Eq.~\eqref{eq:flag-quadratic-action}, written in terms of the spin-wave fields, which we collectively denote by \(\pi\) for notational simplicity. According to Eq.~\eqref{eq:T-separation}, the coupling between the two sectors is
\begin{equation} S_{\sigma\pi} = -\lambda_{\sigma\pi} \sum_{a,b} \int d\tau\,dx\; M_{ab}(\sigma) P^a_s P^{\bar b}_s,
\end{equation}
with
\begin{equation} \lambda_{\sigma\pi}=\frac{1-p}{(Q-1)!}. \end{equation}
Again, due to the identification in Eq.~\eqref{eq:Ps-expansion}, the coupling \(\lambda_{\sigma\pi}\) is irrelevant whenever the background sector is described by a \(z=2\) spin-wave theory. Consequently, even at the spin-sharpening transition, the critical theory factorizes in the infrared into the pairing sector and a spectator diffusive sector. One can therefore first integrate out the diffusive background, treating its effect as the induced effective interaction between the pairing variables.

We next ask whether the algebraic tail of the coupling in Eq.~\eqref{eq:A-tail} changes the universality class of the order-disorder transition. An algebraic tail changes the critical behavior only if it decays slowly enough to be a relevant perturbation of the short-range critical point. For a tail that produces a nonanalytic term \(|k|^{\lambda}\) in the coupling, this requires \(\lambda<2-\eta_{\rm SR}\), with \(\eta_{\rm SR}\) the anomalous dimension at the short-range critical point~\cite{FisherMaNickel1972,Sak1973}. The induced interaction \(A(x,\tau)\), though algebraic,
has finite second moments,
\begin{equation}
    \sum_{x,\tau} x^2 A(x,\tau)<\infty,
    \qquad
    \sum_{x,\tau} \tau^2 A(x,\tau)<\infty.
\end{equation}
So, the leading nonconstant terms of its Fourier transform at small momentum and frequency are \(k^2\) and \(\omega^2\), and nonanalyticities appear only
at subleading orders. In fact, one can check that, at small momentum and frequency, the leading nonanalytic terms in the Fourier transform of \(A(x,\tau)\) occur only at high powers, namely
\begin{equation}
    |\omega|^{\lambda_\tau},\qquad \lambda_\tau=\frac{9}{2};
    \qquad
    |k|^{\lambda_x},\qquad \lambda_x=9 .
    \label{eq:SRSak}
\end{equation}
Since both exponents are larger than \(2-\eta_{\rm SR} = d - 2\Delta_\sigma\), where
\(\eta_{\rm SR}\) (\(\Delta_\sigma\)) is the anomalous dimension (scaling dimension) of the
\(S_Q\) order-parameter field at the corresponding transition with only short-range
interaction, the interaction in Eq.~\eqref{eq:A-tail} lies on the short-range side of the Sak
criterion~\cite{Sak1973} for any positive \(\Delta_\sigma\) in \(d=2\). We therefore conclude that the spin-sharpening transition is controlled by the short-range order--disorder transition of the \(S_Q\) pairing field. It is then natural to conjecture that the transition is in the universality class of the generic measurement-induced transition of monitored circuits without a global symmetry, namely the order--disorder transition of the \(S_Q\) pairing field with a \(S_Q^F\times S_Q^B\) symmetry~\cite{ZhouNahum2019,Bao2020,Jian2020}, tensored with a critical spin-wave sector. 

Finally, we ask whether this fixed point is stable against the slow fluctuations of the conserved charge. It is known that a diffusing conserved density can make the local couplings of the pairing field fluctuate slowly in space and time, and thereby sometimes change the universality class of the entanglement transition. For a \(U(1)\) charge, this effect enters the effective action as
\begin{equation}
\delta S\;\sim\;\kappa\int\dd x\,\dd t\;\delta\rho\,\varepsilon_\sigma,
\label{eq:coupling-modulation}
\end{equation}
with \(\delta\rho\) the local charge density fluctuation and \(\varepsilon_\sigma\) the energy
operator of the pairing field. As shown in Ref.~\cite{ha2024measurement}, this coupling is relevant at the measurement-induced phase transition (MIPT). For \(SU(2)\), this effect also appears, at \(O(\nTL^{-5})\), one order higher in \(1/\nTL\) than the induced interaction. However, we argue that the corresponding coupling is irrelevant, for a symmetry reason. The
spin density is an \(SU(2)\) vector, so it cannot couple linearly to \(\varepsilon_\sigma\). The
leading coupling is instead to the replica-diagonal singlet density, which is
\(E\sim|\partial_x\pi|^{2}\) in the spin-wave variables. This composite is not conserved, and
its correlations decay faster in time [Eq.~\eqref{eq:CE-derivative-app}]. Repeating the argument of
Ref.~\cite{ha2024measurement}, one finds that in the \(SU(2)\) case such a coupling is relevant only when
\(\nu<1\), which is not satisfied at the MIPT (\(\nu\approx 1.3\)~\cite{Zabalo2020}). The
perturbation is therefore irrelevant, and we conclude that the fixed point identified above is
stable against the fluctuations of the $SU(2)$ conserved density. We emphasize that our conclusion on
universality follows from the leading large-\(\nTL\) expansion, and we leave a more detailed
study of the transition to future work.

\section{Learning times}
\label{sec:learning}

\subsection{\texorpdfstring{{ The two parts of the record fidelity: a steady rate and a transient}}{The two parts of the record fidelity: a steady rate and a transient}}
\label{subsec:F-separation}

We finally explain how one can deduce the spin-learning time scales in the two phases observed in Ref.~\cite{Majidy2023}. We start from the R\'enyi record overlap in Eq.~\eqref{eq:replicatedoverlap}. The replicated tensor network then contains \(2q\) replicas, each with a forward and a backward branch. Let
\begin{equation}
    Z_{\bm S}(t)
    =
    \mathbb E_{\rm circ}
    \sum_{\bfm}
    \prod_{a=1}^{2q}p_{\bfm}(S_a,t),
    \qquad
    \bm S=(S_1,\ldots,S_{2q}),
    \label{eq:Z-sector-vector}
\end{equation}
where the total-spin sector label \(S_a\) is assigned to replica \(a\). We denote the transfer matrix in the time direction, averaged over one brickwork period containing two layers of gates, by \(\mathbb T\), so that \(Z_{\bm S}(t)=\langle e|\mathbb T^{t}|\rho_{\bm S,0}\rangle\). Here \(\langle e|\) is the identity pairing introduced in Sec.~\ref{sec:loop-model}, which glues each forward branch \(a\) to its own backward partner \(\bar a\) and thereby implements the trace \(\Tr\rho_{\bfm}\) that produces the Born probability in each replica. It is in particular a product over the \(2q\) replicas,
\begin{equation}
    \langle e|=\bigotimes_{a=1}^{2q}\langle e^{a,\bar a}| ,
    \label{eq:e-factorized}
\end{equation}
The initial state is likewise a product over the replicas,
\begin{equation}
    \rho_{\bm S,0}=\bigotimes_{a=1}^{2q}\rho_{0,S_a}^{a,\bar a},
    \label{eq:rho0-factorized}
\end{equation}
one factor per replica, with \(\rho_{0,S_a}\) the initial state in sector \(S_a\) (in Ref.~\cite{Majidy2023} the system is scrambled under unitary-only dynamics before the monitored circuit begins, so that \(\rho_{0,S_a}\) is effectively the maximally mixed state within its total-spin sector).  

Note that by definition
\begin{equation}
-\log F^{(q)}_t
=
\tfrac12\log Z^{(q)}_{S,S}
+\tfrac12\log Z^{(q)}_{S',S'}
-\log Z^{(q)}_{S,S'}
\label{eq:F-cancellation}
\end{equation}
vanishes identically, at every \(t\), whenever the partition function factorizes over the \(2q\) replicas,
\(Z_{\bm S}(t)=\prod_{a=1}^{2q} z_{S_a}(t)\): every replica-local
contribution cancels between the mixed assignment \((S,S')\) and the two
uniform ones, \((S,S)\) and \((S',S')\). The record fidelity is
therefore a measure of \emph{inter-replica correlation}. We call combinations with this
cancellation property, such as $\mathcal T(t)$ in Eq.~\eqref{eq:F-decomposition} below,
\emph{replica connected}.

To organize the full time dependence, we  separate, for each assignment \(\bm S=(S,S)\), \((S',S')\) or \((S,S')\), the contribution of the leading eigenvector from the rest,
\begin{equation}
\begin{split}
Z^{(q)}_{\bm S}(t)
&=
a^{\bm S}_0\,\lambda_{\bm S}^{\,t}\,
\mathcal N_{\bm S}(t),
\\
\mathcal N_{\bm S}(t)
&=
1+\sum_{n\ge1}\tilde a^{\bm S}_n\,e^{-\Omega^{\bm S}_nt},
\qquad
\Omega^{\bm S}_n\equiv\log\frac{\lambda_{\bm S}}{\lambda^{\bm S}_n}.
\end{split}
\label{eq:block-transient}
\end{equation}
Here \(\lambda_{\bm S}\equiv\lambda^{\bm S}_0\) is the leading eigenvalue of \(\mathbb T\) for the assignment \(\bm S\), \(\lambda^{\bm S}_n\) are the subleading ones, and \(\Omega^{\bm S}_n\) are the corresponding gaps. The amplitudes are \(a_n^{\bm S}=\langle e|b^{\bm S}_n\rangle\langle b^{\bm S}_n|\rho_{\bm S,0}\rangle\), with \(|b^{\bm S}_n\rangle\) the eigenvectors and \(\rho_{\bm S,0}\) the initial state of Eq.~\eqref{eq:rho0-factorized}. We write \(|b_{\bm S}\rangle\equiv|b^{\bm S}_0\rangle\) for the leading eigenvector, as in the End Matter, and \(\tilde a^{\bm S}_n=a^{\bm S}_n/a^{\bm S}_0\). The factor \(\mathcal N_{\bm S}\) depends only on the gaps, and \(\mathcal N_{\bm S}(t\to\infty)=1\). At \(t=0\) no record has been produced, so \(Z^{(q)}_{\bm S}(0)=1\) and \(\mathcal N_{\bm S}(0)=1/a^{\bm S}_0\). Substituting into Eq.~\eqref{eq:F-cancellation}, the record fidelity separates exactly into a steady part and a transient part,
\begin{equation}
\begin{split}
-\log F^{(q)}_t
&=
\Delta\,t
+
\big[\mathcal T(0)-\mathcal T(t)\big],
\\
\mathcal T(t)
&\equiv
\Big[\log\mathcal N_{S,S'}-\tfrac12\log\mathcal N_{S,S}
-\tfrac12\log\mathcal N_{S',S'}\Big](t),
\end{split}
\label{eq:F-decomposition}
\end{equation}
where
\(\Delta=\tfrac12\big(\log\lambda_{S,S}+\log\lambda_{S',S'}\big)-\log\lambda_{S,S'}\)
is the replica-connected combination of the leading eigenvalues. The first term grows steadily in time. The bracket is a step. It rises from zero at \(t=0\) to the plateau
\begin{equation}
\mathcal T(0)=-\log\frac{a_0^{(S,S')}}{\sqrt{a_0^{(S,S)}\,a_0^{(S',S')}}},
\label{eq:T0-SM}
\end{equation}
the replica-connected amplitude of the leading eigenvectors, on the time scale set by the smallest gap of the three assignments, \(\Omega_1\equiv\min_{\bm S}\Omega^{\bm S}_1\). Since the background sector has \(z=2\), \(\Omega_1=O(1/L^2)\). After a Thouless time \(O(L^2)\), the bracket is simply the constant \(\mathcal T(0)\). The learning time is therefore set by the competition between the steady growth \(\Delta t\) and the step height \(\mathcal T(0)\).

\subsection{\texorpdfstring{{ Steady rate in the fuzzy phase: $\Delta\sim p/L^3$}}{Steady rate in the fuzzy phase: Delta ~ p/L3}}
\label{subsec:fuzzy-learning}

We now compute the replica-connected combination of the leading eigenvalues, i.e., the learning rate \(\Delta\) of Eq.~\eqref{eq:F-decomposition}, in the weak-measurement, \(S_Q\)-ordered phase, where one can perform perturbation theory about the purely unitary point \(p=0\).  For the purpose of this subsection, it is more convenient to work directly at the physical value \(\nTL=2\).  Each of the \(2q\) replicas \(a=1,\ldots,2q\) of Eq.~\eqref{eq:Z-sector-vector} has a forward branch \(a\) and a
backward branch \(\bar a\).  The spin-sector label \(S_a\) is assigned to both the forward branch \(a\)
and the backward branch \(\bar a\), and is determined by the initial state
of the monitored dynamics.

At \(p=0\) nothing is measured, and the replicated unitary dynamics has many steady states. There is one for each way of pairing forward with backward branches of equal total spin, and in each of them every paired couple of branches is maximally mixed over the internal states of its spin sector (the orientation of the total spin, is also preserved). We now derive this. We first identify the \(p=0\) steady manifold of the transfer matrix of the replicated dynamics
\(\mathbb T_u\) (the subscript \(u\) denotes the unitary-only limit \(p=0\)) in the presence of
\(2q\) replicas.  Consider an operator
\(X_a\) that intertwines the forward branch \(a\) with the backward branch
\(\overline{\sigma(a)}\), where \(\sigma\in S_{2q}\) is a pairing permutation.
For \(X_a\) to be a steady operator, namely $\mathbb T_u |X_a\rangle = |X_a\rangle$, it must satisfy
\begin{equation}
    U_x^{(a)} X_a = X_a U_x^{(\overline{\sigma(a)})},
    \qquad \forall x,\theta ,
\end{equation}
with
\begin{equation}
    U_x(\theta)=\exp(i\theta P_{t,x})
\end{equation}
being a local \(SU(2)\)-symmetric gate, up to an irrelevant overall phase.
Using
\begin{equation}
    P_{t,x}=\frac{1+R_x}{2},
    \qquad
    P_{s,x}=\frac{1-R_x}{2},
    \label{eq:swaprep}
\end{equation}
this becomes the intertwining condition
\begin{equation}
    R_x^{(a)} X_a = X_a R_x^{(\overline{\sigma(a)})},
    \qquad \forall x .
    \label{eq:p0-intertwiner-condition}
\end{equation}
Now decompose the Hilbert space of the spin chain as
\begin{equation}
    \mathcal H_L
    =
    (\mathbb C^2)^{\otimes L}
    =
    \bigoplus_S \mathcal V_S\otimes \mathcal M_S ,
\end{equation}
where \(\mathcal V_S\) is the spin-\(S\) irrep of \(SU(2)\), and
\(\mathcal M_S\) is the corresponding multiplicity space.  In this
decomposition, the adjacent swaps \(R_x\) generate the site-permutation
algebra, which acts irreducibly on \(\mathcal M_S\), while different
values of \(S\) correspond to inequivalent irreducible representations of
this algebra.  Therefore Eq.~\eqref{eq:p0-intertwiner-condition} forces the
steady manifold of \(\mathbb T_u\) to be spanned by
\begin{equation}
\begin{split}
    \mathrm{Fix}(\mathbb T_u)
    &=
    \mathrm{span}_{\sigma\in S_{2q}}
    \left\{
    P_\sigma
    \bigotimes_{a=1}^{2q}
    \left[
    \bigoplus_{S_a}
    Y_{S_a}
    \otimes
    \mathbf 1_{\mathcal M_{S_a}}
    \right]_{a,\bar a}
    \right\},\\
    Y_{S_a} &\in \mathrm{End}(\mathcal V_{S_a}).
\end{split}
    \label{eq:p0-steady-manifold}
\end{equation}
Here $P_\sigma$ is the replica permutation operator on the $2q$-fold replicated Hilbert space, which, under the state-operator mapping, pairs each forward branch $a$ with the backward branch $\overline{\sigma(a)}$. Once the total-spin sector of each replica is fixed by the initial
state, Eq.~\eqref{eq:p0-steady-manifold} implies that the pairing \(\sigma\in S_{2q}\) can only pair forward and backward
branches with identical total spin. This is the
\(SU(2)\)-symmetry-resolved version of the standard replicated Haar-unitary
fixed point: averaging projects onto the commutant of the tensor-power
unitary action, whose basis is given by permutation pairings by
Schur--Weyl duality~\cite{li2024statistical}. If the initial state is chosen randomly within a fixed total-spin sector
(for example, with each magnetic quantum number appearing with the same
probability), one may take \(Y_{S_a}\) in
Eq.~\eqref{eq:p0-steady-manifold} to be proportional to the identity.

The fact that the steady manifold of \(\mathbb T_u\) contains a span over
\(\sigma\in S_{2q}\) reflects the \(S_{2q}\)-ordered fixed point at \(p=0\).
The identity top boundary \(\langle e|\) then acts as a boundary
symmetry-breaking field, which selects the identity-paired
broken-symmetry sector \(\sigma=e\).  Indeed, one can check that, after
taking the overlap with \(\langle e|\), all other pairings in
Eq.~\eqref{eq:p0-steady-manifold} are suppressed by powers of
\(d_{S_a}^{-1}\), where \(d_{S_a}\) is the Hilbert-space dimension of the
total-spin-\(S_a\) sector and is generally exponentially large in system
size.  Therefore, denoting by
\begin{equation}
    \tilde\Pi_{S}=\frac{\Pi_{S}}{d_{S}},
    \qquad
    d_{S}=\Tr \Pi_{S}
    \label{eq:p0-rhoS-boundary-visible}
\end{equation}
the maximally mixed state within the total-spin-\(S\) sector, with \(\Pi_S\) the projector onto
that sector, the relevant leading eigenvector whose perturbed eigenvalue we need is
\begin{equation}
    |b^{\bm S}_0\rangle=|\tilde\Pi_{\bm S}\rangle,
    \qquad
    \tilde\Pi_{\bm S}
    =
    \bigotimes_{a=1}^{2q} \tilde\Pi_{S_a}^{a,\bar a},
    \label{eq:p0-leading-eigenvector}
\end{equation}
where \(\tilde\Pi_{S_a}^{a,\bar a}\) denotes the density matrix whose
ket and bra live in the branches \(a\) and \(\bar a\), respectively.

We now compute this perturbed eigenvalue at small \(p\).  At a single
spacetime vertex \(v\), the local tensor in Eq.~\eqref{eq:localtensorp} can be expressed as
\begin{equation}
\begin{split}
    \cT_v
    &=
    \cT_{u,v}
    +
    p(\cT_{m,v}-\cT_{u,v})
    =
    \cT_{u,v}
    -
    p\,\mathfrak M_v, \\
    \mathfrak M_v
    &:=
    \cT_{u,v}-\cT_{m,v}.
\end{split}
    \label{eq:lt-small-p-tensor}
\end{equation}
Thus \(\mathfrak M_v\) is the difference between the unitary and
measurement tensors at the vertex \(v\).  To first order in \(p\), the
leading eigenvalue \(\lambda_{\bm S}\) of \(\mathbb T\), which governs
\(Z_{\bm S}(t)=\langle e|\mathbb T^{t}|\rho_{\bm S,0}\rangle
\sim\lambda_{\bm S}^{t}\) at large \(t\), is
\begin{equation}
    \begin{split}
        \lambda_{\bm S}
    &=
    \bra{e}
    \left(\mathbb T_u-p\sum_v \mathfrak M_v\right)
    \ket{\tilde\Pi_{\bm S}}
    +O(p^2) \\
    &=
    1
    -
    p\sum_v
    \langle e | \mathfrak M_v | \tilde\Pi_{\bm S} \rangle
    +O(p^2),
    \end{split}
    \label{eq:lambda-small-p}
\end{equation}
where we have used
\(\langle e|\tilde\Pi_{\bm S}\rangle=1\) and
\(\mathbb T_u|\tilde\Pi_{\bm S}\rangle=|\tilde\Pi_{\bm S}\rangle\). Using the form of the measurement tensor [Eq.~\eqref{eq:Tm}], we find
\begin{equation}
    \begin{split}
        \langle e | \mathfrak M_v | \tilde\Pi_{\bm S} \rangle
        &=
        1
        -
        \prod_{a=1}^{2q}
        \langle P_{s,v}\rangle_{S_a}
        -
        \prod_{a=1}^{2q}
        \langle P_{t,v}\rangle_{S_a} \\
        &=
        1
        -
        \frac{1}{2^{2q}}
        \left[
        \prod_{a=1}^{2q}
        \left(1-r_{S_a}\right)
        +
        \prod_{a=1}^{2q}
        \left(1+r_{S_a}\right)
        \right],
    \end{split}
    \label{eq:M-expectation-general}
\end{equation}
where
\begin{equation}
    r_{S_a}:=\langle R_v\rangle_{S_a}
    =
    \Tr(R_v\tilde\Pi_{S_a}).
    \label{eq:rS}
\end{equation}
Here \(\langle \cdots\rangle_{S_a}\) denotes the expectation value with
respect to the maximally mixed state in the total-spin sector \(S_a\),
defined in Eq.~\eqref{eq:p0-rhoS-boundary-visible}.  We have also used
Eq.~\eqref{eq:swaprep}.

Specializing to the three assignments in Eq.~\eqref{eq:F-cancellation}, the steady rate \(\Delta\) in Eq.~\eqref{eq:F-decomposition} becomes
\begin{equation}
\begin{split}
    \Delta
    &=
    \frac12
    \left(
    \log\lambda_{S,S}
    +
    \log\lambda_{S',S'}
    \right)
    -
    \log\lambda_{S,S'} \\
    &=
    \frac{p}{2}
    \sum_v
    \left[
    \left(u_S^q-u_{S'}^q\right)^2
    +
    \left(v_S^q-v_{S'}^q\right)^2
    \right]
    +O(p^2),
\end{split}
\label{eq:DeltaQ-small-p}
\end{equation}
where
\begin{equation}
    u_S=\frac{1-r_S}{2},
    \qquad
    v_S=\frac{1+r_S}{2}.
\end{equation}
For any fixed \(q\), this gives
\begin{equation}
    \Delta
    =
    p\sum_v
    C_q(\bar r)\,(r_S-r_{S'})^2
    +O\!\left(p(r_S-r_{S'})^3\right)
    +O(p^2),
    \label{eq:DeltaQ-r-expand}
\end{equation}
with
\begin{equation}
\begin{split}
    C_q(\bar r)
    &=
    \frac{q^2}{8}
    \left[
    \left(\frac{1-\bar r}{2}\right)^{2q-2}
    +
    \left(\frac{1+\bar r}{2}\right)^{2q-2}
    \right], \\
    \bar r & =\frac{r_S+r_{S'}}{2}.
\end{split}
\label{eq:CQ-def}
\end{equation}
The quantity \(r_S\) is computed in Sec.~\ref{app:weak-learning}.  For two sectors \(S,S'=O(1)\), for example \(S=0\) and \(S'=1\) as in
the setup of Ref.~\cite{Majidy2023}, one has
\begin{equation}
    r_S-r_{S'}
    =
    \frac{2\left[S(S+1)-S'(S'+1)\right]}{L(L-1)}
    =
    O(L^{-2}) .
    \label{eq:rssector}
\end{equation}
The physical fidelity corresponds to $q=1/2$, where Eq.~\eqref{eq:CQ-def} continues to
\begin{equation}
    C_{1/2}(\bar r)=\frac{1}{8\left(1-\bar r^{\,2}\right)}>0 .
    \label{eq:C-half}
\end{equation}
For $S=0$ and $S'=1$, Eq.~\eqref{eq:rssector} gives $r_S-r_{S'}=-4/[L(L-1)]$, while
$\bar r=(L-2)^{2}/[2L(L-1)]\to\tfrac12$, so that $C_{1/2}\to\tfrac16$ at large $L$.
Since the sum over \(v\) within one period of the circuit gives \(O(L)\)
local vertices, we obtain
\begin{equation}
    \Delta
    \sim
    pL\,(r_S-r_{S'})^2
    =
    O\!\left(\frac{p}{L^3}\right).
\label{eq:Delta-fuzzy-final}
\end{equation}
Physically, a measured bond distinguishes the two sectors only through the difference of its singlet probability between them, proportional to \(r_S-r_{S'}\), which we call the bias. The information gained from one outcome is of second order in the bias, \(O(L^{-4})\). After the circuit average the system is homogeneous, so each of the \(O(pL)\) outcomes per time step contributes the same amount, and \(\Delta\) is their sum. The observer therefore needs \(O(L^{4})\) outcomes to tell the two sectors apart.

\subsection{The transient part: how $S_Q$ order hides the diffusive sector}
\label{subsec:sharp-learning}

We now examine the transient part in Eq.~\eqref{eq:F-decomposition}, which comes from the $z=2$ background sector. These diffusive modes exist throughout the phase diagram,
on both sides of the $S_Q$-ordering transition, so by themselves they cannot distinguish the two phases.
What distinguishes the phases, we argue in this section, is whether the \(z=2\) modes are \emph{visible} to the record
fidelity, namely whether the transient term survives the replica-connected combination in the
second line of Eq.~\eqref{eq:F-decomposition}.

In the fuzzy phase, the factorization of $a_0^{\bm S}$, enforced by the broken replica-permutation symmetry, protects the cancellation of the transient term. It is illuminating to first examine the unitary-only limit, $p\to 0$. As derived in the previous subsection, the symmetry-breaking boundary \(\langle e|\) selects the identity pairing
\(\sigma=e\) [Eq.~\eqref{eq:p0-steady-manifold}], and the leading eigenvector is the replica-factorized
vacuum of Eq.~\eqref{eq:p0-leading-eigenvector},
\begin{equation}
|b^{\bm S}_0\rangle=|\tilde\Pi_{\bm S}\rangle
=
\bigotimes_{a=1}^{2q}
|\tilde\Pi^{a,\bar a}_{S_a}\rangle,
\label{eq:factorized-vacuum}
\end{equation}
the product over replicas of the maximally mixed states within
the fixed total-spin sectors.  The saturation value
\(\mathcal T(0)\) in Eq.~\eqref{eq:F-decomposition} thus vanishes: both \(\langle e|\)
[Eq.~\eqref{eq:e-factorized}] and \(|b^{\bm S}_0\rangle\) are products over the replicas, so that
\begin{equation}
a_0^{\bm S}=\langle e|b^{\bm S}_0\rangle\langle b^{\bm S}_0|\rho_{\bm S,0}\rangle =\prod_{a=1}^{2q}c_{S_a},
\label{eq:a0-factorized}
\end{equation}
with \(c_{S_a}\) a single-replica factor depending only on the sector label, and the replica-connected
combination cancels term by term, \(\mathcal T(0)=0\).\footnote{As in the numerics of Ref.~\cite{Majidy2023}, where the state is
scrambled by unitary-only dynamics before monitoring begins, one may instead take as initial
state the corresponding steady state of the replicated unitary-only dynamics, derived in
Sec.~\ref{subsec:fuzzy-learning}:
\begin{equation}
|\rho_{\bm S,0}\rangle
=
\sum_{\sigma}\Big|P_\sigma\textstyle\prod_{a=1}^{2q}\tilde\Pi_{S_a}\Big\rangle ,
\end{equation}
where the sum runs over all pairings \(\sigma\) between replicas with the same total spin,
\(P_\sigma\) is the replica permutation of Eq.~\eqref{eq:p0-steady-manifold}, and an overall
factor \(1+O(1/d_S)\) has been dropped.  The conclusion is unchanged: in the overlap
\(\langle b^{\bm S}_0|\rho_{\bm S,0}\rangle\) the \(\sigma=e\) term dominates, all others being
suppressed by powers of \(d_S^{-1}\), exponentially small in \(L\), so the factorization over
replicas, as well as \(\mathcal T(0)=0\), holds up to exponentially small corrections.}  This is consistent with the fact that a
unitary-only circuit produces no record and can never distinguish the two spin sectors.  As we
argue below, even away from \(p=0\), in the fuzzy phase \(\mathcal T(0)\) remains bounded by
\(O(1/L)\).  In this sense the \(S_Q\) order \emph{protects} the fuzzy phase: the
order pins a pairing in which the \(S\) and \(S'\) replicas are not coupled, and the record is blind to the diffusive background.  On the other hand, in the sharp phase the \(S_Q\) symmetry is
restored: no pairing is selected, and all replicas are correlated strongly through the shared measurement record, so that \(|b^{\bm S}_0\rangle\), and
therefore \(a_0^{\bm S}\), does not factorize.  Nothing then protects the cancellation of
\(\mathcal T(0)\) that occurs in the fuzzy phase.  We now make this dichotomy quantitative.

Let us focus on $S=0$ and $S'=1$, as in the numerics of
Ref.~\cite{Majidy2023}.  We make three arguments.

(1) \emph{Within the expansion, \(\Delta=O(1/L^{3})\) in both phases.}  It is illuminating to view
\(Z^{(q)}_{\bm S}\) as a partition function whose initial boundary condition is the spin-sector assignment $\bm S$. \(\Delta\,t\) is then the bulk contribution, extensive
in spacetime, to the replica-connected combination of the free energies. Physically, \(\Delta\) is the rate at which the local \(SU(2)\)-symmetric measurements accumulate
information distinguishing the steady states of the two spin sectors.  It must therefore take the
form
\begin{equation}
\Delta\;\propto\;L\,\big(\bar O_S-\bar O_{S'}\big)^{2},
\label{eq:Delta-form}
\end{equation}
where \(\bar O_S\) is the expectation value of a local \(SU(2)\)-symmetric operator in the
sector-\(S\) steady state of a single replica.  Two features fix this form.  (a)~The
square: by definition, \(\Delta\) vanishes identically for \(S=S'\), while \(F^{(q)}_t(S,S')\) is manifestly
symmetric under \(S\leftrightarrow S'\), so a term linear in the bias, being odd under
the exchange, cannot appear.  (b)~Per unit time the dynamics performs \(O(L)\) local
measurements, giving the factor \(L\). Formally, starting from $2Q$ weakly coupled branches (the identity pairing at $p=0$, or $2Q$ ferromagnets at $p=1$) and turning on a generic interaction between pairs of branches, $V\sim\sum_{x}\sum_{a,b}O^{a}_{x}O^{\bar b}_{x}$, the cumulant expansion always returns Eq.~\eqref{eq:Delta-form} at first order in $V$.

Crucially, since \(O_x\) is a local \(SU(2)\) scalar, it cannot be linear in the spin
operator, so its expectation value depends on the sector only through the quadratic invariant, i.e., the Casimir density, which scales as \(S(S+1)/L^{2}\) in a homogeneous system (after circuit average).  For \(S,S'=O(1)\) this gives
\(\bar O_S-\bar O_{S'}=O(1/L^{2})\), and hence \(\Delta=O(1/L^{3})\) in both phases within our large-$\nTL$ controlled framework.  This is the
physics underlying the explicit weak-measurement computation of the
previous subsection, Eqs.~\eqref{eq:DeltaQ-r-expand}--\eqref{eq:Delta-fuzzy-final}.

This argument is, however, perturbative in the coupling between the replicas, and is carried out at integer \(Q>1\). We believe it is valid in the fuzzy phase, where the correlations between the replicas generated by the shared measurement record are weak. In the sharp phase, on the other hand, the numerics of Ref.~\cite{Majidy2023} show no saturation of spin sharpening after $t\sim L^{2}$, pointing to a nonperturbative enhancement of $\Delta$ to $O(1/L^{2})$ that lies beyond the scope of the present analytical treatment (see the Discussion in the main text).

(2) \emph{The saturation value \(\mathcal T(0)\) is where the two phases differ.}  In the fuzzy
phase the factorization of Eq.~\eqref{eq:a0-factorized} is exact at \(p=0\), so
\(\mathcal T(0)=0\) there.  At finite \(p\) it can be estimated as follows.  First observe that
\(\mathcal T(0)\) is the constant, boundary contribution to the replica-connected free energy, as
the extensive bulk contribution is already contained in the \(\Delta\,t\) piece, so it is
generated only within one relaxation time of the temporal boundaries, and since the slow modes
have \(z=2\), that boundary layer has temporal thickness \(O(L^{2})\).  Consequently, perturbing around the replica-factorized vacuum, the boundary contribution to the
replica-connected free energy (restricted to an \(O(L^{2})\) window in time and \(O(L)\) sites in
space) carries the same squared bias as in argument (1), so that
\begin{equation}
\mathcal T(0)\;\sim\;L^{2}\times L\times\big(\bar O_S-\bar O_{S'}\big)^{2}
\;=\;O\!\left(\frac{1}{L}\right)
\end{equation}
within the fuzzy phase.  Since this saturation value vanishes as \(L\to\infty\), learning must
wait for the steady part, \(t_{\rm learn}\sim\Delta^{-1}\sim L^{3}/p\), as found in the previous subsection.

On the other hand, this estimate relies on the \(S_Q\) order, which selects the factorized vacuum as the correct starting point and thereby protects the cancellation of $\mathcal{T}(0)$.  In the sharp phase the pairing field is disordered, and the expansion around the factorized vacuum breaks down. No symmetry then forces $|b_{\bm S}\rangle$ to factorize over the replicas, so \(a_0^{\bm S}\) no longer factorizes
as in Eq.~\eqref{eq:a0-factorized}. Physically, in the sharp phase all replicas stay correlated through the shared measurement record, and $a_0$ retains these correlations. Without the protection of the pairing order, \(\mathcal T(0)\) is therefore generically \(O(1)\). As we will see shortly, $\mathcal{T}(0)$ can be understood as discrete information stored in the longest-wavelength mode, and it receives an $O(1)$ contribution once the $z=2$ imaginary-time dynamics of the background sector has resolved that mode. 

(3) \emph{Finally, we explain why the \(O(1)\) saturation only occurs at \(t\sim L^{2}\), and not
already at \(t=O(1)\).}  The total spin information is encoded in the
uniform (\(k=0\)) component of the spin texture [cf.~Eqs.~\eqref{eq:n2-spin-constraint} and~\eqref{eq:spiral}]. A region of size \(\ell\) contains only a fraction
\((\ell/L)^{2}\) of this information: an \(SU(2)\)-symmetric operator supported on that region is sensitive to the spin-Casimir only, at order
\((\ell/L)^{2}\). In the sharp phase the background sector undergoes a $z=2$ imaginary-time evolution, with an effective inverse temperature proportional to \(t\). A mode is resolved once it has frozen, because it then contributes deterministically to the statistics of the outcomes. The modes that have frozen by time \(t\) are those of wavelength shorter than \(\ell(t)\sim\sqrt{Dt}\), with \(D\) a diffusion constant. In Eq.~\eqref{eq:block-transient} these are the modes with \(\Omega_n\gtrsim1/t\), the only ones that have contributed to \(\mathcal T(0)-\mathcal T(t)\). The record therefore resolves the background sector from short wavelengths to long. By time \(t\) the measurements can
resolve a fraction
\begin{equation}
\left(\frac{\ell(t)}{L}\right)^{2}\sim\frac{Dt}{L^{2}}
\end{equation}
of the spin-sector information, which becomes of order unity only once the diffusive dynamics
has explored the entire chain, at \(t\sim L^{2}/D\), the inverse of the smallest gap \(\Omega_{1}\). Hence in the sharp phase we have
\begin{equation}
t_{\rm learn}\sim L^{2},
\label{eq:tlearn-L2}
\end{equation}
the Thouless time of the diffusive dynamics.

\subsection{Derivation of \texorpdfstring{\(r_S\)}{rS}}
\label{app:weak-learning}

We now derive the quantity $r_S$ defined in Eq.~\eqref{eq:rS}.  Let \(s_i=\bm\sigma_i/2\).  In a total-spin-\(S\) sector,
\begin{equation}
\bm S^2=\sum_i s_i^2+2\sum_{i<j}s_i\cdot s_j
=\frac{3L}{4}+2\sum_{i<j}s_i\cdot s_j.
\end{equation}
Therefore
\begin{equation}
\left\langle\bm\sigma_i\cdot\bm\sigma_j\right\rangle_S
=\frac{4S(S+1)-3L}{L(L-1)}
\end{equation}
for any pair \(i\ne j\) in the maximally mixed state within the spin-\(S\) sector.  Here we have used the permutation symmetry among all sites. Since
\begin{equation}
R_{ij}=\frac12\left(1+\bm\sigma_i\cdot\bm\sigma_j\right),
\end{equation}
we obtain
\begin{equation}
r_S=\avg{R_{ij}}_S
=\frac{L^2-4L+4S(S+1)}{2L(L-1)}.
\end{equation}

\fi

\end{document}